\documentclass[12pt]{article}
\usepackage[utf8]{inputenc}
\usepackage[english, russian]{babel}
\usepackage{amsmath}
\usepackage{amsfonts}
\usepackage{amssymb}

\def\lb{\label}
\def\be{\begin{equation}}
\def\ee{\end{equation}}
\def\qed{\rule{5pt}{5pt}}

\newcommand{\norm}{z}
\newcommand{\normt}{z^*}
\newcommand{\normm}{|\mu^\rho\lambda_\rho|}
\newtheorem{proposition}{Proposition}

\begin{document}

\vspace{3cm}

	\begin{center}
		{\LARGE {Two-spinor description of massive particles

\vspace{0.2cm}

 and relativistic spin projection operators}}

 \vspace{1cm}

\large \sf
A.P.~Isaev$^{a,b,c,}$\footnote{\sf e-mail: isaevap@theor.jinr.ru},
M.A.~Podoinitsyn$^{a,b,c,}$\footnote{\sf e-mail: mikhailpodoinicin@gmail.com} \\

\vspace{1cm}

\begin{itemize}
\item[$^a$]
{\it Bogoliubov Laboratory of Theoretical Physics, JINR, Dubna, Russia}
\item[$^b$]
{\it State University of Dubna, University street, 19, Dubna, Russia}
\item[$^c$]
{\it St. Petersburg Department of
V.A. Steklov Institute of Mathematics of
the Russian Academy of Sciences,
27 Fontanka, St. Petersburg, Russia}
\end{itemize}
\end{center}

\vspace{2cm}
\begin{otherlanguage}{english}
\begin{abstract}
\noindent
On the basis of the Wigner unitary representations of the covering group
$ISL(2,\mathbb{C})$ of the Poincar\'{e} group, we obtain spin-tensor wave
functions of free massive particles with arbitrary spin.
The wave functions automatically satisfy the
Dirac-Pauli-Fierz equations. In the framework of the two-spinor formalism we
construct spin-vectors of polarizations and obtain conditions that fix
the corresponding relativistic spin projection operators
(Behrends-Fronsdal projection
operators). With the help of these conditions we find explicit
expressions for relativistic spin projection operators
 for integer spins (Behrends-Fronsdal
projection operators) and then find relativistic spin projection operators
  for half integer spins. These 
  projection operators
determine the nominators in the propagators of fields of
 relativistic particles.  We deduce generalizations of the Behrends-Fronsdal projection operators for
arbitrary space-time dimensions $D>2$.
\end{abstract}
\end{otherlanguage}

\vspace{3cm}

\newpage

\section{Introduction}
\setcounter{equation}0

In this paper, using the Wigner unitary representations $\;$\cite{WignI}$\;$
of the group $ISL(2,C)$, which covers the Poincaré group,
we construct spin-tensor wave functions of a special form.
These spin-tensor wave functions form spaces of
 irreducible representations of the group $ISL(2,C)$
 and automatically satisfy the Dirac-Pauli-Fierz wave equations
 \cite{Dirac36}, \cite{Fierz}, \cite{PauFir}
for free massive particles of arbitrary spin.
In our work, we use the approach set forth in the book \cite{IR}.
The construction is carried out with the help of Wigner operators
(a similar construction was developed in \cite{Weinb2}; see also
\cite{Novozhil}), which translate unitary massive
 representation of the group  $ISL(2,C)$
(induced from the irreducible representation of the stability
 subgroup $SU(2)$) acting in the space of Wigner wave functions
to a representation of the group  $ISL(2,C)$, acting in the space of
special spin-tensor fields of massive particles.
In our paper, following \cite {IR},
a special parametrization of Wigner operators is proposed,
with the help of which the momenta
of particles on the mass shell and
solutions of the Dirac-Pauli-Fierz wave equations are rewritten
 in terms of a pair of Weyl spinors
({\it two-spinor formalism}  \cite{KPTod}, \cite{Hugh}; see  also
 \cite{Fed1}, \cite{CoMa} and references therein).
The expansion of a completely symmetric Wigner wave function over
 a specially chosen basis
provides a natural recipe for describing polarizations
 of massive particles with arbitrary spins.
As the application of this formalism, a
 generalization of the Behrends-Fronsdal projection operator is constructed,
which determines the spin-tensor structures of the two-point Green function (propagator)
of massive particles with any higher spins in the case
of arbitrary space-time dimension $D$. We would also like to stress here
that spin projection operators are employed for analysis of
the high energy scattering amplitudes, differential cross sections, etc.
(\cite{BF}; see also \cite{FFR}, \cite{Chu}, \cite{ShizH} and
references therein).

 \vspace{0.2cm}

The work is organized as follows. In Section 2, we recall the
definition of the group $ISL(2,\mathbb{C})$,
 the universal covering of the Poincaré group, and
build spin-tensor wave functions
 $\psi^{_{(r)}(\dot{\beta_1}...\dot{\beta_r})}
_{(\alpha_1...\alpha_p)}(k)$ (in the momentum representation)
 for free massive particles of arbitrary spin.
 Further in this section we prove that the constructed spin-tensor wave
functions satisfy the Dirac-Pauli-Fierz equation system and show that
these wave functions have a natural parametrization in terms of
a pair of Weyl spinors ({\it two-spinor formalism}).
At the end of Section 2, we prove that the wave functions
 $\psi^{_{(r)}(\dot{\beta_1}...\dot{\beta_r})}
_{(\alpha_1...\alpha_p)}(k)$
are eigenvectors for the Casimir operator $\hat{W}_n \hat{W}^n$
of the Poincaré group ($\hat{W}_n$ are the components
of the Pauli-Lubanski vector)
with eigenvalues proportional to $j(j+1)$,
 where the parameter $j=(p+r)/2$ is called spin.
In Section 3, as examples, we discuss in detail the construction of
spin-tensor wave functions for spins
$j=1/2,1,3/2$ and $2$. In particular, we show how to derive
contributions from different polarizations
and calculate the density matrices (the relativistic
spin projectors) for particles with spins $j=1/2,1,3/2$ and $2$ as sums over the polarizations of
 quadratic combinations of polarization spin-tensors.
In Section 4, we find the general form of the polarization tensors for arbitrary integer spin $j$,
and we also establish the conditions which uniquely fix the form
 of density matrices
 (the relativistic spin projectors or the Behrends-Fronsdal
projection operators) for integer spin $j$.
In Section 5, in the case of
integer spins and arbitrary space-time dimensions $D>2$,
an explicit formula of the density matrices is derived.
 This formula is a generalization of the Behrends-Fronsdal formula
 for the projection operator known (see \cite {Fronsd},\cite {BF})
 for $D = 4$.
At the end of Section 5, an explicit expression for
 the density matrix of relativistic particles with
arbitrary half-integer spin $j$ is deduced.
 This expression will be obtained
 as the solution of the conditions to which the density matrix
  (the sum over
 the quadratic combinations of polarization spin-tensors) obeys.


\section{Massive unitary representations of the group $ISL(2,\mathbb{C})$}
\setcounter{equation}0

\subsection{Covering group $ISL(2,\mathbb{C})$ of the Poincaré group.\label{2.1}}

To fix the notation, we recall the definition of the covering group
$ISL(2, \mathbb {C})$ of the Poincare group $ISO^{\uparrow}(1,3)$ and
introduce its Lie algebra $is\ell(2,\mathbb{C})$ (see, for example, \cite {BLOT}).
 The group $ISL(2,\mathbb{C})$ is the set of all pairs
 $(A, X)$, where $A \in SL(2,\mathbb{C})$, and
  $X$ is any Hermitian $2\times 2$ matrix which
  can  always be represented in the form
  \be
  \lb{herM}
  X  = x_0 \, \sigma^0 + x_1 \, \sigma^1 +
   x_2 \, \sigma^2 + x_3 \, \sigma^3= x_k \, \sigma^k =
  \left(
\begin{array}{cc}
x_0 + x_3 \! & \! x_1 - i x_2 \\
x_1 + i x_2 \! & \! x_0 - x_3
\end{array}
\right) \; , \;\;\; x_m \in \mathbb{R} \; ,
 \ee
 \be
 \lb{pauli}
 {\scriptsize
 \sigma^0 =
\left(\!
\begin{array}{cc}
1 & 0 \\
0 & 1
\end{array}
\!\right) \equiv I_2 \, , \;\;
\sigma^1 =
\left(\!
\begin{array}{cc}
0 & 1 \\
1 & 0
\end{array}
\!\right)  , \;\;
\sigma^2 =
\left(\!
\begin{array}{cc}
0 & -i \\
i & 0
\end{array}
\!\right)  , \;\;
\sigma^3 =
\left(\!
\begin{array}{cc}
1 & 0 \\
0 & -1
\end{array}
\!\right) }.
 \ee
With the use of (\ref{herM}) each Hermitian matrix $X$ is uniquely associated
with the four-vector
$x = (x_0, x_1, x_2, x_3)$ in the Minkowski space $\mathbb {R} ^ {1,3}$ .
  Sometimes below we use the notation $(A, x)$ instead of $(A, X)$.

  The product in the group $ISL (2, \mathbb {C}) $ is given by the formula
  $$
 (A', Y') \cdot (A, Y) =
 (A' \cdot A, \; A' \cdot Y \cdot A^{\prime \, \dagger} + Y')  \; ,
 $$
which defines the rule of the $ISL(2,\mathbb{C})$ group action
in the Minkowski space $\mathbb{R}^{1,3}={\bf H}$
\begin{equation}
\label{act1}
(A, Y)\cdot  X = A \cdot X
\cdot A^{\dagger}+Y \; \in \; {\bf H} \; , \;\;\;\;\;
 \forall \; X,Y \in \bf H \; .
\end{equation}
It is obvious that the set of pairs $(A, 0)$ forms a subgroup
 $SL(2,\mathbb{C}) \subset ISL(2,\mathbb{C})$ which is
 the covering group of the Lorentz group $SO^{\uparrow}(1,3)$.
From eq. (\ref{act1}) one can deduce the action
 of the $SL(2,\mathbb{C})$ group
on the vectors $x$ in the Minkowski space $\bf H=\mathbb{R}^{1,3}$:
 \be
 \lb{lorenz}
 \begin{array}{c}
 X \; \to \; X' = A \cdot X
\cdot A^{\dagger}  \;\; \Rightarrow \;\;
 \sigma^k  x_{k}' =
 \sigma^k \; \Lambda_k^{\;\; m}(A) \; x_m
 \;\; \Rightarrow \;\;
  x_{k}' =  \Lambda_k^{\;\; m}(A) \; x_m  \; , 
  \end{array}
 \ee
 where
 \begin{equation}\label{zam1f}
X_{\alpha \dot{\beta}}=x_{k} \; \sigma^k_{\alpha \dot{\beta}}\;,
\;\;\;\;\; (\alpha,\dot{\beta} =1,2) \; ,
\end{equation}
 and the $(4 \times 4)$ matrix $||\Lambda^m_{\;\; k}(A)|| \in
 SO^{\uparrow}(1,3)$ is determined from the relations
 \be
 \lb{Alambd}
  A \cdot \sigma^m \cdot A^{\dagger} =
 \sigma^k \; \Lambda_k^{\;\; m}(A) \;\; \Leftrightarrow \;\;
 A_\xi^{\;\;\alpha} \; A_{\dot{\gamma}}^{* \;\;\dot{\beta}}
 \; \sigma^m_{\alpha \dot{\beta}} = \sigma^k_{\xi \dot{\gamma}} \;
 \Lambda_k^{\;\; m}(A)  \; .
 \ee

 \noindent
 {\bf Remark 1.}
  The matrix $||A_\xi^{\;\;\alpha}|| \in SL(2,\mathbb{C})$
  and its complex conjugate matrix $||A_{\dot{\gamma}}^{* \;\;\dot{\beta}}||$
  can be considered, respectively, as defining and its conjugate
 representations of the same element $A \in SL(2,\mathbb{C})$. It is
  known that these representations are nonequivalent.
  In order to distinguish these representations,
  we put dots over the indices of the matrices $A^*$. \\
{\bf Remark 2}. To determine the covariant product of the	
$\sigma$-matrices (\ref{pauli}), it is necessary to introduce the dual set:
 \be
 \lb{dusig}
 (\tilde{\sigma}^k)^{\dot{\beta}\alpha} =
  \varepsilon^{\alpha \xi} \;
 \varepsilon^{\dot{\beta} \dot{\gamma}} \;
 \sigma^k_{\xi \dot{\gamma}}
  \; , \;\;\;
 \tilde{\sigma}^k =
 (\sigma^0,-\sigma^1,-\sigma^2,-\sigma^3) \; .
 \ee
The symmetrized products of
 $\sigma_n$ and $\tilde{\sigma}_m$ satisfy the identities
 \be
 \lb{sigsig}
 \begin{array}{c}
(\sigma^n \, \tilde{\sigma}^m +
\sigma^m \, \tilde{\sigma}^n)_{\beta}^{\;\;\; \alpha}=
2 \, \eta^{nm} \; \delta_{\beta}^{\alpha} \; , \;\;\;
(\tilde{\sigma}^n \, \sigma^m +
\tilde{\sigma}^m \, \sigma^n)_{\;\;\; \dot{\beta}}^{\dot{\alpha}}=
2 \, \eta^{nm} \; \delta_{\dot{\beta}}^{\dot{\alpha}}
\; , \\ [0.2cm]
\eta = {\rm diag}(+1,-1,-1,-1) \; .
\end{array}
 \ee
Raising of spinor indices in (\ref{dusig}) is
 carried out by contracting
 with the antisymmetric metric
\begin{equation}
\begin{array}{c}
\lb{epsilon}
||\varepsilon^{\gamma_1 \gamma_2}||=
||\varepsilon^{\dot{\gamma}_1 \dot{\gamma}_2}||=
\begin{pmatrix}
0 \! & \! -1 \\ 1 \! & \! 0
\end{pmatrix}  ,
\;
||\varepsilon_{\gamma_1 \gamma_2}||=
||\varepsilon_{\dot{\gamma}_1 \dot{\gamma}_2}||=
\begin{pmatrix}
0 \! & \! 1 \\ -1 \! & \! 0
\end{pmatrix} , 
 \;\;
\varepsilon^{\alpha \gamma} \varepsilon_{\gamma \beta} = \delta^\alpha_\beta
\; , \;\; \varepsilon^{\dot{\alpha} \dot{\gamma}}
\varepsilon_{\dot{\gamma} \dot{\beta}} =
\delta^{\dot{\alpha}}_{\dot{\beta}} \, .
\end{array}
\end{equation}

\noindent
{\bf Remark 3.} The generators $P_n$ and $M_{mn}$
($m,n=0,1,2,3$) of the Lie algebra
$iso(1,3)$ of the Poincaré group
(and its covering $ISL(2,\mathbb{C})$) obey the
commutation relations
 \be
\lb{genP}
[P^n , \, P^m] = 0 \; , \;\;\; [P^n , \, M^{mk}] = \eta^{mn} P^k - \eta^{kn} P^m \; ,
\ee
\be
\lb{lilo2}
[M^{nm}, \, M^{k\ell}] = \eta^{mk} M^{n\ell} + \eta^{n\ell} M^{mk}
 - \eta^{nk} M^{m\ell} - \eta^{m\ell} M^{nk} \; .
\ee
The elements $P_n$ and $M_{nm}$ generate translations and Lorentz rotations in
$\mathbb{R}^{1,3}$, respectively.
We note that antisymmetrized products of the matrices
 $\sigma_n$ and $\tilde{\sigma}_m$:
 \be
\lb{snm2}
(\sigma_{nm})_{\alpha}^{\;\; \beta} =
\frac{1}{4} (\sigma_n \tilde{\sigma}_m - \sigma_m \tilde{\sigma}_n)_{\alpha}^{\;\; \beta} \; , \;\;\;
(\tilde{\sigma}_{nm})^{\dot{\alpha}}_{\;\; \dot{\beta}} =
\frac{1}{4} (\tilde{\sigma}_n \sigma_m -
\tilde{\sigma}_m  \sigma_n )^{\dot{\alpha}}_{\;\; \dot{\beta}} \; ,
\ee
satisfy the commutation relations (\ref{lilo2}) and thus realize spinor representations $\rho$: $M_{nm} \to \sigma_{nm}$ and $\tilde{\rho}$:
 $M_{nm} \to \tilde{\sigma}_{nm}$ of the
 subalgebra $so(1,3) \subset iso(1,3)$.
We stress that the tensor
 $\sigma_{nm}$ is self-dual and the tensor $\tilde{\sigma}_{nm}$
 is anti-self-dual
 \be
 \lb{asmd01}
 \frac{i}{2} \varepsilon^{k \ell nm} \sigma_{nm} =  \sigma^{k \ell} \; ,  \;\;\;\;
  \frac{i}{2} \varepsilon^{k \ell nm} \tilde{\sigma}_{nm} =
  -  \, \tilde{\sigma}^{k \ell} \; ,
 \ee
 where $\varepsilon_{mnpb}$ is a
completely antisymmetric tensor
 $(\varepsilon_{0123}=-\varepsilon^{0123}=1)$.

 Now we define the Pauli-Lubanski vector $W$ with the components
 \begin{equation}\label{plv}
W_m=\frac{1}{2}\varepsilon_{mnij}M^{ij}P^n
 \; , \;\;\;\; (m=0,1,2,3) \; .
\end{equation}
It is known that for the Lie algebra $iso(1,3)$ of the Poincaré group
with the structure relations (\ref{genP}), (\ref{lilo2}) one can define
 only two Casimir operators: $(P)^2 :=P_m P^m= \eta^{mn} P_m P_n$ and
$(W)^2 :=W_m W^m$. The eigenvalues of these operators characterize
irreducible representations of the algebra $iso(1,3)$
(and the group $ISL(2,\mathbb{C})$). The eigenvalue of the operator
 $(P)^2$ is written as ${\sf m}^2$ and for ${\sf m}^2\geq 0$ the parameter
 ${\sf m} \in \mathbb{R}$  is called mass.
The case when ${\sf m} > 0$ is called massive.
In the massive case, according to the
classification of all irreducible unitary representations of the group
  $ISL(2,\mathbb{C})$ and its Lie algebra $iso(1,3)$, the eigenvalue of the operator
 $(W)^2$ is equal to $-{\sf m}^2\, j(j+1)$,
where the parameter $j$ is called spin and can take only
non-negative integer or half-integer values  \cite{WignI}
 (see also \cite{Novozhil}, \cite{BuchKuz},
 \cite{BLOT}, and references therein).

\subsection{Spin-tensor representations of group $ISL(2,\mathbb{C})$
and Dirac-Pauli-Fierz equations\label{2.2}}

Further in this paper we will consider only the massive case when
${\sf m} >0$.
In this case the unitary irreducible
representations of the group $ISL(2,\mathbb{C})$ are characterized by spin
$j=0,\frac{1}{2},1,\frac{3}{2},\dots$ and act in the spaces of Wigner wave functions
$\phi_{(\alpha_1 \dots \alpha_{2j})}(k)$, which are components of a completely symmetric
 $SU(2)$-tensor of rank $2j$. Here
 the brackets $(.)$ in the notation of multi-index
$(\alpha_1 \dots \alpha_{2j})$ indicate the full
symmetry in permutations of indices
$\alpha_\ell$, and $k=(k_0,k_1,k_2,k_3)$
 denotes the four-momentum of a particle with mass ${\sf m}$:
$$
(k)^2 =
k^n k_n = k_r \eta^{rn} k_n = k_0^2 - k_1^2 - k_2^2 - k_3^2 = {\sf m}^2 \; , \;\;\;
$$

 Let us
fix some test momentum $q=(q_0,q_1,q_2,q_3)$
such that  $(q)^2 = {\sf m}^2$, $q_0 > 0$.
 For each momentum $k$
 belonging to the orbit $(k)^2 = {\sf m}^2$, $k^0 > 0$
 of the Lorentz transformations (\ref{lorenz})
 which transfer the test momentum $q$ to the momentum $k$,
 we choose representative $A_{(k)} \in SL(2,\mathbb{C})$:
 \begin{equation}
 \label{genK}
 (k \sigma) = A_{(k)} \; (q \sigma) \; A_{(k)}^\dagger \;\;\; \Leftrightarrow \;\;\;
 k_m = (\Lambda_k)^{\;\; n}_m \; q_n \;  ,
 \end{equation}
 where $(k \sigma) = k^n \sigma_n$, $(q \sigma) = q^n \sigma_n$.
 The relation between the matrices $A_{(k)}$ and $\Lambda_k \equiv \Lambda(A_{(k)})$
 is standard (see (\ref{Alambd})). One can rewrite the Lorentz transformation (\ref{genK}) in an equivalent form
 \be
 \lb{genK1}
 (k\,  \tilde{\sigma}) = A_{(k)}^{-1 \dagger} \cdot (q\,\tilde{\sigma}) \cdot A_{(k)}^{-1} \; ,
 \ee
 where $(k \,\tilde{\sigma}) = (k^n \tilde{\sigma}_n)$ and
 $(q \,\tilde{\sigma}) = (q^n \tilde{\sigma}_n)$.
This form of the transformation will be needed later.

Define a {\it stability subgroup}
  ({\it little group})
 $G_q \subset SL(2,\mathbb{C})$ of the momentum $q$
  as the set of matrices $A \in SL(2,\mathbb{C})$ satisfying the condition
 \begin{equation}
 \label{ssm0}
 A \cdot (q \sigma) \cdot A^\dagger = (q \sigma) \;\; \Leftrightarrow \;\;
A_\alpha^{\;\; \gamma} \; (q^n\sigma_n)_{\gamma \dot{\alpha}} \;
(A^{*})^{\;\; \dot{\alpha}}_{\dot{\gamma}}
=(q^n\sigma_n)_{\alpha \dot{\gamma}} \; ,
\end{equation}
which, by means of the identity $(q \tilde{\sigma})(q \sigma)  = {\sf m}^2$,
is equivalently rewritten as
\begin{equation}\label{ssm7}
 A = (q \, \sigma) \cdot (A^{-1})^\dagger \cdot (q \, \sigma)^{-1} =
  (q \tilde{\sigma})^{-1} \cdot (A^{-1})^\dagger \cdot (q \tilde{\sigma}) \; .
\end{equation}
In the massive case $(q)^2 = {\sf m}^2$, ${\sf m} >0$,
 one can prove that the stability subgroup $G_q$
is isomorphic to $SU(2)$ regardless of the choice of
 test momenta $q$.
 Now we note that the matrix
  $A_{(k)} \in SL(2,\mathbb{C})$, which transfers
   the test momentum $q$ to momentum $k$
   is not determined by (\ref{genK}) uniquely. Indeed, $A_{(k)}$
   can be multiplied by any element $U$ of the stability subgroup
 $G_q=SU(2)$ from the right since we have
   $$
   (A_{(k)} \cdot  U) \cdot (q\sigma) \cdot ( A_{(k)} \cdot  U)^\dagger =
    A_{(k)} \cdot  (U \cdot (q\sigma) \cdot U^\dagger)
    \cdot  A_{(k)}^\dagger = (k\sigma) \; .
   $$
For each $k$ we  fix a unique
  matrix $A_{(k)}$ satisfying (\ref{genK}). The fixed matrices $A_{(k)}$
 numerate left cosets in
 $SL(2,\mathbb{C})$ with respect to the subgroup $G_q=SU(2)$, i.e.
 they numerate points in the coset space $SL(2,\mathbb{C})/SU(2)$.

Let $T^{(j)}$ be a finite-dimensional irreducible $SU(2)$ representation
 with spin $j$, acting in the space of symmetric  spin-tensors
 of the rank $2j$ with the components $\phi_{(\alpha_1 \dots \alpha_{2j})}$.
The Wigner unitary irreducible representations $U$
of the group $ISL(2,\mathbb{C})$ with spin $j$
 are defined in \cite{WignI} (see also \cite{Novozhil}, \cite{BuchKuz},
\cite{IR} and references therein) by the following action of the element
$(A,a) \in ISL(2,\mathbb{C})$ in the space of wave functions
$\phi_{(\alpha_1,\dots,\alpha_{2j})}(k)$:
 \begin{equation}
\label{fie03}
 \begin{array}{c}
 [U(A, a) \cdot \phi]_{\bar{\alpha}}(k) \equiv
 \phi^{\, \prime}_{\bar{\alpha}}(k) =   e^{i a^m k_m} \; T^{(j)}_{\bar{\alpha} \bar{\alpha}'}
 (h_{A, \Lambda^{-1} \cdot k})
  \; \phi_{\bar{\alpha}'}(\Lambda^{-1} \cdot k)  \; .
  \end{array}
  \end{equation}
 Here we use the concise notation
 \be
 \lb{mual}
 \phi_{\bar{\alpha}}(k) \equiv \phi_{(\alpha_1 \dots \alpha_{2j})}(k) \; ,
 \ee
 the indices $\bar{\alpha}, \bar{\alpha}'$ must be understood as multi-indices
 $(\alpha_1 \dots \alpha_{2j})$, $(\alpha_1' \dots \alpha_{2j}')$,
 the matrix $\Lambda \in SO^{\uparrow}(1,3)$
 is related to $A \in SL(2,\mathbb{C})$
  by eq. (\ref{Alambd}), and the element
 \be
 \lb{fieSU}
   h_{A, \Lambda^{-1} \cdot k}  = A_{(k)}^{-1} \cdot A \cdot
   A_{(\Lambda^{-1} \cdot k)}  \;\; \in \;\; SU(2) \; ,
 \ee
 belongs to the stability subgroup $SU(2) \subset SL(2,\mathbb{C})$.
In formula (\ref{fie03})
the element $h_{A, \Lambda^{-1} \cdot k}$ of the stability subgroup
is taken in the representation $T^{(j)}$ as matrix
 $||T^{(j)}_{\bar{\alpha} \bar{\alpha}'}(h_{A, \Lambda^{-1} \cdot k})||$ which
 can be represented in a factorized form
 \be
\lb{matrT4}
\begin{array}{l}
T^{(j)}_{\bar{\beta} \bar{\alpha}}\bigl(h_{A, \Lambda^{-1} \cdot k}\bigr)  =
\Bigl[
\bigl(h_{A, \Lambda^{-1} \cdot k}\bigr)_{\beta_1}^{\;\; \alpha_1} \cdots
 \bigl(h_{A, \Lambda^{-1} \cdot k}\bigr)_{\beta_{p+r}}^{\;\; \alpha_{p+r}} \Bigr]=
\Bigl[\bigl(h_{A, \Lambda^{-1} \cdot k}\bigr)_{\beta_1}^{\;\; \alpha_1} \cdots
 \bigl(h_{A, \Lambda^{-1} \cdot k}\bigr)_{\beta_{p}}^{\;\; \alpha_{p}} \Bigr]
 \; \cdot \\ [0.2cm]
 \;\;\;\;\;  \cdot \;
 \Bigl[ \bigl( (q \tilde{\sigma})^{-1} \cdot
 h^{\dagger \; -1}_{A, \Lambda^{-1} \cdot k} \cdot (q \tilde{\sigma}) \bigr)_{
 \beta_{p+1}}^{\;\; \alpha_{p+1}} \cdots
 \bigl((q \tilde{\sigma})^{-1} \cdot
 h^{\dagger \; -1}_{A, \Lambda^{-1} \cdot k} \cdot (q \tilde{\sigma}) \bigr)_{
 \beta_{p+r}}^{\;\; \alpha_{p+r}} \Bigr] \; .
  \end{array}
\ee
 Here we split the tensor product of
  $2j=(p+r)$ factors $h_{A, \Lambda^{-1} \cdot k}$
  into two groups. The first group consists of the
 $p$ factors $h_{A, \Lambda^{-1} \cdot k}$, and in the second
 group we use the identity (\ref{ssm7}) and write
  $r$ multipliers $h_{A, \Lambda^{-1} \cdot k}$
in the form  $\bigl((q \tilde{\sigma})^{-1} \cdot
 h^{\dagger \; -1}_{A, \Lambda^{-1} \cdot k} \cdot (q \tilde{\sigma}) \bigr)$.
Further, we substitute (\ref{fieSU}) into (\ref{matrT4}) and
 split the result as follows:
  \be
\lb{matrT3}
\begin{array}{c}
T^{(j)}_{\bar{\beta} \bar{\alpha}}\bigl(h_{A, \Lambda^{-1} \cdot k}\bigr)
 = \bigl(A_{(k)}^{-1}\bigr)^{\;\; \kappa_1 ... \kappa_p}_{\beta_1 ... \beta_p}
 \; \bigl(A \bigr)^{\;\; \gamma_1 ... \gamma_p}_{\kappa_1 ... \kappa_p}
 \; \bigl(A_{(\Lambda^{-1}\cdot k)}\bigr)^{\;\;
 \alpha_1 ... \alpha_p}_{\gamma_1 ... \gamma_p}
 \;\; \cdot \\ [0.2cm]
 \cdot \;  \bigl((q \tilde{\sigma})^{-1} \;
 A_{(k)}^{\dagger}\bigr)_{\beta_{p+1} ... \beta_{p+r};
 \dot{\kappa}_{p+1} ... \dot{\kappa}_{p+r}}
 \; \bigl(A^{\dagger -1} \bigr)_{\;\; \dot{\gamma}_{p+1} ... \dot{\gamma}_{p+r}}
 ^{\dot{\kappa}_{p+1} ... \dot{\kappa}_{p+r}}
 \; \bigl(A^{\dagger -1}_{(\Lambda^{-1}\cdot k)}(q \tilde{\sigma})\bigr)^{\dot{\gamma}_{p+1} ... \dot{\gamma}_{p+r};
 \alpha_{p+1} ... \alpha_{p+r}} \; ,
 \end{array}
\ee
where we introduced the concise notation
\be
\lb{matrT1}
\begin{array}{c}
\bigl( X \bigr)_{\alpha_1...\alpha_p}^{\;\; \beta_1...\beta_p}  =
X_{\alpha_1}^{\;\; \beta_1} \cdots
X_{\alpha_{p}}^{\;\; \beta_{p}} , \;\;
\bigl( Y \bigr)^{\dot{\alpha}_1...\dot{\alpha}_r}
_{\;\; \dot{\beta}_1...\dot{\beta}_r}  =
Y^{\dot{\alpha}_1}_{\;\; \dot{\beta}_1} \cdots
Y^{\dot{\alpha}_{r}}_{\;\; \dot{\beta}_{r}} , \;\; \\ [0.3cm]
\bigl(Z\bigr)^{\dot{\alpha}_1...\dot{\alpha}_r;
\beta_1...\beta_r}  =
Z^{\dot{\alpha}_1\beta_1} \cdots
Z^{\dot{\alpha}_{r}\beta_{r}} \; .
\end{array}
\ee
In the operator form the matrix (\ref{matrT3}) is represented as
\be
\lb{matrT}
\begin{array}{c}
T^{(j)}\bigl(h_{A, \Lambda^{-1} \cdot k}\bigr) =
\bigl( A_{(k)}^{-1} \cdot  A \cdot
   A_{(\Lambda^{-1} \cdot k)} \bigr)^{\otimes p} \otimes
   \bigl( (q \tilde{\sigma})^{-1} \,
 A_{(k)}^{\dagger} \cdot  A^{\dagger -1} \cdot
 A^{\dagger -1}_{(\Lambda^{-1}\cdot k)} \,  (q \tilde{\sigma})
  \bigr)^{\otimes r} = \\ [0.2cm]
 = \Bigl( A_{(k)}^{\otimes p} \otimes
 \bigl(A^{\dagger -1}_{(k)}(q \tilde{\sigma})
 \bigr)^{\otimes r} \Bigr)^{-1} \Bigl(  A^{\otimes p} \otimes
 \bigl(A^{\dagger -1} \bigr)^{\otimes r} \Bigr)
   \Bigl( A_{(\Lambda^{-1} \cdot k)}^{\otimes p}
 \otimes \bigl(A^{\dagger -1}_{(\Lambda^{-1}\cdot k)}(q \tilde{\sigma})
 \bigr)^{\otimes r} \Bigr) \; ,
 \end{array}
\ee
 Now we use factorized representation
(\ref{matrT}) for the matrix $T^{(j)}\bigl(h_{A, \Lambda^{-1} \cdot k}\bigr)$
and rewrite the $ISL(2,\mathbb{C})$-transformation (\ref{fie03})
 in the form
\begin{equation}
\lb{fie33}
 \begin{array}{c}
 [U(A, a) \cdot \psi^{_{(r)}}]^{(\dot{\beta_1}...\dot{\beta_r})}
 _{(\alpha_1...\alpha_p)}(k) =  e^{i a^m k_m} \;\; A^{\;\; \gamma_1...\gamma_p}_{\alpha_1...\alpha_p} \;\;
 \bigl(A^{\dagger -1} \bigr)^{\dot{\beta_1}...\dot{\beta_r}}
 _{\;\; \dot{\kappa}_1...\dot{\kappa}_r}
  \;\; \psi^{_{(r)}(\dot{\kappa}_1...\dot{\kappa}_r)}
  _{(\gamma_1...\gamma_p)}(\Lambda^{-1} \cdot k)  \; ,
  \end{array}
  \end{equation}
where instead of the Wigner wave functions
 $\phi_{(\delta_1... \delta_{p+r})}(k)$
we introduced spin-tensor wave functions
 of $(\frac{p}{2},\frac{r}{2})$-type
 (with $r$ dotted and $p$ undotted indices):
\begin{equation}
 \label{tp}
\psi^{_{(r)}(\dot{\beta_1}...\dot{\beta_r})}
_{(\alpha_1...\alpha_p)}(k)=\frac{1}{{\sf m}^r}
(A_{(k)})^{\;\;\; \delta_1 ... \delta _p}_{\alpha_1 ... \alpha_p}
\cdot
 \bigl(A^{-1\dagger}_{(k)}\cdot (q \tilde{\sigma})
 \bigr)^{\dot{\beta}_{p+1}...\dot{\beta}_{p+r};
 \delta _{p+1}...\delta _{p+r}}
 \phi_{(\delta _1...\delta _p \delta_{p+1} ...\delta_{p+r} )}(k) \; .
\end{equation}
The upper index $(r)$ of the spin-tensors $\psi^{_{(r)}}$
 distinguishes these spin-tensors with respect to the number of dotted indices.
 The operators $A_{(k)}^{\otimes p} \otimes
 \bigl(A^{\dagger -1}_{(k)}(q \tilde{\sigma})
 \bigr)^{\otimes r}$, used in (\ref{tp}) to translate the Wigner wave functions into spin-tensor functions of
 $(\frac{p}{2},\frac{r}{2})$-type, are called
 {\it the Wigner operators}.

In the massive case, the test momentum can be conveniently chosen
 in the form $q=({\sf m},0,0,0)$.
Then, relations (\ref{genK}) and (\ref{genK1}) for
the elements $A_{(k)},A_{(k)}^{\dagger -1}\in SL(2,\mathbb{C})$ are
written as:
\begin{equation}\label{ssm1}
{\sf m} \,
(A_{(k)})_\alpha^{\;\; \beta}\; (\sigma_0)_{\beta \dot{\gamma}}=
(\sigma_n k^n)_{\alpha \dot{\beta}}
(A^{\dagger-1}_{(k)})^{\dot{\beta}}_{\;\; \dot{\gamma}}
\;\; \Leftrightarrow \;\;
 {\sf m} \, (A^{\dagger-1}_{(k)})^{\dot{\beta}}_{\;\; \dot{\gamma}} \;
(\tilde{\sigma}_0)^{\dot{\gamma} \alpha}=
(\tilde{\sigma}_n k^n)^{\dot{\beta} \gamma}(A_{(k)})^{\;\;\alpha}_\gamma \, .
\end{equation}
 Since $\sigma_0$ and $\tilde{\sigma}_0$ are the unit matrices,
 eqs. (\ref{ssm1}) have a concise form:
\begin{equation}\label{ssm2}
{\sf m} \, A_{(k)}=(\sigma_n k^n) \, A^{\dagger-1}_{(k)}
\Leftrightarrow {\sf m} \, A^{\dagger-1}_{(k)}=
 (\tilde{\sigma}_n k^n) \, A_{(k)} \; ,
\end{equation}
 but we should stress here that
 the balance of dotted and undotted indices
 is violated in (\ref{ssm2}).
\begin{proposition}\label{prop1}. Let us choose the test momentum
 as $q=({\sf m},0,0,0)$. Then the wave functions
 $\psi^{_{(r)}}$
defined in (\ref{tp}) satisfy the Dirac-Pauli-Fierz equations
\cite{Dirac36}, \cite{Fierz}, \cite{PauFir}:
\begin{equation}\label{osudp}
\begin{array}{l}
k^m(\tilde{\sigma}_m)^{\dot{\gamma}_1 \alpha_1}\psi^{_{(r)}(\dot{\beta_1}...\dot{\beta_r})}_{(\alpha_1...\alpha_p)}(k)=
{\sf m} \, \psi^{_{(r+1)}(\dot{\gamma}_1 \dot{\beta_1}...\dot{\beta_r})}_{(\alpha_2...\alpha_p)}(k) \; ,
\;\;\;\; (r=0,\dots,2j-1) \; , \\[0.2cm]
k^m(\sigma_m)_{\gamma_1 \dot{\beta}_1}\psi^{_{(r)}(\dot{\beta_1}...\dot{\beta_r})}_{(\alpha_1...\alpha_p)}(k)
= {\sf m} \, \psi^{_{(r-1)}(\dot{\beta_2}...\dot{\beta_r})}
_{(\gamma_1 \alpha_1...\alpha_p)}(k) \; ,
\;\;\;\; (r=1,\dots,2j) \; ,
\end{array}
\end{equation}
which describe the dynamics of a massive relativistic
 particle with spin $j = (p+r)/2$.
The compatibility conditions for the system of equations
 (\ref{osudp}) are given by the mass shell relations
 $(k^n k_n -m^2) \, \psi^{_{(r)}}(k)=0$.
\end{proposition}
{\bf Proof.} The proof of the
 first equation in (\ref{osudp}) is given by the chain of relations:
$$
\begin{array}{c}
k^m(\tilde{\sigma}_m)^{\dot{\gamma}_1 \alpha_1}\; \psi^{_{(r)}(\dot{\beta_1}.. \dot{\beta_r})}_{(\alpha_1...\alpha_p)}(k)=k^m(\tilde{\sigma}_m)^{\dot{\gamma}_1 \alpha_1}(A_{(k)})^{\delta^\prime_1}_{\alpha_1}
(A_{(k)})^{\delta^\prime_2...\delta^\prime_p}_{\alpha_2...\alpha_p}
\; \cdot \\[0.2cm]
\cdot \; (A^{-1 \dagger}_{(k)} \;
\tilde{\sigma}_0)^{\dot{\beta}_{p+1}...\dot{\beta}_{p+r};
\delta_{p+1}^\prime...\delta_{p+r}^\prime} \;
\phi_{(\delta^\prime_1...\delta^\prime_p \delta_{p+1}^\prime...\delta_{p+r}^\prime)}(k)
={\sf m} \, (A^{-1 \dagger}_{(k)} \; \tilde{\sigma}_0)^{\dot{\gamma}_1 \delta^\prime_1}
 (A_{(k)})^{\;\; \delta^\prime_2...\delta^\prime_p}_{\alpha_2...\alpha_p} \cdot
\\[0.2cm]
 (A^{-1 \dagger}_{(k)} \; \tilde{\sigma}_0)^{\dot{\beta}_{p+1}...\dot{\beta}_{p+r};
\delta_{p+1}^\prime...\delta_{p+r}^\prime} \; \phi_{(\delta^\prime_1...
 \delta_{p+r}^\prime)}(k) = {\sf m} \, \psi^{_{(r+1)}(\dot{\gamma}_1 \dot{\beta_1}...\dot{\beta_r})}_{(\alpha_2...\alpha_p)}(k),
\end{array}
$$
where we applied
the second formula of (\ref{ssm1}) and used the symmetry of the
  Wigner wave functions
$\phi_{(\delta^\prime_1... \delta_{p+r}^\prime)}(k)$  with respect to
 any permutations of indices $\delta^\prime_k$.
The second equation in (\ref{osudp}) is proved analogously (we need to use
 the first relation in (\ref{ssm1})). The consistence conditions for
 the system (\ref{osudp}) follow from the chain of relations
 \be
 \lb{k2m2}
 \begin{array}{c}
 (k^n k_n) \; \psi^{_{(r)}(\dot{\beta_1}...\dot{\beta_r})}
 _{(\tau \alpha_2...\alpha_p)}(k) =
(k \sigma)_{\tau \dot{\gamma}_1}(k \tilde{\sigma})^{\dot{\gamma}_1 \alpha_1}
\, \psi^{_{(r)}(\dot{\beta_1}...\dot{\beta_r})}_{(\alpha_1...\alpha_p)}(k) =
\\[0.2cm]
= {\sf m} \; k^n(\sigma_n)_{\tau \dot{\gamma}_1}\psi^{_{(r+1)}(\dot{\gamma}_1 \dot{\beta_1}...\dot{\beta_r})}_{(\alpha_2...\alpha_p)}(k) =
{\sf m}^2 \;
\psi^{_{(r)}(\dot{\beta_1}...\dot{\beta_r})}_{(\tau \alpha_2...\alpha_p)}(k) \;  ,
 \end{array}
 \ee
where we used the identities (\ref{sigsig}) and equations
 (\ref{osudp}).
 Comparing the left and right parts in (\ref{k2m2}),
 we obtain the mass-shell condition
 $(k^nk_n - {\sf m}^2)\psi^{_{(r)}}(k)=0$. \hfill \qed

 \vspace{0.2cm}

 We now return back to the discussion of the
  matrices $A_{(k)} \in SL(2,\mathbb{C})$
 which, according to (\ref{genK}), transfer the test momentum
  $q$ to the momentum $k$. The matrices $A_{(k)}$
 numerate points of the coset space
 $SL(2,\mathbb{C})/SU(2)$. The left action of the group $SL(2,\mathbb{C})$
on the coset space $SL(2,\mathbb{C})/SU(2)$ is given by the formula
 \be
 \lb{fieSU1}
 A \cdot A_{(k)}  = A_{(\Lambda \cdot k)} \cdot U_{A,k} \; ,
 \ee
where the matrices $A \in SL(2,\mathbb{C})$ and $\Lambda \in SO^{\uparrow}(1,3)$
 are related by condition (\ref{Alambd}) and the element
 $U_{A,k} \in SU(2)$ depends on the matrix $A$ and momentum $k$.
Under this action the point $A_{(k)} \in SL(2,\mathbb{C})/SU(2)$
is transformed to the point $A_{(\Lambda \cdot k)} \in SL(2,\mathbb{C})/SU(2)$.
We note that formula (\ref{fieSU1}) is equivalent to the definition
 (\ref{fieSU}) of the element $h_{A, k}$
 of the stability subgroup in Wigner's representation (\ref{fie03}).

The left action (\ref{fieSU1})
of the element $A \in SL(2,\mathbb{C})$ transforms
 two columns of the matrix $A_{(k)}$ as Weyl spinors.
Therefore, it is convenient to represent the matrix $A_{(k)}$
 by using two Weyl spinors $\mu$ and $\lambda$ with components
 $\mu_\alpha, \; \lambda_\alpha$
(the matrix $A^{\dagger}_{(k)}$
 will be correspondingly expressed in terms of the conjugate
spinors $\overline{\mu}$ and $\overline{\lambda}$)
in the following way \cite{IR}:
\begin{equation}
 \label{achs}
 \begin{array}{c}
(A_{(k)})_{\alpha}^{\;\; \beta}=
 \frac{1}{z}
\begin{pmatrix} \mu_1 & \lambda_1 \\ \mu_2 & \lambda_2 \end{pmatrix} \; , \;
(A^{\dagger -1}_{(k)})^{\dot{\alpha}}_{\;\; \dot{\beta}} =
\frac{1}{z^*}
\begin{pmatrix} \overline{\lambda}_{\dot{2}} & -\overline{\mu}_{\dot{2}} \\ -\overline{\lambda}_{\dot{1}} & \overline{\mu}_{\dot{1}} \end{pmatrix}
 , \;\;\; \\ [0.4cm]
(z)^2 = \mu^\rho \, \lambda_\rho \, , \;\; (z^{*})^{2} =
\overline{\mu}^{\dot{\rho}}\, \overline{\lambda}_{\dot{\rho}} \, ,
\end{array}
\end{equation}

\begin{equation}
 \label{vs1}
\mu=
\begin{pmatrix}
\mu_1 \\ \mu_2
\end{pmatrix} \; ,
\;
\lambda=
\begin{pmatrix}
\lambda_1 \\ \lambda_2
\end{pmatrix} \; ,
\;
\overline{\mu} =
\begin{pmatrix}
\overline{\mu}_{\dot{1}} \\ \overline{\mu}_{\dot{2}}
\end{pmatrix} \; ,
\;
\overline{\lambda} =
\begin{pmatrix}
\overline{\lambda}_{\dot{1}} \\ \overline{\lambda}_{\dot{2}}
\end{pmatrix} \, , \;\;\;\;
\overline{\lambda}_{\dot{\alpha}}=(\lambda_\alpha)^* \, , \;\;\;\;
\overline{\mu}_{\dot{\alpha}}=(\mu_\alpha)^* \; .
\end{equation}
In eqs. (\ref{achs}) we fix the normalization of matrices
 $A_{(k)},A^{\dagger -1}_{(k)} \in SL(2,\mathbb{C})$
 so that $\det(A_{(k)})=1=\det(A^{\dagger -1}_{(k)})$.
From formulas (\ref{ssm2}) it follows that the momentum $k$ is expressed in terms of the spinors
 $\mu_{\alpha} ,  \lambda_{\beta}, \overline{\mu}_{\dot{\beta}}, \overline{\mu}^{\dot{\beta}}$ as follows:
\begin{equation}\label{ichs1}
\frac{\sf m}{\normm}(\mu_{\alpha}\overline{\mu}_{\dot{\beta}}+
\lambda_\alpha\overline{\lambda}_{\dot{\beta}})=
(k^n\sigma_n)_{\alpha\dot{\beta}} \; , \;\;\;
\frac{\sf m}{\normm}(\mu^{\alpha}\overline{\mu}^{\dot{\beta}}
 +\lambda^\alpha\overline{\lambda}^{\dot{\beta}})
  =(k^n\widetilde{\sigma}_n)^{\dot{\beta}\alpha} \; ,
\end{equation}
where $\normm= z \, z^*$. Thus, in view of (\ref{tp})
 and (\ref{ichs1}) the wave functions of massive
 relativistic particles, which are functions of the four-momentum
 $k$, can be considered as functions of two Weyl
  spinors $\lambda$ and $\mu$.
The two-spinor expression (\ref{ichs1}) for
 the four-vector $k$ ($k^2={\sf m}^2$ and $k_0 > 0$)
is a generalization of the well-known twistor representation for
 momentum $k$ of a massless particle \cite{PenrMac}.
We will see below that
the two-spinor description of massive particles (about
two-spinor formalism see also papers
 \cite{KPTod} --
 \cite{BrBiD}, \cite{CoMa})
 based on the representation (\ref{ichs1})
proves to be extremely convenient in describing
polarization properties of massive particles with arbitrary spin $j$.
In the next Section, we apply this two-spinor
formalism to describe relativistic particles with spins $j=1/2,1,3/2$ and $j=2$.

\vspace{0.2cm}

At the end of this Subsection, we demonstrate why,
in the case of $p+r=2j$, the system of spin-tensor wave functions (\ref{tp}), which
 obey the Dirac-Pauli-Fierz  equations (\ref{osudp}),
 does describe relativistic particles with spin $j$.

Take the representation (\ref{fie33}) of the group $ISL(2,\mathbb{C})$,
acting in the space of spin-tensor wave functions
 $\psi^{_{(r)}(\dot{\beta}_1...\dot{\beta}_r)}
 _{(\alpha_1...\alpha_p)}(k)$, and consider
 this representation in the special case when the element
 $(A,a) \in ISL(2,\mathbb{C})$ is close to the unit element $(I_2,0)$,
 i.e., we fix the vector $a=0$ and take the matrices
 $A,A^{\dagger -1} \in SL(2,\mathbb{C})$ such that
 $$
A_{\alpha}^{\;\; \beta} = \delta_{\alpha}^{\;\; \beta} +
\frac{1}{2} (\omega^{nm} \sigma_{nm})_{\alpha}^{\;\; \beta} + \dots \; ,
\;\;\;\;\;
(A^{\dagger-1})^{\dot{\alpha}}_{\;\; \dot{\beta}} =
\delta^{\dot{\alpha}}_{\;\; \dot{\beta}} +
\frac{1}{2} (\omega^{nm} \tilde{\sigma}_{nm})^{\dot{\alpha}}_{\;\; \dot{\beta}} + \dots \; ,
 $$
where $\omega^{nm} = - \omega^{mn}$ are small real parameters, and
 the matrices $\sigma_{nm}$ and $\tilde{\sigma}_{nm}$
  are the spinor representations (\ref{snm2})
 of the generators $M_{nm} \in so(1,3)$.
 As a result, we obtain for (\ref{fie33}) the expansion
 \be\lb{SLor1}
  \bigl[U\bigl(I +\frac{1}{2} \omega^{mn}\sigma_{mn}+... \, , \; 0\bigr) \cdot \psi^{_{(r)}}\bigr]^{\bar{\dot{\beta}}}
 _{\bar{\alpha}}(k) =   {\psi^{_{(r)}}}^{\bar{\dot{\beta}}}_{\bar{\alpha}}(k)
  +  \frac{1}{2} \; \omega^{mn} \; (M_{mn})^{\bar{\dot{\beta}} \; \bar{\gamma}}
  _{\bar{\alpha}\; \bar{\dot{\kappa}}}
  \;\; {\psi^{_{(r)}}}^{\bar{\dot{\kappa}}}
  _{\bar{\gamma}}(k)  + \dots \; ,
 \ee
 where we used the notation $\bar{\alpha}$ and
 $\bar{\dot{\beta}}$ for multi-indices $(\alpha_1...\alpha_p)$ and
 $(\dot{\beta_1}...\dot{\beta_r})$. In eq. (\ref{SLor1}) the
 operators $M_{mn} = \bigl( (k_m \frac{\partial}{\partial k^n} -
  k_n \frac{\partial}{\partial k^m}) + \hat{\Sigma}_{mn}\bigr)$ are the
 generators of the algebra $so(1,3)$ in the representation $U$ and
 the matrices
  \be
 \lb{genPLS}
  \hat{\Sigma}_{mn} = \sum\limits_{a=1}^p (\sigma_{mn})_a
  +   \sum\limits_{b=1}^r (\tilde{\sigma}_{mn})_b \; ,
 \ee
describe the spin contribution to the components of the angular momentum $M_{mn}$.
 The operators
 $(\sigma_{mn})_a$ and $(\tilde{\sigma}_{mn})_b$
 in (\ref{genPLS}) are defined as follows:
 $$
 \begin{array}{c}
 [(\sigma_{mn})_a
 \psi^{_{(r)}}]^{\bar{\dot{\beta}}}
 _{\bar{\alpha}}(k) =
 (\sigma_{mn})_{\alpha_a}^{\;\; \gamma_a} \;
 {\psi^{_(r)}}^{\bar{\dot{\beta}}}
 _{\bar{\alpha}_a}(k) \; , \;\;\;\; 
 [(\tilde{\sigma}_{mn})_b
 \psi^{_{(r)}}]^{\bar{\dot{\beta}}}
 _{\bar{\alpha}}(k) =
 (\tilde{\sigma}_{mn})
 ^{\dot{\beta}_b}_{\;\; \dot{\gamma}_b}\;
 {\psi^{_(r)}}^{\bar{\dot{\beta}}_b}
 _{\bar{\alpha}}(k) \; ,
 \end{array}
 $$
 where we used the notation $\bar{\alpha}_a = (\alpha_1...\alpha_{a-1}\gamma_a\alpha_{a+1}
 ...\alpha_p)$
 and  $\bar{\dot{\beta}}_b =(\dot{\beta_1}...\dot{\beta}_{b-1}\dot{\gamma}_b
\dot{\beta}_{b+1}...\dot{\beta_r})$.
 \begin{proposition}\label{W2pq}.
Spin-tensor wave functions $\psi_{(\alpha_1 \dots \alpha_p)}
^{_{(r)}(\dot{\beta}_1 \dots \dot{\beta}_r)}(k)$
 of type $(\frac{p}{2},\frac{r}{2})$,
which obey the Dirac-Pauli-Fierz equations (\ref{osudp}),
automatically satisfy the equations
  \be
 \lb{genPW2j}
 [(\hat{W}^m \, \hat{W}_m) \; \psi]_{(\alpha_1 \dots \alpha_p)}
 ^{_{(r)}(\dot{\beta}_1 \dots \dot{\beta}_r)}(k) = - {\sf m}^2 j(j+1) \;
 \psi_{(\alpha_1 \dots \alpha_p)}^{_{(r)}(\dot{\beta}_1 \dots \dot{\beta}_r)}(k) \; ,
 \ee
 where $j = (\frac{p}{2} + \frac{r}{2})$, $\hat{W}_m$
are the components of the Pauli-Lubanski vector (\ref{plv}) and
$\hat{W}_m \hat{W}^m$ is the casimir operator for the group $ISL(2,\mathbb{C})$.
 \end{proposition}
 {\bf Proof.}
 The components (\ref{plv}) of the Pauli Lubanski vector
 are equal to
 \begin{equation}\lb{plv1}
W_m=\frac{1}{2}\varepsilon_{mnij}M^{ij}P^n
= \frac{1}{2}\varepsilon_{mnij}\hat{\Sigma}^{ij}P^n  \; ,
\end{equation}
 and depend only on the spin part (\ref{genPLS})
of the components $M^{ij}$ of the total angular momentum.
To prove formula (\ref{genPW2j}), it is convenient to write down the
symmetrized spin-tensor
 wave functions $\psi^{_{(r)}(\dot{\beta_1}...\dot{\beta_r})}
 _{(\alpha_1...\alpha_p)}(k)$ in the form of generating functions
 \be
 \lb{psixu}
 \psi^{_{(r)}}(k; u, \; \bar{u}) =
 \psi_{(\alpha_1 \dots \alpha_p)}^{_{(r)}(\dot{\beta}_1 \dots \dot{\beta}_r)}(k)
 \cdot
 u^{\alpha_1} \cdots u^{\alpha_p} \cdot \bar{u}_{\dot{\beta}_1} \cdots
 \bar{u}_{\dot{\beta}_r}  \; ,
 \ee
 where $u^{\alpha}$ and $\bar{u}_{\dot{\beta}}$ are
auxiliary Weyl spinors. Then the
  action of the spin operators (\ref{genPLS}) on the functions
  $\psi^{_{(r)}(\dot{\beta_1}...\dot{\beta_r})}
 _{(\alpha_1...\alpha_p)}(k)$
is equivalent to the action of the differential operators
  \be
 \lb{Sigmuu}
 \hat{\Sigma}_{mn}(u, \; \bar{u}) =
 u^\alpha \; (\sigma_{mn})_\alpha^{\;\; \beta} \; \partial_{\beta} +
 \bar{u}_{\dot{\beta}} \;
  (\tilde{\sigma}_{mn})_{\;\;  \dot{\alpha}}^{\dot{\beta}} \; \bar{\partial}^{\dot{\alpha}}
  \; ,
 \ee
  on the generating functions (\ref{psixu}). In eq. (\ref{Sigmuu})
 we have used the notation
 $\partial_{\beta} =\partial_{u^\beta}$
 and $\bar{\partial}^{\dot{\beta}} = \partial_{\bar{u}_{\dot{\beta}}}$.
We set $W_m(u,\bar{u})
= \frac{1}{2}\varepsilon_{mnij}P^n \hat{\Sigma}^{ij}(u,\bar{u})$
 and use the identity
 \be
\lb{centWS}
 \hat{W}_n(u,\bar{u}) \, \hat{W}^n(u,\bar{u}) =
 - \frac{1}{2}
  \bigl(\hat{\Sigma}_{nm}(u,\bar{u}) \, \hat{\Sigma}^{nm}(u,\bar{u}) \bigr)\; \hat{P}^2  +
\bigl(\hat{\Sigma}_{nm}(u,\bar{u}) \hat{P}^n \bigr)
\bigl(\hat{\Sigma}^{km}(u,\bar{u}) \hat{P}_k \bigr) \; ,
\ee
which is obtained from (\ref{plv1}) by direct computation.
The terms  $\bigl(\hat{\Sigma}_{nm} \, \hat{P}^n \bigr)
\bigl(\hat{\Sigma}^{km} \, \hat{P}_k \bigr)$
and $\bigl(\hat{\Sigma}_{nm} \, \hat{\Sigma}^{nm} \bigr)$
 in the right-hand side of (\ref{centWS}) are reduced to the forms:
  \be
\lb{centWS1}
 \begin{array}{c}
 \hat{\Sigma}_{nm}(u,\bar{u}) \; \hat{\Sigma}^{nm}(u,\bar{u}) =
 2 \, (u^\alpha   \partial_{\alpha}) + (u^\alpha   \partial_{\alpha})^2 +
  2 \, (\bar{u}_{\dot{\alpha}} \bar{\partial}^{\dot{\alpha}}) +
  (\bar{u}_{\dot{\alpha}} \bar{\partial}^{\dot{\alpha}})^2
  \equiv D(u,\bar{u}) \; ,
 \end{array}
 \ee
 \be
\lb{centWS2}
 \bigl(\hat{\Sigma}_{nm} \, \hat{P}^n \bigr)
\bigl(\hat{\Sigma}^{km} \, \hat{P}_k \bigr) =
  \frac{\hat{P}^2}{4} \Bigl( D(u,\bar{u})
     - 2\, (u^\alpha   \partial_{\alpha})
 (\bar{u}_{\dot{\gamma}} \bar{\partial}^{\dot{\gamma}}) \Bigr) +
 (\hat{P} \, \sigma)_{\alpha \dot{\gamma}}
 \, (\hat{P} \, \tilde{\sigma})^{\dot{\delta} \beta}
 u^\alpha \partial_{\beta} \; \bar{u}^{\dot{\gamma}}
 \bar{\partial}_{\dot{\delta}} \; ,
 \ee
 where $(\hat{P} \sigma)=(\hat{P}_n \sigma^n)$ and
 $(\hat{P} \tilde{\sigma})=(\hat{P}_n \tilde{\sigma}^n)$.
 To obtain (\ref{centWS1}) and (\ref{centWS2}), it is necessary
 to apply identities of the following type:
 \be
 \lb{stsg1}
 (\sigma_m)_{\alpha_1 \dot{\beta}_1} \; (\sigma^m)_{\alpha_2 \dot{\beta}_2} =
  2 \, \varepsilon_{\alpha_1 \alpha_2} \,
  \varepsilon_{\dot{\beta}_1 \dot{\beta}_2}    \; .
 \ee
 The substitution of (\ref{centWS1}) and
 (\ref{centWS2}) into the right-hand side of (\ref{centWS}) gives
 \be
 \lb{Wubu}
 \bigl(\hat{W}(u,\bar{u})\bigr)^2 =
 - \frac{\hat{P}^2}{4} \Bigl( (u^\alpha   \partial_{\alpha} +
 \bar{u}_{\dot{\gamma}} \bar{\partial}^{\dot{\gamma}})^2 +
      2\, (u^\alpha   \partial_{\alpha} +
 \bar{u}_{\dot{\gamma}} \bar{\partial}^{\dot{\gamma}}) \Bigr) +
 ( (\hat{P} \, \sigma)_{\alpha}^{\;\; \dot{\gamma}}
 \, u^\alpha   \bar{u}_{\dot{\gamma}})
 \, ((\hat{P} \, \tilde{\sigma})_{\dot{\delta}}^{\;\; \beta}
 \partial_{\beta}  \bar{\partial}^{\dot{\delta}}) .
 \ee
Finally, the result of the action of the operator (\ref{Wubu})
 on the generating function (\ref{psixu})
of the spin-tensor fields of $(\frac{p}{2},\frac{r}{2})$-type
can be calculated directly and we have
  \be
 \lb{sk05w}
 \hat{W}^2 \; \psi^{_{(r)}}(x,u,\bar{u}) = - \frac{{\sf m}^2}{4} \;
  \left( (p + r)^2 + 2(p+r)  \right) \;
  \psi^{_{(r)}}(x,u,\bar{u}) \; ,
 \ee
 that is equivalent to (\ref{genPW2j}) for
 $j = \frac{p+r}{2}$. To obtain (\ref{sk05w}),
we used the equality
 $\hat{P}_n\, \psi^{_{(r)}}(k,u,\bar{u}) =
 k_n \, \psi^{_{(r)}}(k,u,\bar{u})$,
 the Dirac-Pauli-Fierz equations (\ref{osudp}),
 and the relations:
 $$
 (u^\alpha   \partial_{\alpha})
\, \psi^{_{(r)}}(k,u,\bar{u}) = p \, \psi^{_{(r)}}(k,u,\bar{u}) \; , \;\;\;
 (\bar{u}_{\dot{\gamma}} \bar{\partial}^{\dot{\gamma}}) \,
 \psi^{_{(r)}}(k,u,\bar{u}) =
 r \, \psi^{_{(r)}}(k,u,\bar{u}) \; ,
 $$
 following from the calculation of
  the degree of homogeneity of polynomials (\ref{psixu}). \hfill \qed

  \vspace{0.2cm}

  Formula (\ref{genPW2j}) shows that the spin-tensor wave functions
 $\psi^{_{(r)}}(k)$ are eigenvectors for the
 Casimir operator $(\hat{W}^m \, \hat{W}_m)$ of the algebra $iso(1,3)$.
 According to Remark {\bf 3} of
 Section {\bf \ref{2.1}}, these wave functions generate
 the space of unitary representation of the Lie algebra
 $iso(1,3)$ (of the group $ISL(2,\mathbb{C})$) with spin $j$.

In Section {\bf \ref{3examp}},
we consider in detail four special examples
$j=1/2$, $j=1$, $j=3/2$ and $j=2$
of the general construction presented above.
 These examples are important from the point of view of applications
 in physics.

\section{Spin-tensor representations of group $ISL(2,\mathbb{C})$
  for $j=1/2,1,3/2$ and $2$ \label{3examp}}
 \setcounter{equation}0

\subsection{Spin $j=1/2$.\label{2.2jp}}

According to the general construction developed in Subsection
{\bf \ref{2.2}}, the
unitary Wigner representation (\ref{fie03}) of the group $SL(2,\mathbb{C})$
with spin $j=1/2$
is realized in the space of $SU(2)$ spinors $\phi$:
\begin{equation}\label{vtve1}
\phi(k)= \begin{pmatrix}
\phi_1(k) \\ \phi_2(k)
\end{pmatrix} =
\phi_1(k)
\epsilon^{+}
+ \phi_2(k)\epsilon^{-} , \;\;\;\;
 \epsilon^{+}=\begin{pmatrix}
1 \\ 0
\end{pmatrix}, \; \epsilon^{-}=\begin{pmatrix}
0 \\ 1
\end{pmatrix} \; ,
\end{equation}
where the components $\phi_{\alpha}(k)$ are functions of the $4$-momentum $k$.
 Further it is convenient to write these components as follows:
\begin{equation}
 \lb{psm03}
\phi_\beta(k)=\phi_1(k) \, \epsilon^{+}_\beta +
\phi_2(k) \, \epsilon^{-}_\beta \; .
\end{equation}
Then, taking into account (\ref{tp}) and (\ref{achs}), we obtain that
for $j=1/2$ the spin-tensor representation (\ref{fie33})
 is realized in the spaces of Weyl spinors:
\begin{equation}
\label{psm01}
\psi^{(0)}_\alpha(k)
=(A_{(k)})^{\; \beta}_\alpha \phi_\beta
 =\frac{1}{\norm}(\phi_{(1)}\mu_\alpha+\phi_{(2)}\lambda_\alpha)
 \; ,
\end{equation}
\begin{equation}
\label{psm02}
\psi^{(1)\dot{\alpha}}(k)=
(A_{(k)}^{\dagger -1}\tilde{\sigma}_0)
^{\dot{\alpha}\beta} \phi_\beta
 = \frac{1}{\normt}(\phi_{(2)}\overline{\mu}{}^{\dot{\alpha}}-
 \phi_{(1)}\overline{\lambda}{}^{\dot{\alpha}}) \; ,
\end{equation}
 where $\tilde{\sigma}_0$ is the unit matrix $I_2$ (see (\ref{dusig}))
and we used the identities that follow from (\ref{achs}):
 \be
 \lb{psm05}
\begin{array}{c}
 \mu_\alpha = z \, (A_{(k)})^{\; \beta}_\alpha \,
 \epsilon^{+}_\beta  \, , \;\;
 \lambda_\alpha = z \, (A_{(k)})^{\; \beta}_\alpha \, \epsilon^{-}_\beta \, , \;\;
\\[0.2cm]
 \overline{\mu}{}^{\dot{\alpha}} = z^* \,
 (A_{(k)}^{\dagger -1}\; \tilde{\sigma}_0)^{\dot{\alpha} \beta} \, \epsilon^{-}_\beta \, , \;\;
 \overline{\lambda}{}^{\dot{\alpha}} = - z^* \,
 (A_{(k)}^{\dagger -1}\; \tilde{\sigma}_0)^{\dot{\alpha} \beta} \,
  \epsilon^{+}_\beta \, .
\end{array}
 \ee
As it was shown in Proposition
{\bf \em \ref{prop1}}, the Weyl spinors $\psi^{(0)}_\alpha(k)$, $\psi^{(1)\dot{\alpha}}(k)$
satisfy the system of equations
(a particular case of equations (\ref{osudp})):
\begin{equation}\label{bwe}
\begin{array}{c}
(k^n\sigma_n)_{\gamma \dot{\alpha}}\psi^{_{(1)}\dot{\alpha}}=
{\sf m}\psi^{_{(0)}}_\gamma \; , \;\;\;\;
(k^n\tilde{\sigma}_n)^{\dot{\alpha}\gamma }\psi^{_{(0)}}_{\gamma}={\sf m}\psi^{_{(1)}\dot{\alpha}}  \; .
\end{array}
\end{equation}
It is well known that this system
is equivalent to the Dirac equation:
\begin{equation}\label{fde}
(k^n\gamma_n - {\sf m})\Psi(k) =0, \;
\end{equation}
where $\gamma_n$ are the Dirac $(4 \times 4)$ matrices and $\Psi$ is the Dirac bispinor:
 \begin{equation}\label{fde1}
 \gamma_n=\begin{pmatrix}
0 & \sigma_n \\
\tilde{\sigma}_n & 0
\end{pmatrix} \; , \;\;\;
\Psi(k) =\begin{pmatrix}
\psi^{_{(0)}}_\alpha(k) \\ \psi^{_{(1)}\dot{\beta}}(k)
\end{pmatrix} \; .
 \end{equation}
We note that the verification of equations (\ref{bwe}) in the
 framework of the two-spinor formalism (\ref{ichs1}), (\ref{psm01}),
 (\ref{psm02}) is a simple, purely algebraic, exercise
\begin{equation}
\begin{array}{c}
{\displaystyle
\frac{\sf m}{\normm\normt}
(\mu_{\alpha}\overline{\mu}_{\dot{\beta}}+
\lambda_\alpha\overline{\lambda}_{\dot{\beta}})
(\phi_{(2)}\overline{\mu}{}^{\dot{\beta}}-
\phi_{(1)}\overline{\lambda}{}^{\dot{\beta}})=
\frac{\sf m}{\norm}(\phi_{(1)}\mu_\alpha+
\phi_{(2)}\lambda_\alpha) \; , } \\ [0.2cm]
{\displaystyle \frac{\sf m}{\normm\norm}
(\mu^{\alpha}\overline{\mu}{}^{\dot{\beta}}
+\lambda^\alpha\overline{\lambda}{}^{\dot{\beta}})
(\phi_{(1)}\mu_\alpha+\phi_{(2)}\lambda_\alpha)=
\frac{\sf m}{\normt}(\phi_{(2)}\overline{\mu}{}^{\dot{\beta}}-
\phi_{(1)}\overline{\lambda}{}^{\dot{\beta}}) } \; ,
\end{array}
\end{equation}
since here we only used the properties of the commuting Weyl spinors:
 $\lambda_\alpha \lambda^\alpha = 0$,
 $\overline{\mu}_{\dot{\beta}} \overline{\mu}^{\dot{\beta}}= 0$, etc.

Taking into account (\ref{psm01}) and (\ref{psm02}),
the spinor $\Psi$ given in ({\ref{fde1}) can be written in the form
 \be
 \lb{psm04}
 \Psi(k) =\phi_1(k)\, e^{(+)}+\phi_2(k)\, e^{(-)} \; ,
 \ee
 where we introduced two Dirac spinors:
\begin{equation}
 \lb{epem}
e^{(+)}=\begin{pmatrix}
\frac{1}{\norm} \mu_\alpha \\[0.1cm]
-\frac{1}{\normt} \overline{\lambda}{}^{\dot{\beta}}
\end{pmatrix} \; , \;
e^{(-)}=\begin{pmatrix}
\frac{1}{\norm}\lambda_{\beta} \\[0.1cm] \frac{1}{\normt}\overline{\mu}{}^{\dot{\alpha}}
\end{pmatrix} \; .
\end{equation}
We denote the components of these spinors as $e^{(+)}_A$ and
$e^{(-)}_A$ ($A=1,2,3,4$) and
 define Dirac adjoint spinors with components $\overline{e}^{(+)A}$
  and $\overline{e}^{(-)A}$ in the standard way:
\begin{equation}
\overline{e}^{(+)}=(e^{(+)})^{\dagger}\gamma_0=\begin{pmatrix}
- \frac{1}{\norm} \, \lambda^\alpha \; , &
\frac{1}{\normt} \, \overline{\mu}_{\dot{\alpha}}
\end{pmatrix} \; , \;\;\;
\overline{e}^{(-)}=(e^{(-)})^{\dagger}\gamma_0=\begin{pmatrix}
\frac{1}{\norm}\, \mu^\alpha \; , &
\frac{1}{\normt} \, \overline{\lambda}_{\dot{\alpha}}
\end{pmatrix} \; .
\end{equation}
In view of (\ref{achs}) it is easy to verify that spinors (\ref{epem})
are normalized as follows:
\begin{equation}
\lb{prh7}
\overline{e}^{(+)A} \, e^{(-)}_A=
\overline{e}^{(-)A} \, e^{(+)}_A = 0  \;\; , \;\;\;\;
\overline{e}^{(+)A} \, e^{(+)}_A=
\overline{e}^{(-)A} \, e^{(-)}_A = 2  \; .
\end{equation}

We note that the coefficients
 $\phi_1(k)$ and $\phi_2(k)$ of the Wigner wave function (\ref{psm03})
  (in the representation space of the stability subgroup
 $SU(2)$ with spin $j=1/2$) correspond
 to the projections $+1/2$ and $-1/2$ of the operator of the third spin component
 $S_3 = \frac{1}{2}\sigma_3$. Therefore, comparing the expansions
 (\ref{psm03}) and (\ref{psm04}),  it is natural to interpret the spinors
 $e^{(+)}$ and $e^{(-)}$ in the
 expansion of the Dirac spinor $\Psi(k)$
 as ''vectors'' of polarization of a particle with spin $j=1/2$.

The sum over all polarizations
($+1/2$ and $-1/2$) of the products of
components of the spinors $e^{(\pm)}$ and $\overline{e}^{(\pm)}$
gives the $4 \times 4$ matrix:
\begin{equation}\label{prh1}
\begin{array}{c}
( \Theta^{(1/2)} )^B_{A} = \frac{1}{2}
\Bigl( e^{(+)}_A \, \overline{e}^{(+)B} +
e^{(-)}_A \, \overline{e}^{(-)B} \Bigr) =
\frac{1}{2} \begin{pmatrix}
\delta_\alpha^\beta & \frac{1}{\normm}(\mu_{\alpha}\overline{\mu}_{\dot{\beta}}
+\lambda_{\alpha}\overline{\lambda}_{\dot{\beta}}) \\
\frac{1}{\normm}(\mu^{\beta}\overline{\mu}{}^{\dot{\alpha}}+
\lambda^\beta\overline{\lambda}{}^{\dot{\alpha}}) & \delta_{\dot{\alpha}}^{\dot{\beta}}
\end{pmatrix} =\\[0.6cm]
\displaystyle{
=\frac{1}{2{\sf m}}\begin{pmatrix}
{\sf m}\delta_\alpha^\beta  &
 (k^n\sigma_n)_{\alpha\dot{\beta}}\\
(k^n \widetilde{\sigma}_n)^{\dot{\alpha}\beta} &
{\sf m} \delta_{\dot{\alpha}}^{\dot{\beta}}
\end{pmatrix}
=\frac{1}{2{\sf m}}(k^n \, \gamma_n + m\, I_4) \; , }
\end{array}
\end{equation}
where we used the identities
$(\lambda_\alpha\mu^\beta-\mu_\alpha\lambda^\beta)
=\delta_\alpha^\beta \, (\mu^\rho\lambda_\rho)$ and
$(\overline{\lambda}_{\dot{\alpha}}
\overline{\mu}{}^{\dot{\beta}}-
\overline{\mu}_{\dot{\alpha}}\overline{\lambda}{}^{\dot{\beta}})=
\delta_{\dot{\alpha}}^{\dot{\beta}} \,
(\overline{\mu}{}^{\dot{\rho}}\overline{\lambda}_{\dot{\rho}})$.
The matrix $\Theta^{(1/2)}$ is sometimes called the density
 matrix of a particle with spin $j = \frac{1}{2}$. This matrix by construction
(in view of relations (\ref{prh7}))
is a projection operator:
 $(\Theta^{(1/2)})^2=\Theta^{(1/2)}$. Moreover, expression (\ref{prh1})
determines the numerator of the propagator for massive
   particles with spin $j=1/2$.

\vspace{0.2cm}

\subsection{Spin $j=1$.\label{2.2j1}}

As it was shown in Section {\bf \ref{2.2}},
the unitary representation of the group $ISL(2,\mathbb{C})$
with spin $j=1$ acts in the space of spin-tensor wave functions
$\psi^{_{(2)}\dot{\alpha} \dot{\beta}}(k)$,
$\psi_{\alpha}^{_{(1)}\dot{\beta}}(k)$ and  $\psi^{_{(0)}}_{\alpha \beta}(k)$.
These functions satisfy the system
of Dirac-Pauli-Fierz equations (\ref{osudp}):
\begin{equation}\label{u1}
(k \tilde{\sigma})^{\dot{\beta}_1 \alpha_1}\psi_{(\alpha_1 \alpha_2)}^{_{(0)}}(k)={\sf m}\psi_{\alpha_2}^{_{(1)} \dot{\beta}_1}(k) \; ,
\end{equation}
\begin{equation}\label{u2}
(k \tilde{\sigma})^{\dot{\beta}_1 \alpha_2}\psi_{\alpha_2}^{_{(1)} \dot{\beta}_2}={\sf m}\psi^{_{(2)} \dot{\beta}_1 \dot{\beta}_2}(k) \; ,
\end{equation}
\begin{equation}\label{u3}
(k \sigma)_{ \alpha_1 \dot{\beta}_1}\psi^{_{(2)} \dot{\beta}_1 \dot{\beta}_2}(k)={\sf m}\psi_{\alpha_1}^{_{(1)} \dot{\beta}_2}(k) \; ,
\end{equation}
\begin{equation}\label{u4}
(k \sigma)_{ \alpha_1 \dot{\beta}_2}\psi_{\alpha_2}^{_{(1)} \dot{\beta}_2}(k)={\sf m}\psi_{(\alpha_1 \alpha_2)}^{_{(0)}}(k) \; .
\end{equation}
\begin{proposition}\label{prop3}
 The system of equations
  (\ref{u1})--(\ref{u4}) is equivalent to the Proca equation for the
 massive vector field:
\begin{equation}
\lb{proka}
k^n\Bigl(k_n A_m(k)-k_m A_n(k)\Bigr) - {\sf m}^2A_m(k)=0,
\end{equation}
where the vector field components $A_m(k)$ are defined by the spin-tensor
wave function $\psi^{_{(1)}}(k)$:
\begin{equation}\label{vp12}
A_m(k)=\frac{1}{2}\, (\sigma_m)_{\alpha \dot{\beta}} \,
\varepsilon^{\alpha \gamma}\, \psi_{\gamma}^{_{(1)} \dot{\beta}}(k) \; .
\end{equation}
\end{proposition}
{\bf Proof.} The spin-tensor wave functions $\psi_{(\alpha_1 \alpha_2)}^{_{(0)}}(k)$, $\psi_\alpha^{_{(1)} \dot{\beta}}$,
 $\psi^{_{(2)} \dot{\beta}_1 \dot{\beta}_2}(k)$ are
in one-to-one correspondence with the vector-tensors
$F^{(+)}_{mn}(k)$, $A_{m}(k)$,
 $F^{(-)}_{mn}(k)$ in the Minkowski space:
\begin{equation}\label{st0}
\psi^{_{(0)}}_{(\alpha \beta)}(k)=(\sigma_{mn})_{\alpha \beta}F^{(+)mn} (k) \; ,
\end{equation}
\begin{equation}\label{st1}
\psi^{_{(1)} \dot{\beta}}_{\alpha}(k)=\varepsilon^{\dot{\beta} \dot{\gamma}} (\sigma^{m})_{\alpha \dot{\gamma} }A_m(k) \; ,
\end{equation}
\begin{equation}\label{st3}
\psi^{_{(2)} (\dot{\alpha} \dot{\beta)}}(k)=(\tilde{\sigma}_{mn})^{\dot{\alpha} \dot{\beta}}F^{(-)mn}(k) \; ,
\end{equation}
where the matrices $\sigma_{mn}, \; \tilde{\sigma}_{mn}$  were
 introduced in (\ref{snm2}) and we use them in the forms
$$
(\sigma_{mn})_{\alpha\beta} = \varepsilon_{\beta\gamma}  \;
(\sigma_{mn})_{\alpha}^{\;\; \gamma} \; , \;\;\;\;
 (\tilde{\sigma}_{mn})^{\dot{\alpha}\dot{\beta}} =
  \varepsilon^{\dot{\beta}\dot{\gamma}} \;
 (\tilde{\sigma}_{mn})^{\dot{\alpha}}_{\;\;\dot{\gamma}}\; ,
$$
which are symmetric under permutations
$\alpha \leftrightarrow \beta$ and
 $\dot{\alpha} \leftrightarrow \dot{\beta}$. Note that equation
 (\ref{st1}) is equivalent to the definition (\ref{vp12}).
We also note that in view of the properties
 (\ref{asmd01}) of the matrices $\sigma_{mn}$
and $\tilde{\sigma}_{mn}$ the antisymmetric vector-tensors
 $F^{(+)}_{mn}(k)$ and $F^{(-)}_{mn}(k)$ are self-dual
  and anti-self-dual, respectively:
 \be
 \lb{asmd03}
 \frac{i}{2} \varepsilon^{k \ell nm} F^{(+)}_{nm} =  F^{(+) k \ell} \; ,  \;\;\;\;
  \frac{i}{2} \varepsilon^{k \ell nm} F^{(-)}_{nm} = -  \,
  F^{(-) k \ell} \; .
 \ee
Now we substitute (\ref{st0}) and (\ref{st1}) into
equation (\ref{u1})
\begin{equation}\label{vu1}
k^b(\tilde{\sigma}_b)^{\dot{\beta} \beta}(\sigma_{mn})_{\alpha \beta}
F^{(+)mn}(k)={\sf m}
\varepsilon_{\alpha \gamma} (\tilde{\sigma}^{r})^{\dot{\beta} \gamma }A_r(k),
\end{equation}
then multiply both sides of (\ref{vu1}) by
$\varepsilon^{\delta \alpha} (\sigma_p)_{\delta \dot{\beta}}$
and contract the indices $\alpha$ and $\dot{\beta}$.
 As a result, we have the relation
\begin{equation}\label{vu3}
k^b \; {\rm Tr}(\tilde{\sigma}_b \, \sigma_{mn} \, \sigma_p)
\; F^{(+)mn}(k) = {\sf m} \,
{\rm Tr}(\tilde{\sigma}^{r}\, \sigma_p) \, A_r(k) \; ,
\end{equation}
which in view of the identities
${\rm Tr}(\tilde{\sigma}^{r}\, \sigma_p) = 2 \delta^r_p$,
${\rm Tr}(\tilde{\sigma}_b \, \sigma_{mn}\sigma_p)=
(\eta_{np}\eta_{mb}-\eta_{nb}\eta_{mp} - i \, \varepsilon_{mnpb})$
and (\ref{asmd03})
is written as the equation
\begin{equation}\label{u1v}
 2 \, k^\ell \, F^{(+)}_{\ell p}(k) ={\sf m} \, A_p(k) \; .
\end{equation}
In the same way, the remaining equations of the system (\ref{u1})-(\ref{u4})
can be transformed into the following equations:
\begin{equation}\label{u2v}
k_\ell \, A_p (k)-k_p  \, A_\ell(k) + i \, \varepsilon_{\ell pbr} \, k^b \, A^r(k)
 = 2 \, {\sf m} \, F^{(+)}_{\ell p}(k) \; ,
\end{equation}
\begin{equation}\label{u3v}
k_\ell \, A_p (k)-k_p  \, A_\ell(k) - i \, \varepsilon_{\ell pbr} \, k^b \, A^r(k)
 = 2 \, {\sf m} \, F^{(-)}_{\ell p}(k) \; ,
\end{equation}
\begin{equation}\label{u4v}
2 \, k^\ell \,  F^{(-)}_{\ell p}(k) ={\sf m} \; A_p(k) \; .
\end{equation}
It follows from equations
 (\ref{u2v}) and (\ref{u3v}) that
 $F^{(+)}_{\ell p}(k)$ and $F^{(-)}_{\ell p}(k)$
 are expressed in terms of the vector field components $A_r(k)$.
Summing equations (\ref{u2v}) and (\ref{u3v}), we obtain
  $$
 k_\ell \, A_p (k)-k_p  \, A_\ell(k) = F_{\ell p}(k) \; ,
  $$
  where we introduced the notation $F_{\ell p}(k) =
  {\sf m} \, (F^{(+)}_{\ell p}(k) + F^{(-)}_{\ell p}(k))$
 for the strength of the vector field $A_r(k)$. In view of
 (\ref{asmd03}) the tensors
  $({\sf m} \, F^{(+)}_{\ell} p)$ and $({\sf m} \, F^{(-)}_{\ell} p)$
  determine the self-dual and anti-self-dual
parts of the stress tensor $F_{\ell p}(k)$, respectively.
Finally, substituting expression (\ref{u2v}) for $F^{(+)}_{\ell p}(k)$ into (\ref{u1v})
 (or expression (\ref{u3v}) for $F^{(-)}_{\ell p}(k)$ into (\ref{u4v})),
 we deduce the Proca equation (\ref{proka})
which describes the dynamics of free spin-1 particles with mass ${\sf m}$.
  \hfill \qed

\vspace{0.2cm}

According to formula (\ref{tp}),
the spin-tensor wave functions
  $\psi_\alpha^{_{(1)} \dot{\beta}}$, which
  are related to the vector fields $A_m$
  (see equations (\ref{vp12}) and (\ref{st1})),
   are determined by the corresponding Wigner wave function $\phi$:
\begin{equation}\label{ns11}
\psi_\alpha^{_{(1)} \dot{\beta}}(k) =\frac{1}{\sf m} \;
(A_{(k)})^{\gamma_1}_\alpha \; (A^{\dagger -1}_{(k)} \,
 (q\tilde{\sigma}))^{\dot{\beta} \gamma_2} \; \phi_{(\gamma_1 \gamma_2)}(k) \; .
\end{equation}
 In the space of $SU(2)$ symmetric tensors of the second rank,
 we introduce the normalized basis vectors
 $$
 \epsilon^{(+1)} = \epsilon^+\otimes \epsilon^+ \; , \;\;\;
 \epsilon^{(0)} = \frac{1}{\sqrt{2}} \, (
\epsilon^+ \otimes \epsilon^-
+ \epsilon^-\otimes \epsilon^+) \; , \;\;\;
 \epsilon^{(-1)} = \epsilon^- \otimes \epsilon^- \; ,
 $$
 where spinors
 $\epsilon^+$ and $\epsilon^-$ were defined in (\ref{vtve1}),
and expand the symmetric Wigner wave function $\phi(k)$
with the components $\phi_{(\alpha_1 \alpha_2)}(k)$
 over these basis vectors
\begin{equation}\label{tpen1}
\phi(k)=
  \phi_{(11)}(k) \, \epsilon^{(+1)}
 +  \sqrt{2} \; \phi_{(12)}(k) \,  \epsilon^{(0)}   + \phi_{(22)}(k) \,
 \epsilon^{(-1)} \; .
\end{equation}
Note that $\epsilon^{(+1)}$, $\epsilon^{(0)}$,
$\epsilon^{(-1)}$  are eigenvectors with eigenvalues
 $+1$, $0$, $-1$ of the operator $S_3$
of the third spin component in the representation
of the group $SU(2)$ for spin $j=1$.

Now we substitute the decomposition (\ref{tpen1}) into (\ref{ns11}) and fix
 the test momentum as $q=({\sf m},0,0,0)$. As a result,
 expansion (\ref{ns11}) is written as
\begin{equation}\label{psiA}
 \psi_\alpha^{_{(1)} \dot{\beta}}=
 \Bigl( \phi_{(11)}(k) \, \overset{_{(+)}}{e}{}_\alpha^{\;\dot{\beta}}(k)+
 \sqrt{2} \; \phi_{(12)}(k) \, \overset{_{(0)}}{e}{}_\alpha^{\;\dot{\beta}}(k)+
 \phi_{(22)}(k) \, \overset{_{(-)}}{e}{}_\alpha^{\;\dot{\beta}}(k) \Bigr) \; ,
\end{equation}
where, according to (\ref{achs}),
 the components of the spin-tensors
 $\overset{_{(+)}}{e}(k),\;\overset{_{(0)}}{e}(k)$ and $\overset{_{(-)}}{e}(k)$
have the following form:
\begin{equation}\label{kew1}
\overset{_{(+)}}{e}{}_\alpha^{\;\dot{\beta}}(k)=(A_{(k)})^{\gamma_1}_\alpha(A^{\dagger -1}_{(k)}\; \tilde{\sigma}_0)^{\dot{\beta} \gamma_2}
\epsilon^+_{\gamma_1}\epsilon^+_{\gamma_2}=
-\frac{\mu_\alpha\overline{\lambda}{}^{\dot{\beta}}}{\norm\normt} \; ,
\end{equation}
\begin{equation}\label{kew2}
\overset{_{(0)}}{e}{}_\alpha^{\;\dot{\beta}}(k)=
 (A_{(k)})^{\gamma_1}_\alpha(A^{\dagger -1}_{(k)}\; \tilde{\sigma}_0)^{\dot{\beta} \gamma_2}
\frac{1}{\sqrt{2}}(\epsilon^+_{\gamma_1}\epsilon^-_{\gamma_2}+
\epsilon^-_{\gamma_1}\epsilon^+_{\gamma_2})=\frac{1}{\sqrt{2}}
\frac{(\mu_\alpha\overline{\mu}{}^{\dot{\beta}}
-\lambda_\alpha\overline{\lambda}{}^{\dot{\beta}})}{\norm\normt} \; ,
\end{equation}
\begin{equation}\label{kew3}
\overset{_{(-)}}{e}{}_\alpha^{\;\dot{\beta}}(k)=(A_{(k)})^{\gamma_1}_\alpha(A^{\dagger -1}_{(k)}\; \tilde{\sigma}_0)^{\dot{\beta} \gamma_2}
\epsilon^-_{\gamma_1}\epsilon^-_{\gamma_2}=
\frac{\lambda_\alpha\overline{\mu}{}^{\dot{\beta}}}{\norm\normt} \; .
\end{equation}

In the space of spin-tensors
 $\psi_\alpha^{_{(1)} \dot{\beta}}$ we define the Hermitian scalar product
\begin{equation}
\lb{scalpr}
(\xi^{_{(1)}}, \, \psi^{_{(1)}})= \overline{\xi}_{\dot{\beta}}^{_{(1)} \alpha}
\, \psi_\alpha^{_{(1)} \dot{\beta}} =
 (\psi^{_{(1)}}, \, \xi^{_{(1)}})^* \; ,
\end{equation}
where $\overline{\xi}_{\dot{\beta}}^{_{(1)} \alpha}
 =(\xi_{\beta}^{_{(1)} \dot{\alpha}})^*$ is the complex conjugate
 spin-tensor.
 With respect to the scalar product (\ref{scalpr}) the spin-tensors
 (\ref{kew1}) -- (\ref{kew3}) form the orthonormal system:
 \begin{equation}
 \label{dgfs2}
(\overset{_{(+)}}{e} \, , \, \overset{_{(+)}}{e})=
(\overset{_{(0)}}{e} \, , \, \overset{_{(0)}}{e})=
(\overset{_{(-)}}{e} \, , \, \overset{_{(-)}}{e})=1 \; ,\;\;\;
(\overset{_{(\pm)}}{e} \, , \, \overset{_{(0)}}{e})=
(\overset{_{(+)}}{e} \, , \, \overset{_{(-)}}{e})=0 \; .
\end{equation}
 One can deduce
 relations (\ref{dgfs2}) by using the
 normalization (\ref{achs}) of the spinors  $\lambda$ and $\mu$.

Let us find the expansion of the vector potential $A_m(k)$
 over the components $\phi_{(\alpha\beta)}$
of the Wigner wave function $\phi$.
To do this, we substitute the expansion (\ref{psiA})
 into formula (\ref{vp12}) and obtain
\begin{equation}\label{raA}
A_m= \frac{1}{\sqrt{2}} \left(\phi_{(11)}(k) \,
{\sf e}^{(+)}_m(k)+  \sqrt{2} \, \phi_{(12)}(k) \, {\sf e}^{(0)}_m(k)
+\phi_{(11)}(k) \, {\sf e}^{(-)}_m(k) \right) \; ,
\end{equation}
where the vectors
\begin{equation}\label{fpsl1}
\begin{array}{c}
{\sf e}^{(+)}_m(k)=\frac{1}{\sqrt{2}} \, (\sigma_m)_{\alpha \dot{\beta}} \,
 \varepsilon^{\alpha \gamma} \, \overset{_{(+)}}{e}{}_\gamma^{\;\dot{\beta}}(k)
 \; ,\;\;\;
{\sf e}^{(0)}_m(k)=\frac{1}{\sqrt{2}} \, (\sigma_m)_{\alpha \dot{\beta}} \, \varepsilon^{\alpha \gamma} \, \overset{_{(0)}}{e}{}_\gamma^{\;\dot{\beta}}(k) \; ,
\\ [0.2cm]
{\sf e}^{(-)}_m(k)=\frac{1}{\sqrt{2}} \, (\sigma_m)_{\alpha \dot{\beta}} \,
\varepsilon^{\alpha \gamma} \, \overset{_{(-)}}{e}{}_\gamma^{\;\dot{\beta}}(k) \; ,
\end{array}
\end{equation}
can be naturally interpreted (bearing in mind that the
 components  $\phi_{(11)},\phi_{(12)},\phi_{(22)}$ of the Wigner
 wave function (\ref{tpen1}) correspond to the projections
 $+1,0,-1$  of the spin generator $S_3$ of the stability
 subgroup $SU(2)$) as polarization
vectors of a massive vector particle.

The polarization vectors
${\sf e}^{_{(+)}}(k), {\sf e}^{_{(0)}}(k),{\sf e}^{_{(-)}}(k)$ with
components (\ref{fpsl1}) are orthonormal and transverse to the momentum $k$.
 The normalization and orthogonality of these vectors follow
from the conditions (\ref{dgfs2})
for the corresponding spin-tensors.
Indeed, for any two complex four-vectors
 ${\sf e}_m$ and ${\sf e}_m'$, which
 are related  to spin-tensors
  $e_{\gamma}^{\;\dot{\beta}}$ and $e_{\gamma}^{\,\prime\;\dot{\beta}}$
  by means of equations (\ref{fpsl1}), we have the following identity
   between two scalar products:
\begin{equation}
\lb{ideneE}
\begin{array}{c}
{\sf e}_m \; \overline{\sf e}^{\, \prime \, m} = \frac{1}{2}
 \, (\sigma_m)_{\alpha_1 \dot{\beta}_1} \, (\sigma^m)_{\beta_2 \dot{\alpha}_2}
  \, \varepsilon^{\alpha_1 \gamma_1}
\, e_{\gamma_1}^{\;\; \dot{\beta}_1} \,
\varepsilon^{\dot{\alpha}_2 \dot{\gamma}_2} \,
 \overline{e}_{\dot{\gamma}_2}^{\; \prime \, \beta_2} = \\[0.5cm]
=\varepsilon_{\alpha_1\beta_2} \, \varepsilon_{\dot{\beta}_1 \dot{\alpha}_2} \,
 \varepsilon^{\alpha_1 \gamma_1}\, \varepsilon^{\dot{\alpha}_2 \dot{\gamma}_2}
\, e_{\gamma_1}^{\;\dot{\beta}_1} \,
 \overline{e}_{\dot{\gamma}_2}^{\;\prime \, \beta_2} =
- \delta^{\gamma_1}_{\beta_2} \,
\delta^{\dot{\gamma}_2}_{\dot{\beta}_1}\, e_{\gamma_1}^{\;\dot{\beta}_1} \,
 \overline{e}_{\dot{\gamma}_2}^{\, \prime \;\beta_2} = - (e' \, , \, e) \; ,
\end{array}
\end{equation}
where we put $\overline{\sf e}_{m}=({\sf e}_{m})^*$
and applied formula (\ref{stsg1}). Then from identity (\ref{ideneE}) and the normalization (\ref{dgfs2}) of the spin-tensors
$\overset{_{(+)}}{e}(k),\;\overset{_{(0)}}{e}(k)$ and $\overset{_{(-)}}{e}(k)$
we deduce
 \be
 \lb{norme}
{\sf e}^{_{(+)} m} \,  \overline{\sf e}^{_{(+)}}_{m}=
{\sf e}^{_{(0)} m} \,  \overline{\sf e}^{_{(0)}}_{m} =
{\sf e}^{_{(-)} m} \,  \overline{\sf e}^{_{(-)}}_{m} = - 1 \; ,\;\;\;
{\sf e}^{_{(\pm)} m} \,  \overline{\sf e}^{_{(0)}}_{m}=
{\sf e}^{_{(+)} m} \,  \overline{\sf e}^{_{(-)}}_{m} =0 \; .
 \ee

The property that the vectors ${\sf e}^{_{(\pm)}}_m$,
${\sf e}^{_{(0)}}_m$ are  transverse to the momentum $k_m$
can be easily proved in the framework of the two-spinor formalism by using the representations (\ref{ichs1}) and (\ref{kew1}) --  (\ref{kew3}).
For example, the transversality of the vector ${\sf e}^{_{(+)}}_m$
follows from the chain of equalities:
\begin{equation}
k^m {\sf e}^{_{(+)}}_m(k)=\frac{1}{\sqrt{2}}\, (k^m\sigma_m)_{\alpha \dot{\beta}}
\, \varepsilon^{\alpha \gamma}\,
 \overset{_{(+)}}{e}{}_\gamma^{\;\dot{\beta}}(k)=
-\frac{{\sf m}}{\sqrt{2}\, {\norm}^2{\normt}^2} \,
(\mu_{\alpha} \, \overline{\mu}_{\dot{\beta}}+
\lambda_\alpha \, \overline{\lambda}_{\dot{\beta}}) \,
\mu^\alpha \, \overline{\lambda}{}^{\dot{\beta}}=0 \; ,
\end{equation}
where we used formulas (\ref{ichs1}),
 (\ref{kew1}) and the properties of the commuting Weyl spinors:
 $\mu_\alpha \mu^\alpha = \varepsilon^{\alpha\beta} \mu_\alpha \mu_\beta = 0\;$,
 $\;\overline{\lambda}_{\dot{\beta}} \overline{\lambda}{}^{\dot{\beta}}=
 \varepsilon^{\dot{\beta}\dot{\alpha}} \overline{\lambda}_{\dot{\beta}} \overline{\lambda}_{\dot{\alpha}}=0$,
 which follow from the definition (\ref{epsilon})
 of the metrics $||\varepsilon^{\alpha\beta}||$
 and $||\varepsilon^{\dot{\alpha}\dot{\beta}}||$.
  The transversality of the vectors
  ${\sf e}^{_{(0)}}_m$ and ${\sf e}^{_{(-)}}_m$ is proved analogously.
Thus,
${\sf e}^{_{(+)}}(k), {\sf e}^{_{(0)}}(k)$ and ${\sf e}^{_{(-)}}(k)$ are
indeed interpreted as  polarization vectors, and
expression (\ref{raA}) gives
 the expansion of the vector potential with the components $A_m(k)$ over
 these polarization vectors.

 \vspace{0.3cm}

At the end of this Subsection devoted to a detailed discussion of unitary
representations of the group $ISL(2,\mathbb{C})$ with spin $j=1$ ,
 we calculate the sum over polarizations of the product of vectors
 (\ref{fpsl1}):
\begin{equation}\label{sp01}
\Theta^{(1)}_{nm}(k)= - \bigl( {\sf e}^{_{(+)}}_n(k) \, \overline{\sf e}^{_{(+)}}_m(k)
+{\sf e}^{_{(-)}}_n(k) \, \overline{\sf e}^{_{(-)}}_m(k)+
{\sf e}^{_{(0)}}_n(k) \, \overline{\sf e}^{_{(0)}}_m(k)
\bigr) \; .
\end{equation}
Here the common minus sign is chosen in accordance with the
normalization (\ref{norme}) of the vectors ${\sf e}^{_{(+)}}(k), {\sf e}^{_{(0)}}(k)$ and ${\sf e}^{_{(-)}}(k)$ so that the matrix
 $(\Theta^{(1)})^r_{m} = \eta^{rn} \, \Theta^{(1)}_{nm}(k)$
  satisfies the projection property
  \be
  \lb{thepr}
  (\Theta^{(1)})^n_{r}(k) \, (\Theta^{(1)})^r_{m}(k)
  = (\Theta^{(1)})^n_{m}(k) \; .
  \ee
The matrix
 $||\Theta^{(1)}_{nm}(k)||$ is called the density matrix of massive
vector particles and plays an important role in the relativistic theory.
To calculate the sum (\ref{sp01}),  we use the two-spinor formalism.
We substitute formulas (\ref{fpsl1}) and (\ref{kew1})--(\ref{kew3})
 into (\ref{sp01}) and obtain:
\begin{equation}\label{sp1}
\Theta^{(1)}_{nm}(k)=
-\frac{1}{4}\frac{(\sigma_n)_{\beta\dot{\beta}}
(\sigma_m)_{\alpha\dot{\alpha}}}{(\mu^\rho\lambda_\rho)
(\overline{\mu}{}^{\dot{\rho}}\overline{\lambda}_{\dot{\rho}})}
\Bigl(2\mu^\beta\overline{\lambda}{}^{\dot{\beta}}\lambda^\alpha
\overline{\mu}{}^{\dot{\alpha}}
+2\lambda^\beta\overline{\mu}{}^{\dot{\beta}}\mu^\alpha
\overline{\lambda}{}^{\dot{\alpha}}
+(\mu^\beta\overline{\mu}{}^{\dot{\beta}}-\lambda^\beta
\overline{\lambda}{}^{\dot{\beta}})
(\mu^\alpha\overline{\mu}{}^{\dot{\alpha}}-\lambda^\alpha
\overline{\lambda}{}^{\dot{\alpha}})\Bigr) \; .
\end{equation}
Then we expand the brackets in the right-hand side of (\ref{sp1})
and group terms in a different way:
\begin{equation}
\Theta^{(1)}_{nm}(k)=\frac{1}{4}\frac{(\sigma_n)_{\beta\dot{\beta}}
(\sigma_m)_{\alpha\dot{\alpha}}}
{(\mu^\rho\lambda_\rho)(\overline{\mu}{}^{\dot{\rho}}
\overline{\lambda}_{\dot{\rho}})}
\Bigl((\overline{\mu}{}^{\dot{\beta}}\overline{\lambda}{}^{\dot{\alpha}}-
\overline{\mu}{}^{\dot{\alpha}}\overline{\lambda}{}^{\dot{\beta}})
(\mu^\beta\lambda^\alpha-\mu^\alpha\lambda^\beta)
- (\mu^\beta\overline{\mu}{}^{\dot{\alpha}}+\lambda^\beta
\overline{\lambda}{}^{\dot{\alpha}})
(\mu^\alpha\overline{\mu}{}^{\dot{\beta}}+
\lambda^\alpha\overline{\lambda}{}^{\dot{\beta}})\Bigr) \, ,
\end{equation}
whereupon we use the two-spinor representations (\ref{ichs1})
for the momentum $k$ and the identities:
\begin{equation}
\bigl(\overline{\mu}{}^{\dot{\beta}}\, \overline{\lambda}{}^{\dot{\alpha}}-
\overline{\mu}{}^{\dot{\alpha}} \,
\overline{\lambda}{}^{\dot{\beta}}\bigr)=\varepsilon^{\dot{\alpha}\dot{\beta}}
\, (\overline{\mu}{}^{\dot{\rho}} \, \overline{\lambda}_{\dot{\rho}})
\; , \;\;\;
\bigl(\mu^\beta \, \lambda^\alpha-\mu^\alpha \, \lambda^\beta \bigr) =
\varepsilon^{\alpha\beta} \, (\mu^\rho \, \lambda_\rho) \;
\end{equation}
to deduce the final expression for $\Theta^{(1)}_{nm}(k)$:
 \be
 \lb{theta1}
 \begin{array}{c}
 {\displaystyle
\Theta^{(1)}_{nm}(k)=\frac{1}{4}\, (\sigma_n)_{\beta\dot{\beta}} \;
(\sigma_m)_{\alpha\dot{\alpha}}
\Bigl(\varepsilon^{\dot{\alpha}\dot{\beta}}\varepsilon^{\alpha\beta}-
\frac{(k^s\tilde{\sigma}_s)^{\dot{\alpha}\beta}
(k^l\tilde{\sigma}_l)^{\dot{\beta}\alpha}}{{\sf m}^2} \Bigr) =}
 \\ [0.4cm]
 {\displaystyle
=\frac{1}{2}\, \Bigl( \eta_{nm}-\frac{k^s k^l
Tr(\sigma_n\tilde{\sigma}_l\sigma_m\tilde{\sigma}_s)}{2\, {\sf m}^2}\Bigr) =
\Bigl(\eta_{nm}-\frac{k_n k_m}{{\sf m}^2}\Bigr)  =
\Bigl(\eta_{nm}- \frac{k_n k_m}{k^2}
\Bigr) } \; .
 \end{array}
 \ee
Here we take into account the identities
\begin{equation}\label{tssr4}
{\rm Tr}(\tilde{\sigma}_n\sigma_m)= 2 \eta_{nm} \; , \;\;\;
 {\rm Tr}(\sigma_n\tilde{\sigma}_l\sigma_m\tilde{\sigma}_s)=
2(\eta_{nl}\eta_{ms}-\eta_{nm}\eta_{ls}+\eta_{ns}\eta_{ml}
- i\varepsilon_{nlms}) \; ,
\end{equation}
and the
fact that the momentum  $k$ belongs to the mass shell: $(k)^2={\sf m}^2$.

\subsection{Spin $j=3/2$.}\label{2.2j3p}

According to Proposition  {\bf \ref{W2pq}},
 the unitary representation $ISL(2,\mathbb{C})$ with spin  $j=3/2$
is realized in the space of spin-tensor wave functions:
$\psi^{_{(0)}}_{\alpha_1 \alpha_2 \alpha_3}(k),\;
\psi^{_{(1)}\dot{\beta}_1}_{\alpha_1 \alpha_2}(k)$,
$\psi^{_{(2)}\dot{\beta}_1 \dot{\beta}_2}_{\alpha_1}(k)$,
$\psi^{_{(3)}\dot{\beta}_1 \dot{\beta}_2 \dot{\beta}_3}(k)$,
given in (\ref{tp}). In view of the Dirac-Pauli-Fierz equations (\ref{osudp})
 only two functions
$\psi^{_{(1)}\dot{\beta}_1}_{\alpha_1 \alpha_2}(k)$ and
$\psi^{_{(2)}\dot{\beta}_1 \dot{\beta}_2}_{\alpha_1}(k)$
can be regarded as independent.
We transform these two wave functions into spin-vectors
$(\psi^-_n)_{\alpha_2}(k), \; (\psi^+_n)^{\dot{\beta}_2}(k)$
by converting in the standard way one dotted and one undotted indices into a
 vector index $n$:
\begin{equation}
(\psi^-_n)_{\alpha_2}(k)=\frac{1}{2}
\varepsilon^{\gamma_1\alpha_1}(\sigma_n)_{\gamma_1\dot{\beta}_1}
\psi^{_{(1)}\dot{\beta}_1}_{\alpha_1 \alpha_2}(k), \; \;
(\psi^+_n)^{\dot{\beta}_2}(k)=\frac{1}{2}
\varepsilon^{\gamma_1\alpha_1}(\sigma_n)_{\gamma_1\dot{\beta}_1}
\psi^{_{(2)}\dot{\beta}_1 \dot{\beta}_2}_{\alpha_1}(k) \; .
\end{equation}
 From this set of Weyl spinors we form a bispinor wave function
\begin{equation}\label{srshe}
\Psi_n(k)=
\begin{pmatrix}
(\psi^-_n)_{\alpha_2}(k) \\ (\psi^+_n)^{\dot{\beta}_2}(k)
\end{pmatrix}  \; ,
\end{equation}
which simultaneously is a $4$-vector.
 The wave function $(\Psi_n)_A(k)$
 possesses not only the spinor index
 $A=1,2,3,4$ but in addition it also has the vector index $n=0,1,2,3$.
We define
 the Dirac conjugate wave function $\overline{\Psi}_n^B(k)$
 in the standard way:
\begin{equation}
 \overline{\Psi}_n=(\Psi_n)^{\dagger}\gamma_0=
 \begin{pmatrix}
 (\bar{\psi}^+_n)^{\beta_2} \; , & (\bar{\psi}^-_n)_{\dot{\alpha}_2}
\end{pmatrix} \; ,
\end{equation}
where we used the notation:
 $\bar{\psi}^+_n=(\psi^+_n)^*\;$, $\bar{\psi}^-_n=(\psi^-_n)^*$.
\begin{proposition}\label{rshe}
The spin-vector wave function $\Psi_n(k)$,
 defined in (\ref{srshe}), satisfies the Rarita-Schwinger equation
 \be
 \lb{RShw02}
 \gamma^{[m r n]} \, k_r \; \Psi_n(k) +
 {\sf m} \; \gamma^{[m n]} \; \Psi_n(k) = 0 \; ,
 \ee
describing the field of a free
massive relativistic
 particle with spin $j=3/2$. In formula (\ref{RShw02})
we used the notation $\gamma^{[m r s]}$ and $\gamma^{[m n]}$
for antisymmetrized products of $\gamma$-matrices
\be
\lb{clbas1}
\gamma^{[m n]}  = \frac{1}{2} (\gamma^m \gamma^n  - \gamma^n \gamma^m)
\; , \;\;\;
\gamma^{[m r s]} = \frac{1}{3}
(\gamma^m \gamma^{[r s]}  - \gamma^r \gamma^{[m s]} +
\gamma^s \gamma^{[m r]})  \; .
\ee
\end{proposition}
{\bf Proof.} Equation (\ref{RShw02}) is equivalent to a system of three
 equations
$$
  \gamma^n \Psi_n(k) =0 \; , \;\;\; k^n \, \Psi_n(k) = 0 \; , \;\;\;
 (\gamma^{r} \, k_r  - {\sf m}) \, \Psi_n(k) = 0 \; ,
$$
which follow from the Dirac-Pauli-Fierz equations (\ref{osudp})
for the spin-tensors $\psi^{_{(0)}}_{\alpha_1 \alpha_2 \alpha_3}(k),\;
\psi^{_{(1)}\dot{\beta}_1}_{\alpha_1 \alpha_2}(k)$,
$\psi^{_{(2)}\dot{\beta}_1 \dot{\beta}_2}_{\alpha_1}(k)$ and
$\psi^{_{(3)}\dot{\beta}_1 \dot{\beta}_2 \dot{\beta}_3}(k)$.
The proof of these facts is straightforward and employs
the same methods which we have already used in proving
 Proposition {\bf \ref{prop3}}. \hfill \qed

\vspace{0.2cm}

Now by using the two-spinor formalism, we construct
the expansion of the spin-vector wave function $\Psi_n(k)$
over polarizations.
We will make it in the same way as we constructed
 such expansions for spins $j=1/2$ and $j=1$ in Sections
 {\bf \ref{2.2jp}} and {\bf \ref{2.2j1}}.
The spin-tensor wave functions
$\psi^{_{(1)}\dot{\beta}_1}_{\alpha_1 \alpha_2}(k), \;
\psi^{_{(2)}\dot{\beta}_1 \dot{\beta}_2}_{\alpha_1}(k)$, which define
 the spin-vector $\Psi_n(k)$ in (\ref{srshe}), are related to
 the Wigner wave functions $\phi_{(\gamma_1 \gamma_2 \gamma_3)}$
 by formula (\ref{tp}):
\begin{equation}\label{tsb}
\psi^{_{(1)}\dot{\beta}_1}_{\alpha_1 \alpha_2}(k)=(A_{(k)})^{\gamma_1}_{\alpha_1}(A_{(k)})^{\gamma_2}_{\alpha_2}
(A^{\dagger -1}_{(k)}\; \tilde{\sigma}_0)^{\dot{\beta}_1 \gamma_3}
\phi_{(\gamma_1 \gamma_2 \gamma_3)} \; ,
\end{equation}
\begin{equation}\label{dsb}
\psi^{_{(2)}\dot{\beta}_1 \dot{\beta}_2}_{\alpha_1}(k)=(A_{(k)})^{\gamma_1}_{\alpha_1}
(A^{\dagger -1}_{(k)}\; \tilde{\sigma}_0)^{\dot{\beta}_1 \gamma_2}
(A^{\dagger -1}_{(k)}\; \tilde{\sigma}_0)^{\dot{\beta}_2 \gamma_3}
 \phi_{(\gamma_1 \gamma_2 \gamma_3)} \; .
\end{equation}
In the space of symmetric third-rank $SU(2)$-tensors we introduce
  normalized basis vectors
\begin{equation}
\begin{array}{c}
 \epsilon^{(\frac{3}{2})} = \epsilon^+\otimes \epsilon^+\otimes \epsilon^+\; , \;\;\;
 \epsilon^{(\frac{1}{2})} = \frac{1}{\sqrt{3}} \, (
\epsilon^+ \otimes \epsilon^+ \otimes \epsilon^-
+ \epsilon^+ \otimes \epsilon^- \otimes \epsilon^++\epsilon^- \otimes \epsilon^+ \otimes \epsilon^+) \; ,\\[0.5cm]
 \epsilon^{(-\frac{1}{2})} = \frac{1}{\sqrt{3}} \, (
\epsilon^- \otimes \epsilon^- \otimes \epsilon^+
+ \epsilon^- \otimes \epsilon^+ \otimes \epsilon^-+\epsilon^+ \otimes \epsilon^- \otimes \epsilon^-) \; , \;\;\;
\epsilon^{(-\frac{3}{2})} = \epsilon^- \otimes \epsilon^- \otimes \epsilon^-\; ,
\end{array}
\end{equation}
 where the spinors $\epsilon^+$ and $\epsilon^-$ were defined in (\ref{vtve1}),
and we expand the symmetric Wigner wave function $\phi(k)$
with the components $\phi_{(\alpha_1 \alpha_2\alpha_3)}(k)$
over these basis vectors
\begin{equation}\label{tpen3p}
\phi(k)=
  \phi_{(111)}(k) \,  \epsilon^{(\frac{3}{2})}
 +  \sqrt{3} \, \phi_{(112)}(k) \,  \epsilon^{(\frac{1}{2})}   +
 \sqrt{3} \, \phi_{(122)}(k) \,
 \epsilon^{(-\frac{1}{2})} +\phi_{(222)}(k) \,  \epsilon^{(-\frac{3}{2})}\; .
\end{equation}
We substitute expansion (\ref{tpen3p}) into formulas
(\ref{tsb}), (\ref{dsb}), use (\ref{psm05})
and then the result is substituted into (\ref{srshe}).
Finally, we obtain
\begin{equation}
\Psi_n(k)=\frac{1}{\sqrt{2}}\Bigl(\phi_{(111)}(k)e_n^{(\frac{3}{2})}(k)+
\sqrt{3} \, \phi_{(112)}(k)e_n^{_{(\frac{1}{2})}}(k)+\sqrt{3} \, \phi_{(122)}(k)
e_n^{_{(-\frac{1}{2})}}(k)+\phi_{(222)}(k)e_n^{_{(-\frac{3}{2})}}(k)
\Bigr) \, ,
\end{equation}
where we have introduced the notation:
\begin{equation}\label{vp3P}
\begin{array}{c}
e_n^{(\frac{3}{2})}(k)={\sf e}_n^{(+)}(k) e^{(+)}\;, \;\;
e_n^{(\frac{1}{2})}(k)=\sqrt{\frac{2}{3}}{\sf e}_n^{(0)}(k) e^{(+)}+\sqrt{\frac{1}{3}}{\sf e}_n^{(+)} (k)e^{(-)} \;, \\[0.5cm]
e_n^{(-\frac{1}{2})}(k)=\sqrt{\frac{2}{3}}{\sf e}_n^{(0)}(k) e^{(-)}+\sqrt{\frac{1}{3}}{\sf e}_n^{(-)}(k) e^{(+)}\;, \;\;
e_n^{(-\frac{3}{2})}(k)={\sf e}_n^{(-)}(k) e^{(-)}
\end{array}
\end{equation}
and vectors ${\sf e}_n^{(+)}(k), \; {\sf e}_n^{(0)}(k), \; {\sf e}_n^{(-)}(k)$
were defined in (\ref{fpsl1}), while bispinors  $e^{(+)}, e^{(-)}$
were defined  in (\ref{epem}).
As before, it is natural
to assume that the bispinor functions $e_n^{(\frac{3}{2})}(k)$,  $e_n^{(\frac{1}{2})}(k)$,
 $e_n^{(-\frac{1}{2})}(k)$, $e_n^{(-\frac{3}{2})}(k)$
are polarization spin-vectors (below we simply call them polarizations).
 The spin-vector wave function $\Psi_n(k)$ which describes
 particles with spin $j=3/2$ is decomposed into a linear combination
 of polarization spin-vectors (\ref{vp3P}).

Below in this paper, to simplify formulas, we do not often write
 the dependence of polarizations on the momentum $k$.
The bispinors (\ref{vp3P}) are normalized as follows:
\begin{equation}\label{nor3p1}
(\overline{e}_n^{(m)})^A\eta^{nr}(e_r^{(m^\prime)})_A=-2\delta^{m\; m^\prime},\;\;\; m, m^\prime=-\frac{3}{2}, -\frac{1}{2},  \frac{1}{2}, \frac{3}{2}\, ;
\end{equation}
here we used the normalization conditions (\ref{prh7}) and (\ref{norme}).
The sum over the polarizations (the density matrix for particles of spin
 $3/2$)  is defined by the expression:
\begin{equation}\label{pr3pW}
(\Theta^{(\frac{3}{2})}_{nr})_A^{\;\; B}=-\frac{1}{2}\;\; \underset{_{m=-3/2}}{\overset{_{3/2}}{\Sigma}}
(e_n^{(m)})_A(\overline{e}_r^{(m)})^B \; ,
\end{equation}
and in view of (\ref{nor3p1}) satisfies the projector property
 $(\Theta^{(\frac{3}{2})}_{nr})_A^{\;\; B} \, \eta^{r\ell} \,
 (\Theta^{(\frac{3}{2})}_{\ell m})_B^{\;\; C} =
 (\Theta^{(\frac{3}{2})}_{nm})_A^{\;\; C}$.
Finally, we substitute (\ref{vp3P}) into formula (\ref{pr3pW})
and group terms
  so that the density matrix (\ref{pr3pW}) takes the form:
\begin{equation}\label{ers3p}
\begin{array}{c}
(\Theta^{(\frac{3}{2})}_{nr})_{A}^{\;\; B}(k)=
\frac{1}{2}\Bigl( ({\sf e}_n^{_{(+)}}\overline{\sf e}_r^{_{(+)}}
+\frac{2}{3}{\sf e}_n^{_{(0)}}\overline{\sf e}_r^{_{(0)}}
+\frac{1}{3}{\sf e}_n^{_{(-)}}\overline{\sf e}_r^{_{(-)}}) \,
 e^{_{(+)}}_A \, \overline{e}^{^{(+)}B} \; + \\[0.3cm]
+\frac{\sqrt{2}}{3}\, ({\sf e}_n^{_{(0)}}\overline{\sf e}_r^{_{(+)}}
 +{\sf e}_n^{_{(-)}}\overline{\sf e}_r^{_{(0)}})\,
 e^{_{(+)}}_A \, \overline{e}^{^{(-)}B}+
\frac{\sqrt{2}}{3}\, ({\sf e}_n^{_{(+)}}\overline{\sf e}_r^{_{(0)}}
+{\sf e}_n^{_{(0)}}\overline{\sf e}_r^{_{(-)}}) \,
 e^{_{(-)}}_A \, \overline{e}^{^{(+)}B} \; + \\[0.3cm]
+ ({\sf e}_n^{_{(-)}}\overline{\sf e}_r^{_{(-)}}
+\frac{2}{3}{\sf e}_n^{_{(0)}}\overline{\sf e}_r^{_{(0)}}
+\frac{1}{3}{\sf e}^{_{(+)}}_n \overline{\sf e}_r^{_{(+)}})
 \, e^{_{(-)}}_A \, \overline{e}^{^{(-)}B} \Bigr) \; .
\end{array}
\end{equation}
 We will need this form
 of the spin $j=3/2$ density matrix below
 in Section {\bf \ref{PrBeFr}}.

 \subsection{Spin $j=2$.}
 \label{2.2j2}

 According to Proposition {\bf \ref{W2pq}}, the unitary representation
(with spin $j=2$) of the group $ISL(2,\mathbb{C})$  is realized
in the space of spin-tensor wave functions:
 \be
 \lb{spin2}
\psi^{_{(0)}}_{(\alpha_1 \alpha_2 \alpha_3 \alpha_4)}(k) \; ,\;\;\;
\psi^{_{(1)}\dot{\beta}_1}_{(\alpha_1 \alpha_2 \alpha_3)}(k)\; ,\;\;\;
\psi^{_{(2)}(\dot{\beta}_1 \dot{\beta}_2)}_{(\alpha_1 \alpha_2)}(k)\; ,\;\;\;
\psi^{_{(3)}(\dot{\beta}_1 \dot{\beta}_2 \dot{\beta}_3)}_{\alpha_1}(k)\; ,\;\;\;
\psi^{_{(4)}(\dot{\beta}_1 \dot{\beta}_2 \dot{\beta}_3 \dot{\beta}_4)}(k) \; .
\ee
 These wave functions were defined in (\ref{tp}) and satisfy the system of
 Dirac-Pauli-Fierz equations (\ref{osudp}).
  The most important for us spin-tensor wave function is the function
 $\psi^{_{(2)}}(k)$,
  which corresponds to a symmetric second-rank tensor
  $h_{n_1  n_2}(k)$ in the Minkowski space-time. The relation between
  $\psi^{_{(2)}}(k)$
   and $h_{n_1 n_2}(k)$ is given by the standard formula
\begin{equation}\label{grfg}
h_{n_1 n_2}(k)=\frac{1}{4}(\sigma_{n_1})_{\alpha_1 \dot{\beta}_1}\;(\sigma_{n_2})_{\alpha_1 \dot{\beta}_1}
\varepsilon^{\alpha_2 \gamma_2}\varepsilon^{\alpha_1\gamma_1}
\psi^{_{(2)}(\dot{\beta}_1 \dot{\beta}_2)}_{(\gamma_1 \gamma_2)}(k)
 = h_{n_2 n_1}(k) \; .
\end{equation}

\begin{proposition}\label{PaFi-eq}
 The system of
 Dirac-Pauli-Fierz equations (\ref{osudp}) for spin-tensor wave functions
 (\ref{spin2}) is equivalent to the massive Pauli-Fierz equation
 \cite{PauFir} for the wave functions $h_{mn}(k)$
 of massive graviton:
  \be
 \lb{pafi05}
 \begin{array}{c}
  k^2 \, h_{m n}(k) - k_{m} k^{r} \, h_{rn}(k)
  - k_{n} k^{r} \, h_{rm}(k) + \eta_{mn} k^{r} \,
  k^{\ell} \, h_{r\ell}(k) + \\ [0.2cm]
  + k_n k_m h(k) - \eta_{mn} k^{2} \, h(k)
  - {\sf m}^2 ( h_{m n}(k) - \eta_{mn} \, h(k)) = 0 \; ,
  \end{array}
 \ee
 where $h(k) = \eta^{rs} h_{rs}(k)$.
 \end{proposition}
 {\bf Proof.} The proof of this Proposition is straightforward and
is carried out in the same way as the proof of Proposition
{\bf \ref{prop3}}. First, from equations
(\ref{osudp}) it follows that the wave functions $\psi^{_{(r)}}(k)$
for $r=0,1,3,4$ are expressed via one spin-tensor function
 $\psi^{_{(2)}}(k)$. Then one can prove that the symmetric wave function
 $h_{m n}(k)$,
 which is defined in (\ref{grfg}), satisfies the system of equations
 \be
 \lb{pafi04}
  \eta^{mn} \, h_{mn}(k) = 0  \; , \;\;\;\; k^{m} \, h_{mn}(k) = 0  \; , \;\;\;\;
  k^2 \, h_{m n}  = {\sf m}^2 \, h_{m n} \; ,
 \ee
 which is equivalent to the Pauli-Fierz equation (\ref{pafi05}).
\hfill \qed

\vspace{0.2cm}

Now in full analogy with what we have done in Subsections
{\bf \ref{2.2jp}} -- {\bf \ref{2.2j3p}}, we construct the expansion
of the massive graviton wave functions $h_{n m}(k)$
over polarizations (in this case the polarizations
are linearly independent second-rank vector-tensors).
 In accordance with (\ref{tp}), the components of the
spin-tensor wave function
 $\psi^{_{(2)}}(k)$
 are defined via the components
 $\phi_{(\delta_1 \delta_2 \delta_3 \delta_4)}(k)$ of Wigner's wave function
 $\phi(k)$ as follows:
\begin{equation}\label{sgfv0}
\psi^{_{(2)}(\dot{\beta}_1 \dot{\beta}_2)}_{(\alpha_1 \alpha_2)}(k)=
\frac{1}{{\sf m}^2}
(A_{(k)})^{\delta_1}_{\alpha_1}(A_{(k)})^{\delta_2}_{\alpha_2}\;
(A_{(k)}^{\dagger -1} (q \tilde{\sigma}) )^{\dot{\beta}_1 \delta_3}
(A_{(k)}^{\dagger -1} (q \tilde{\sigma}) )^{\dot{\beta}_2 \delta_4}
\phi_{(\delta_1 \delta_2 \delta_3 \delta_4)}(k) \; .
\end{equation}
Introduce basis vectors in the space of completely symmetric
four-rank tensors $\phi(k)$:
\begin{equation}\label{bas4}
\begin{array}{c}
\epsilon^{(2)}=\epsilon^+ \epsilon^+\epsilon^+\epsilon^+, \;
\epsilon^{(1)}=\frac{1}{2}(\epsilon^+ \epsilon^+\epsilon^+\epsilon^- +
\epsilon^+ \epsilon^+\epsilon^-\epsilon^+ +\epsilon^+ \epsilon^-\epsilon^+\epsilon^+
+\epsilon^- \epsilon^+\epsilon^+\epsilon^+), \\[0.4cm]
\epsilon^{(0)}=\frac{1}{\sqrt{6}}
(\epsilon^+ \epsilon^+\epsilon^-\epsilon^-
+\epsilon^+ \epsilon^-\epsilon^+\epsilon^-
+ \epsilon^- \epsilon^+\epsilon^+\epsilon^-
+\epsilon^- \epsilon^-\epsilon^+\epsilon^+
+\epsilon^- \epsilon^+\epsilon^-\epsilon^+
+\epsilon^+ \epsilon^-\epsilon^-\epsilon^+), \\[0.4cm]
\epsilon^{(-1)}=\frac{1}{2}(\epsilon^- \epsilon^-\epsilon^-\epsilon^+ +
\epsilon^- \epsilon^-\epsilon^+\epsilon^- +\epsilon^- \epsilon^+\epsilon^-\epsilon^-
+\epsilon^+ \epsilon^-\epsilon^-\epsilon^-), \;
\epsilon^{(-2)}=\epsilon^-\epsilon^-\epsilon^-\epsilon^- \;,
\end{array}
\end{equation}
where spinors $\epsilon^\pm$ are defined in (\ref{vtve1}),
and to be short, we omit in (\ref{bas4}) the tensor product
signs between multipliers $\epsilon^{\pm}$. The symmetric Wigner wave function
 $\phi(k)$ with the components
$\phi_{(\delta_1 \delta_2 \delta_3 \delta_4)}(k)$ is expanded over
the basis (\ref{bas4}):
\begin{equation} \label{raz4b}
\phi(k)=\phi_{(2)}(k)\epsilon^{(2)}+\phi_{(1)}(k)\epsilon^{(1)}+
\phi_{(0)}(k)\epsilon^{(0)}
+\phi_{(-1)}(k)\epsilon^{(-1)}+\phi_{(-2)}(k)\epsilon^{(-2)},
\end{equation}
where $\phi_{(2)}=\phi_{(1111)}, \; \phi_{(1)}=2\phi_{(1112)}, \; \phi_{(0)}=\sqrt{6}\phi_{(1122)}$,
$\phi_{(-1)}=2\phi_{(1222)}$ and $\phi_{(-2)}=\phi_{(2222)}$.
 We fix as usual the test momentum in the form $q=({\sf m},0,0,0)$.
Then, we substitute the expansion (\ref{raz4b}) for Wigner's wave
function $\phi(k)$ into (\ref{sgfv0}) and obtain
\begin{equation} \label{exsvf}
\psi^{_{(2)}(\dot{\beta}_1 \dot{\beta}_2)}_{(\alpha_1 \alpha_2)}(k)=
\phi_{(2)}\overset{_{(2)}}{e}{}_{(\alpha_1 \alpha_2)}^{\;(\dot{\beta}_1 \dot{\beta}_2)}+
\phi_{(1)}\overset{_{(1)}}{e}{}_{(\alpha_1 \alpha_2)}^{\;(\dot{\beta}_1 \dot{\beta}_2)}+
\phi_{(0)}\overset{_{(0)}}{e}{}_{(\alpha_1 \alpha_2)}^{\;(\dot{\beta}_1 \dot{\beta}_2)}+
\phi_{(-1)}\overset{_{(-1)}}{e}{}_{(\alpha_1 \alpha_2)}^{\;(\dot{\beta}_1 \dot{\beta}_2)}+
\phi_{(-2)}\overset{_{(-2)}}{e}{}_{(\alpha_1 \alpha_2)}^{\;(\dot{\beta}_1 \dot{\beta}_2)},
\end{equation}
 where the coefficients $\phi_{(m)}$ are functions of the momentum $k$ and
 we introduce the notation:
\begin{equation}  \label{rios}
\begin{array}{c}
\overset{_{(2)}}{e}{}_{(\alpha_1 \alpha_2)}^{\;(\dot{\beta}_1 \dot{\beta}_2)}(k)=
\overset{_{(+)}}{e}{}_{\alpha_1}^{\;\dot{\beta}_1}(k)
\overset{_{(+)}}{e}{}_{\alpha_2}^{\;\dot{\beta}_2}(k),\;\;
\overset{_{(1)}}{e}{}_{(\alpha_1 \alpha_2)}^{\;(\dot{\beta}_1 \dot{\beta}_2)}(k)=\frac{1}{\sqrt{2}}
(\overset{_{(+)}}{e}{}_{\alpha_1}^{\;\dot{\beta}_1}(k)
\overset{_{(0)}}{e}{}_{\alpha_2}^{\;\dot{\beta}_2}(k)+
\overset{_{(0)}}{e}{}_{\alpha_1}^{\;\dot{\beta}_1}(k)
\overset{_{(+)}}{e}{}_{\alpha_2}^{\;\dot{\beta}_2}(k))\\[0.4cm]
\overset{_{(0)}}{e}{}_{(\alpha_1 \alpha_2)}^{\;(\dot{\beta}_1 \dot{\beta}_2)}(k)=\frac{1}{\sqrt{6}}
(\overset{_{(+)}}{e}{}_{\alpha_1}^{\;\dot{\beta}_1}(k)
\overset{_{(-)}}{e}{}_{\alpha_2}^{\;\dot{\beta}_2}(k)+
\overset{_{(-)}}{e}{}_{\alpha_1}^{\;\dot{\beta}_1}(k)
\overset{_{(+)}}{e}{}_{\alpha_2}^{\;\dot{\beta}_2}(k)+
2\overset{_{(0)}}{e}{}_{\alpha_1}^{\;\dot{\beta}_1}(k)
\overset{_{(0)}}{e}{}_{\alpha_2}^{\;\dot{\beta}_2}(k))\\[0.4cm]
\overset{_{(-1)}}{e}{}_{(\alpha_1 \alpha_2)}^{\;(\dot{\beta}_1 \dot{\beta}_2)}(k)=\frac{1}{\sqrt{2}}
(\overset{_{(-)}}{e}{}_{\alpha_1}^{\;\dot{\beta}_1}(k)
\overset{_{(0)}}{e}{}_{\alpha_2}^{\;\dot{\beta}_2}(k)+
\overset{_{(0)}}{e}{}_{\alpha_1}^{\;\dot{\beta}_1}(k)
\overset{_{(-)}}{e}{}_{\alpha_2}^{\;\dot{\beta}_2}(k)), \;
\overset{_{(-2)}}{e}{}_{(\alpha_1 \alpha_2)}^{\;(\dot{\beta}_1 \dot{\beta}_2)}(k)=
\overset{_{(-)}}{e}{}_{\alpha_1}^{\;\dot{\beta}_1}(k)
\overset{_{(-)}}{e}{}_{\alpha_2}^{\;\dot{\beta}_2}(k) \, .
\end{array}
\end{equation}
Recall that the
 spin-tensors $\overset{_{(+)}}{e}{}_{\alpha_1}^{\;\dot{\beta}_1}(k),
\overset{_{(-)}}{e}{}_{\alpha_1}^{\;\dot{\beta}_1}(k), \overset{_{(0)}}{e}{}_{\alpha_1}^{\;\dot{\beta}_1}(k)$
were defined in (\ref{kew1})-(\ref{kew3}).
Finally, we substitute expansion (\ref{exsvf}) for
$\psi^{_{(2)}}(k)$
into expression (\ref{grfg}) for the vector-tensors $h_{n_1 n_2}(k)$
 and deduce:
\begin{equation}
h_{n_1 n_2}(k)=\frac{1}{2}(\phi_{(2)}{\sf e}^{_{(2)}}_{n_1 n_2}+\phi_{(1)}{\sf e}^{_{(1)}}_{n_1 n_2}+\phi_{(0)}{\sf e}^{_{(0)}}_{n_1 n_2}
+\phi_{(-1)}{\sf e}^{_{(-1)}}_{n_1 n_2}+\phi_{(-2)}{\sf e}^{_{(-2)}}_{n_1 n_2}),
\end{equation}
where we defined the spin $j=2$ polarization tensors
\begin{equation} \label{rios1}
\begin{array}{c}
{\sf e}^{_{(2)}}_{n_1 n_2}={\sf e}^{_{(+)}}_{n_1}{\sf e}^{_{(+)}}_{n_2}, \;
{\sf e}^{_{(1)}}_{n_1 n_2}=\frac{1}{\sqrt{2}}({\sf e}^{_{(+)}}_{n_1}{\sf e}^{_{(0)}}_{n_2}+
{\sf e}^{_{(0)}}_{n_1}{\sf e}^{_{(+)}}_{n_2}), \\[0.4cm]
{\sf e}^{_{(0)}}_{n_1 n_2}=\frac{1}{\sqrt{6}}({\sf e}^{_{(+)}}_{n_1}{\sf e}^{_{(-)}}_{n_2}+
{\sf e}^{_{(-)}}_{n_1}{\sf e}^{_{(+)}}_{n_2}+2{\sf e}^{_{(0)}}_{n_1}{\sf e}^{_{(0)}}_{n_2}),\\[0.4cm]
{\sf e}^{_{(-1)}}_{n_1 n_2}=\frac{1}{\sqrt{2}}({\sf e}^{_{(-)}}_{n_1}{\sf e}^{_{(0)}}_{n_2}+
{\sf e}^{_{(0)}}_{n_1}{\sf e}^{_{(-)}}_{n_2}), \;
{\sf e}^{_{(-2)}}_{n_1 n_2}={\sf e}^{_{(-)}}_{n_1}{\sf e}^{_{(-)}}_{n_2} .
\end{array}
\end{equation}
 Here
${\sf e}^{_{(+)}}_{n_1}, {\sf e}^{_{(-)}}_{n_1}, {\sf e}^{_{(0)}}_{n_1}$
are the polarization vectors
(for spin $j=1$) which were introduced in (\ref{fpsl1}). The density
matrix for spin $j=2$ is given by the sum over polarizations:
\begin{equation}
\lb{rios05}
\begin{array}{c}
\Theta^{(2)}_{n_1 n_2 r_1 r_2}=
 {\sf e}_{n_1}^{_{(+)}}{\sf e}_{n_2}^{_{(+)}} \overline{\sf e}_{r_1}^{_{(+)}}\overline{\sf e}_{r_2}^{_{(+)}}
+\frac{1}{2}({\sf e}_{n_1}^{_{(+)}}{\sf e}_{n_2}^{_{(0)}}+{\sf e}_{n_1}^{_{(0)}}{\sf e}_{n_2}^{_{(+)}})
(\overline{\sf e}_{r_1}^{_{(+)}}\overline{\sf e}_{r_2}^{_{(0)}}+\overline{\sf e}_{r_1}^{_{(0)}}\overline{\sf e}_{r_2}^{_{(+)}})+\\[0.5cm]
+\frac{1}{6}({\sf e}_{n_1}^{_{(+)}}{\sf e}_{n_2}^{_{(-)}}+2{\sf e}_{n_1}^{_{(0)}}{\sf e}_{n_2}^{_{(0)}}+{\sf e}_{n_1}^{_{(-)}}{\sf e}_{n_2}^{_{(+)}})
(\overline{\sf e}_{r_1}^{_{(+)}}\overline{\sf e}_{r_2}^{_{(-)}}+2\overline{\sf e}_{r_1}^{_{(0)}}\overline{\sf e}_{r_2}^{_{(0)}}+\overline{\sf e}_{r_1}^{_{(-)}}\overline{\sf e}_{r_2}^{_{(+)}})+\\[0.5cm]
+\frac{1}{2}({\sf e}_{n_1}^{_{(0)}}{\sf e}_{n_2}^{_{(-)}}+{\sf e}_{n_1}^{_{(-)}}{\sf e}_{n_2}^{_{(0)}})
(\overline{\sf e}_{r_1}^{_{(0)}}\overline{\sf e}_{r_2}^{_{(-)}}+\overline{\sf e}_{r_1}^{_{(-)}}\overline{\sf e}_{r_2}^{_{(0)}})+
{\sf e}_{n_1}^{_{(-)}}{\sf e}_{n_2}^{_{(-)}} \overline{\sf e}_{r_1}^{_{(-)}}\overline{\sf e}_{r_2}^{_{(-)}} \; .
\end{array}
\end{equation}
We will use this formula below.

\section{The polarization vector for the fields of arbitrary integer spin \label{VectPol}}
\setcounter{equation}0

 The unitary irreducible representation (\ref{fie03})
 of the group $ISL(2,\mathbb{C})$ with spin $j$,
 according to (\ref{mual}) ,acts in the space of
 symmetrized Wigner's wave functions
$\phi_{(\alpha_1\cdots\alpha_{2j})}(k)$.
 It is convenient to write these symmetrized
 wave functions as a generating function:
\begin{equation}\label{opol1}
\phi(k; v)=\phi_{(\alpha_1\cdots\alpha_{2j})}(k)v^{\alpha_1} \cdots v^{\alpha_{2j}},
\end{equation}
where $v^\alpha$ are the components of the auxiliary Weyl spinor $v=(v^1, v^2 )$.
Introduce homogeneous monomials $T_m^j(v)$ in the variables
$v^1$ and $v^2$:
\begin{equation}\label{basm1}
T_m^j(v)=\frac{(v^1)^{j+m}(v^2)^{j-m}}{\sqrt{(j+m)!(j-m)!}},\;\; m=-j, \cdots ,j\, ,
\end{equation}
which can be considered as
$(2j+1)$ basis elements in the space of polynomials (\ref{opol1}) since
any polynomial (\ref{opol1}) can be expanded in terms of
 $T_m^j(v)$:
\begin{equation}\label{rpbv}
\phi(k,v)=\overset{j}{\underset{m=-j}{\Sigma}} \phi_m(k) \; T_m^j(v) \; .
\end{equation}
The relation between the coefficients $\phi_m(k)$ and $\phi_{(\alpha_1\cdots\alpha_{2j})}(k)$ is given by the formula:
\begin{equation}
\phi_m(k)=\frac{(2j)!}{\sqrt{(j+m)!(j-m)!}}
\phi_{(\underbrace{_{1\cdots1}}_{j+m}\underbrace{_{2\cdots2}}_{j-m})}(k) \; .
\end{equation}
In the space of polynomials
(\ref{opol1}) and (\ref{rpbv}) an irreducible representation
 of the algebra $s\ell(2,\mathbb{C})$
is realized with generators:
\begin{equation}\label{trkS}
S_+=v^2\partial_{v^1}, \;\; S_-=v^1\partial_{v^2}, \;\; S_3=\frac{1}{2}(v^1\partial_{v^1}-v^2\partial_{v^2})
\end{equation}
The monomials $T_m^j(v)$ are the eigenvectors of the operator $S_3$
(the third component of the spin vector) given in (\ref{trkS}).
In fact, we have
 \be
 \lb{S3Tm}
 S_3 T_m^j(v) = m \, T_m^j(v) \; ;
 \ee
therefore, the coefficients $\phi_m(k)$
in the expansion (\ref{rpbv}) correspond to the projections $m$ of the operator $S_3$.

Formula (\ref{tp}), which connects the Wigner wave function $\phi_{(\alpha_1\cdots\alpha_{2j})}(k)$
to the spin-tensor wave function
$\psi^{_{(r)}(\dot{\beta_1}...\dot{\beta_r})}_{(\alpha_1...\alpha_p)}(k)$,
can be rewritten in terms of the generating functions (\ref{psixu}), (\ref{opol1}) as follows:
\begin{equation}\label{nopv}
\psi^{_{(r)}}(k;u,\overline{u})= \frac{1}{(2j)! \; {\sf m}^r} \,
 \prod_{i=1}^p
 (u^{\alpha_i} (A_{(k)})^{\;\;\; \rho_i}_{\alpha_i}\partial^{_{(v)}}_{\rho_i})
 \prod_{\ell=1}^r
\Bigl(\overline{u}_{\dot{\beta_\ell}}
\bigl(A^{-1\dagger}_{(k)}\cdot (q \tilde{\sigma})
\bigr)^{\dot{\beta_\ell}\rho_{p+\ell}}
\partial^{_{(v)}}_{\rho_{p+\ell}} \Bigr) \;
\phi(k; v),
\end{equation}
where $p+r=2j\,$ and $\partial^{_{(v)}}_{\rho_i}=\frac{\partial}{\partial v^{\rho_i}}$.
We fix as usual the test momentum $q=({\sf m}, 0,0,0)$ and substitute expression (\ref{rpbv}) for the Wigner wave function
 to the formula (\ref{nopv}).
After that, expanding the left-and right-hand
sides of (\ref{nopv}) over $u$ and $\overline{u}$, we obtain
\begin{equation}\label{dm1}
\psi^{_{(r)}(\dot{\beta_1}...\dot{\beta_r})}
_{(\alpha_1...\alpha_p)}(k)=
\frac{1}{(2j)!} \,
 \sum_{m=-j}^j \phi_m(k) \;
 \overset{p}{\underset{i=1}{\Pi}}
(A_{(k)})^{\;\;\; \rho_i}_{\alpha_i} \;
\overset{r}{\underset{\ell=1}{\Pi}}
\bigl(A^{-1\dagger}_{(k)}\; \tilde{\sigma}_0\bigr)^{\dot{\beta_\ell}\rho_{p+\ell}}
 \; \epsilon^{(m)}_{\rho_1\cdots\rho_{2j}} \; ,
\end{equation}
where
\begin{equation}\label{trns1}
\epsilon^{(m)}_{\rho_1\cdots\rho_{2j}}=\partial^{_{(v)}}_{\rho_1} \cdots \partial^{_{(v)}}_{\rho_{2j}} \; T_m^j(v) \; .
\end{equation}
It is clear that the tensor $\epsilon^{(m)}_{\rho_1\cdots\rho_{2j}}$
does not depend on the components $v^\alpha$ and is symmetric with respect to permutations of the indices $\rho_i$.
In view of the normalization accepted in (\ref{basm1}) equation (\ref{trns1}) gives
\begin{equation}\label{pre1}
\epsilon^{(m)}_{\underbrace{_{1\cdots1}}_{j+m}\underbrace{_{2 \cdots 2}}_{j-m}}=\sqrt{(j+m)!(j-m)!} \; .
\end{equation}
All components $\epsilon^{(m)}_{\rho_1\cdots\rho_{2j}}$
with the number of units in the subscripts differing from $(j+m)$
(and the number of deuces in the subscripts
 differs from $(j-m)$, see (\ref{pre1}))
are equal to zero. Thus, for the symmetric tensor
  $\epsilon^{(m)}_{\rho_1\cdots\rho_{2j}}$,  there exist only
$\frac{(2j)!}{(j-m)!(j+m)!}$  non-zero components equal to each other
(and are equal to (\ref{pre1})).
Define spin-tensors
\begin{equation}\label{npkp}
\overset{_{(m)}}{e}^{(\dot{\beta_1}...\dot{\beta_r})}_{(\alpha_1...\alpha_p)}(k)
=\frac{1}{\sqrt{(2j)}!}\overset{p}{\underset{i=1}{\Pi}}
(A_{(k)})^{\;\;\; \rho_i}_{\alpha_i}\,
\overset{r}{\underset{\ell=1}{\Pi}}
\bigl(A^{-1\dagger}_{(k)}\; \tilde{\sigma}_0\bigr)^{\dot{\beta}_{\ell} \rho_{p+\ell}}\,
\epsilon^{(m)}_{\rho_1\cdots\rho_{2j}},
\end{equation}
where the normalization factor $\frac{1}{\sqrt{(2j)!}}$ is chosen for convenience.
Then, in terms of spin-tensors
 (\ref{npkp}), formula (\ref{dm1}) can be rewritten as:
\begin{equation}\label{vtdsw}
\psi^{_{(r)}(\dot{\beta_1}...\dot{\beta_r})}
_{(\alpha_1...\alpha_p)}(k)=\frac{1}{\sqrt{(2j)!}} \;
 \sum_{m=-j}^j \phi_m(k) \,
\overset{_{(m)}}{e}^{(\dot{\beta_1}...\dot{\beta_r})}_{(\alpha_1...\alpha_p)}(k)
\; .
\end{equation}
\begin{proposition}\label{prop4} The spin-tensor wave functions $\overset{_{(m)}}{e}^{(\dot{\beta_1}...\dot{\beta_r})}_{(\alpha_1...\alpha_p)}(k)$,
 given in (\ref{npkp}),
satisfy the Dirac-Pauli-Fierz equations:
\begin{equation}\label{osudpe}
\begin{array}{l}
k^m(\tilde{\sigma}_m)^{\dot{\gamma}_1 \alpha_1}\overset{_{(m)}}{e}
^{_{(r)}(\dot{\beta_1}...\dot{\beta_r})}_{(\alpha_1...\alpha_p)}(k)=
{\sf m} \; \overset{_{(m)}}{e}^{_{(r+1)}(\dot{\gamma}_1 \dot{\beta_1}...\dot{\beta_r})}_{(\alpha_2...\alpha_p)}(k) \; ,
\;\;\;\; (r=0,\dots,2j-1) \; , \\[0.2cm]
k^m(\sigma_m)_{\gamma_1 \dot{\beta}_1}\overset{_{(m)}}{e}
^{_{(r)}(\dot{\beta_1}...\dot{\beta_r})}_{(\alpha_1...\alpha_p)}(k)
={\sf m} \;\overset{_{(m)}}{e}^{_{(r-1)}(\dot{\beta_2}...\dot{\beta_r})}
_{(\gamma_1 \alpha_1...\alpha_p)}(k) \; ,
\;\;\;\; (r=1,\dots,2j) \; .
\end{array}
\end{equation}
\end{proposition}
{\bf Proof. } The proof is based on the use of the
definition (\ref{npkp}) and is carried out
similarly to the proof of Proposition {\bf \ref{prop1}}. \hfill \qed

\vspace{0.2cm}

In this Section, we will mainly consider spin-tensor
wave functions of type $(\frac{j}{2}, \frac{j}{2})$: $\psi^{_{(j)}(\dot{\beta}_1\cdots \dot{\beta_j})}_{(\alpha_1...\alpha_j)}(k)$
for which the number of dotted and undotted indices is the same.
 It gives us the possibility to suppress sometimes
 the index $(j)$ in the notation of the spin-tensor $\psi^{(j)}$
(we have to restore this index in the proof of Proposition {\bf \ref{svop}}).
The spin-tensor functions of $(\frac{j}{2}, \frac{j}{2})$-type are related to the
 vector-tensors in the Minkowski space by the following formula
 (cf. (\ref{vp12}) and (\ref{grfg})):
\begin{equation}\label{smsv}
f_{n_1\cdots n_j}(k)=\frac{1}{2^j}(\sigma_{n_1})_{\alpha_1\dot{\beta}_1}\cdots
(\sigma_{n_j})_{\alpha_j\dot{\beta}_j}
\varepsilon^{\alpha_1\gamma_1} \cdots \varepsilon^{\alpha_j\gamma_j}
\psi^{(\dot{\beta}_1\cdots \dot{\beta_j})}_{(\gamma_1...\gamma_j)}(k)
\end{equation}
By vector-tensors we call
the tensors with the components $f_{n_1\cdots n_j}(k)$ having only vector indices
 $\{ n_1, \cdots ,n_j\}$.
We note that in view of the symmetry of the components
$\psi^{(\dot{\beta}_1\cdots \dot{\beta_j})}_{(\alpha_1...\alpha_j)}(k)$
under all permutations of the spinor indices
$(\dot{\beta}_1\cdots \dot{\beta_j})$ and $(\alpha_1\cdots \alpha_j)$,
 the vector-tensor $f_{n_1\cdots n_j}(k)$, given in (\ref{smsv}), is completely symmetric with respect to
the permutations of the vector indices.

As in (\ref{scalpr}),
we define the Hermitian scalar product of the spin-tensor functions
 $\psi(k)$ and $\xi(k)$ of type
 $(\frac{j}{2}, \frac{j}{2})$ in the following way:
\begin{equation}\label{nchp}
(\psi(k), \; \xi(k)):=\overline{\psi}^{\;(\alpha_1...\alpha_j)}_{(\dot{\beta}_1\cdots \dot{\beta}_j)}(k)\; \xi^{\;(\dot{\beta}_1\cdots \dot{\beta_j})}_{(\alpha_1...\alpha_j)}(k)
\end{equation}
where $\overline{\psi}^{\;(\alpha_1...\alpha_j)}_{(\dot{\beta}_1\cdots \dot{\beta}_j)}(k):=(\psi^{\;(\dot{\alpha}_1...\dot{\alpha}_j)}_{(\beta_1\cdots \beta_j)}(k))^*$.
 Recall that under complex conjugation $*$
 the dotted indices of the spin-tensors are converted to
 undotted indices
 and vice versa. Now one can check that the
 spin-tensors $\overset{_{(m)}}{e}(k)$ defined in (\ref{npkp})
  are orthonormal with respect to the scalar product (\ref{nchp}).
 This follows from the chain of equations:
\begin{equation}\label{r1ic}
\begin{array}{c}
(\overset{_{(m)}}{e}(k),\overset{_{(\tilde{m})}}{e}(k))=
\frac{1}{(2j)!}\overset{2j}{\underset{i=1}{\Pi}}
\Bigl(\bigl((A_{(k)})^{\;\;\; \gamma_i}_{\beta_i}
\bigl(A^{-1\dagger}_{(k)}\; \tilde{\sigma}_0\bigr)^{\dot{\alpha_i} \gamma_{j+i}}\bigr)^*
(A_{(k)})^{\;\;\; \rho_i}_{\alpha_i}
\bigl(A^{-1\dagger}_{(k)}\; \tilde{\sigma}_0\bigr)^{\dot{\beta_i}\rho_{j+i}}\Bigr) \cdot \\[0.5cm]
\cdot \epsilon^{(m)}_{\gamma_1\cdots \gamma_{2j}}
\epsilon^{(\tilde{m})}_{\rho_1\cdots\rho_{2j}}
=\frac{1}{(2j)!}\epsilon^{(m)}_{\gamma_1\cdots \gamma_{2j}}\epsilon^{(\tilde{m}) \gamma_1\cdots \gamma_{2j}}=\delta^{_{(m) (\tilde{m})}},
\end{array}
\end{equation}
where we used the identities:
\begin{equation}\label{r2ic}
\bigl((A_{(k)})^{\;\;\; \gamma_i}_{\beta_i}
\bigl(A^{-1\dagger}_{(k)}\; \tilde{\sigma}_0\bigr)^{\dot{\alpha_i}\gamma_{j+i}}\bigr)^*=
(A_{(k)}^{\dagger})^{\gamma_i}_{\;\;\; \dot{\beta}_i}
\bigl(A^{-1}_{(k)}\; \tilde{\sigma}_0\bigr)^{\gamma_{j+i}\alpha_i} \; , \;\;\;
\tilde{\sigma}_0 = I_2 \; ,
\end{equation}
\begin{equation}\label{r3ic}
\epsilon^{(m)}_{\gamma_1\cdots \gamma_{2j}} \;
\epsilon^{(\tilde{m}) \gamma_1\cdots \gamma_{2j}}
=(2j)! \; \delta^{_{(m) (\tilde{m})}} \; .
\end{equation}
 Formula (\ref{r3ic}) follows
 directly from the normalization (\ref{pre1})
and the properties of $\epsilon^{(m)}_{\rho_1\cdots\rho_{2j}}$,
 which were discussed after eq. (\ref{pre1}). We stress that
in equations (\ref{r1ic})--(\ref{r3ic}) it is not needed
 to put dots over the indices $\gamma_i$
since these indices correspond to the representations of the group $SU(2)$.

Now we convert the spin-tensor wave functions $\psi^{(\dot{\beta}_1\cdots \dot{\beta_j})}_{(\gamma_1...\gamma_j)}(k)$
 to the vector-tensor functions $f_{n_1\cdots n_j}(k)$
  by means of relation
(\ref{smsv}) and substitute expression (\ref{vtdsw}) for $\psi^{(\dot{\beta}_1\cdots \dot{\beta_j})}_{(\gamma_1...\gamma_j)}(k)$
 in terms of Wigner's coefficients $\phi_m(k)$. As a result, we obtain
 the expansion
\begin{equation}
f_{n_1\cdots n_j}(k)=\frac{1}{\sqrt{2^j (2j)!}} \;
\sum_{m=-j}^j
 \phi_m(k) \; {\sf e}^{_{(m)}}_{n_1\cdots n_j}(k) \; ,
\end{equation}
where we used the notation:
\begin{equation}\label{nvpol1}
{\sf e}^{_{(m)}}_{n_1\cdots n_j}(k)=
\frac{1}{\sqrt{2^j}}(\sigma_{n_1})_{\alpha_1\dot{\beta}_1}
\cdots(\sigma_{n_j})_{\alpha_j\dot{\beta}_j}\,
\varepsilon^{\alpha_1\gamma_1} \cdots \varepsilon^{\alpha_j\gamma_j}\,
\overset{_{(m)}}{e}^{(\dot{\beta_1}...\dot{\beta_j})}_{(\gamma_1...\gamma_j)}(k)
\; .
\end{equation}
The vector-tensors ${\sf e}^{_{(m)}}_{n_1\cdots n_j}(k)$
 will be called polarization tensors for particles with spin $j$.
This terminology is natural since the
 vector-tensors ${\sf e}^{_{(m)}}_{n_1\cdots n_j}(k)$ form the basis
 in the expansion of the fields $f_{n_1\cdots n_j}(k)$
  over Wigner's coefficients $\phi_m(k)$, which in view of (\ref{rpbv})
 and (\ref{S3Tm}) are propotional to contributions of projection $m$
 of the spin component $S_3$.

For further purposes, we need to calculate the normalization of the polarization tensors:
\begin{equation}\label{ctnp}
\begin{array}{c}
\overline{{\sf e}}^{_{(m)}n_1\cdots n_j}(k){\sf e}^{_{(\tilde{m})}}_{ n_1\cdots n_j}(k)=
\frac{1}{2^j}
(\sigma^{n_1})_{\tau_1\dot{\rho}_1}\cdots(\sigma^{n_j})_{\tau_j\dot{\rho}_j}\,
\varepsilon^{\dot{\rho}_1\dot{\nu}_1} \cdots \varepsilon^{\dot{\rho}_j\dot{\nu}_j}\,
\overset{_{(m)}}{\overline{e}}^{(\tau_1...\tau_r)}_{(\dot{\nu}_1...\dot{\nu}_p)}(k)\cdot \\[0.5cm]
\cdot
(\sigma_{n_1})_{\alpha_1\dot{\beta}_1}\cdots(\sigma_{n_j})_{\alpha_j\dot{\beta}_j}\,
\varepsilon^{\alpha_1\gamma_1} \cdots \varepsilon^{\alpha_j\gamma_j}\,
\overset{_{(\tilde{m})}}{e}^{(\dot{\beta_1}...\dot{\beta_r})}_{(\gamma_1...\gamma_p)}(k)
=\\[0.5cm]=
\varepsilon_{\tau_1\alpha_1 }\cdots \varepsilon_{\tau_j\alpha_j }
\varepsilon_{\dot{\rho}_1 \dot{\beta}_1}\cdots \varepsilon_{\dot{\rho}_j\dot{\beta}_j }
\varepsilon^{\alpha_1\gamma_1} \cdots \varepsilon^{\alpha_j\gamma_j}
\varepsilon^{\dot{\rho}_1\dot{\nu}_1} \cdots \varepsilon^{\dot{\rho}_j\dot{\nu}_j}\,
\overset{_{(m)}}{\overline{e}}^{(\tau_1...\tau_j)}_{(\dot{\nu}_1...\dot{\nu}_j)}(k)
\overset{_{(\tilde{m})}}{e}^{(\dot{\beta_1}...\dot{\beta_j})}_{(\gamma_1...\gamma_j)}(k)
=\\[0.5cm]=
(-1)^j\,
\overset{_{(m)}}{\overline{e}}^{(\gamma_1...\gamma_j)}_{(\dot{\beta_1}...\dot{\beta_j})}(k)
\overset{_{(\tilde{m})}}{e}^{(\dot{\beta_1}...\dot{\beta_j})}_{(\gamma_1...\gamma_j)}(k)
=(-1)^j\delta^{_{(m) (\tilde{m})}},
\end{array}
\end{equation}
where $\overline{{\sf e}}^{_{(\tilde{m})} n_1\cdots n_j}(k)=({\sf e}^{_{(\tilde{m})} n_1\cdots n_j}(k))^*$.
 In the calculation (\ref{ctnp}) we used formula (\ref{stsg1}),
the properties of the metrics $\varepsilon_{\alpha \beta}$
and $\varepsilon_{\dot{\alpha} \dot{\beta}}$ (see (\ref{epsilon}))
and the normalization (\ref{r1ic}) of the spin-tensors $\overset{_{(m)}}{e}(k)$.
\newline
{\bf Remark 1.} Using definition (\ref{npkp}) in the case $r=p$ (i.e.
for integer spins $j$) we deduce the recurrence relation:
\begin{equation}\label{rcsd}
\begin{array}{c}
{\sf e}^{(m)}_{n_1 \cdots n_j}=\frac{1}{\sqrt{2j \;(2j-1)}}
\Bigl(\sqrt{(j+m)(j+m-1)}\; {\sf e}^{(m-1)}_{n_1\cdots n_{j-1}} \;
 {\sf e}^{(+)}_{n_j} +\\[0.5cm]
+\sqrt{(j-m)(j-m-1)} \; {\sf e}^{(m+1)}_{n_1\cdots n_{j-1}} \;
{\sf e}^{(-)}_{n_j} +
\sqrt{2(j-m)(j+m)} \; {\sf e}^{(m)}_{n_1\cdots n_{j-1}} \;
{\sf e}^{(0)}_{n_j} \Bigr) \; ,
\end{array}
\end{equation}
which completely determines the polarization tensor
${\sf e}^{(m)}_{n_1 \cdots n_j}$ for any $j$ via the vectors
of polarization ${\sf e}^{(a)}_{n}$ $(a=0,\pm)$ for $j=1$.

\vspace{0.2cm}

We construct the spin projection operator $\Theta(k)$ as the sum of
 products ${\sf e}^{_{(m)}}(k) \cdot \overline{{\sf e}}^{_{(m)}}(k)$
 over all polarizations $m$:
\begin{equation}\label{spp1}
\Theta^{n_1\cdots n_j}_{r_1\cdots r_j}(k):=(-1)^j\overset{j}{\underset{m=-j}{\Sigma}}{\sf e}^{_{(m)}}_{r_1\cdots r_j}(k)\overline{{\sf e}}^{_{(m)}n_1\cdots n_j}(k)
\end{equation}
This operator is sometimes called the density matrix for a massive particle with integer spin $j$, or
the Behrends-Fronsdal projection operator \cite{Fronsd}, \cite{BF}.
For spin $j=1$ the operator $\Theta(k)$ was
explicitly calculated in (\ref{theta1}).
In the case $j=2$, we presented formula (\ref{rios05}) for
 the operator $\Theta(k)$ in terms of polarization vectors (\ref{fpsl1}).
\begin{proposition}\label{svop} The operator $\Theta(k)$, defined in (\ref{spp1}), satisfies the following properties: \newline

1) projective property and reality: $\;\;\;\;\;\Theta^2=\Theta,\;\; \Theta^\dagger=\Theta$; \newline

2) symmetry:	\;\;\;\;\;$\Theta^{n_1\cdots n_j}_{\cdots r_i \cdots r_\ell \cdots}=\Theta^{n_1\cdots n_j}_{\cdots r_\ell \cdots r_i \cdots},\;\;
\Theta^{\cdots n_i\cdots n_\ell \cdots}_{r_1 \cdots r_j}=\Theta^{\cdots n_\ell \cdots n_i\cdots}_{r_1 \cdots r_j}$; \newline

3) transversality:\;\;\;\;\;$k^{r_1}\Theta^{n_1 \cdots n_j}_{r_1 \cdots r_j}=0$,
$\;\;k_{n_1}\Theta^{n_1 \cdots n_j}_{r_1 \cdots r_j}=0$; \newline

4) traceless:\;\;\;\;\; $\eta^{r_1 r_2}\Theta^{n_1 \cdots n_j}_{r_1 r_2\cdots r_j}=0$. \newline
\end{proposition}
{\bf Proof. }
 Two relations from the first property are written
  in the component form as
 $$
 \Theta^{n_1 \cdots n_j}_{r_1 r_2\cdots r_j} \;
 \Theta^{r_1 r_2 \cdots r_j}_{\ell_1 \ell_2\cdots \ell_j}
 =\Theta^{n_1 \cdots n_j}_{\ell_1 \cdots \ell_j} \; ,\;\; (\Theta^{n_1 \cdots n_j}_{r_1 r_2\cdots r_j})^*=
\Theta_{n_1 \cdots n_j}^{r_1 r_2\cdots r_j} \; ,
 $$
 and they are valid in view of definition (\ref{spp1})
of the operator $\Theta(k)$ and the normalization conditions (\ref{ctnp}) for the
 polarization vector-tensors ${\sf e}^{_{(m)}}(k)$.
The second property follows from the symmetry of the vector-tensors
 ${\sf e}^{_{(m)}}_{n_1 \cdots n_j}(k)$ with respect to
 any permutation of vector indices $\{n_1, \cdots, n_j\}$, as it
 follows from formula (\ref{nvpol1}).

The third property is equivalent to the transversality of the vector-tensors
${\sf e}^{_{(m)}}(k)$, i.e., is equivalent
to the condition $k^{r_1} {\sf e}^{_{(m)}}_{r_1 \cdots r_j}(k)=0$.
This condition follows from the chain of relations:
\begin{equation}\label{prsv3}
\begin{array}{c}
k^{r_1}{\sf e}^{_{(m)}}_{r_1 \cdots r_j}(k)= \frac{1}{\sqrt{2^j}}
k^{r_1}(\sigma_{r_1})_{\alpha_1 \dot{\beta}_1}\cdots (\sigma_{r_j})_{\alpha_j \dot{\beta}_j}\,
\varepsilon^{\alpha_1\gamma_1} \cdots \varepsilon^{\alpha_j\gamma_j}\,
\overset{_{(m)}}{e}^{_{(j)}(\dot{\beta_1}...\dot{\beta_j})}
_{(\gamma_1...\gamma_j)}(k)=\\ [0.5cm]=
\frac{{\sf m}}{\sqrt{2^j}}\, (\sigma_{r_2})_{\alpha_2 \dot{\beta}_2}\cdots
(\sigma_{r_j})_{\alpha_j \dot{\beta}_j}\,
\varepsilon^{\alpha_1\gamma_1} \cdots \varepsilon^{\alpha_j\gamma_j}\,
\overset{_{(m)}}{e}^{_{(j-1)}(\dot{\beta_2}...\dot{\beta_j})}_{(\alpha_1 \gamma_1...\gamma_j)}(k)=0 \; ,
\end{array}
\end{equation}
 (here we need to restore the label
 $(j)$ in the notation of the spin-tensors
 $\overset{_{(m)}}{e}(k) \to \overset{_{(m)}}{e}^{_{(j)}}(k)$).
 In  (\ref{prsv3}) we used definition (\ref{nvpol1}) of the vector-tensors
  ${\sf e}^{_{(m)}}_{r_1 \cdots n_j}(k)$, Dirac-Pauli-Fierz
 equations  (\ref{osudpe})
and the fact that the contraction
of a symmetric tensor with an antisymmetric one gives zero.

The fourth property is equivalent to the
 statement that the vector-tensors ${\sf e}^{_{(m)}}(k)$ are traceless,
i.e., $\eta^{r_1 r_2} {\sf e}^{_{(m)}}_{r_1 \cdots r_j}(k)=0$.
This statement can be proven as follows:
\begin{equation}
\begin{array}{c}
\eta^{r_1 r_2}{\sf e}^{_{(m)}}_{r_1 \cdots r_j}(k)=
(\sigma^{r_2})_{\alpha_1 \dot{\beta}_1}(\sigma_{r_2})_{\alpha_2 \dot{\beta}_2}\cdots (\sigma_{r_j})_{\alpha_j \dot{\beta}_j}\,
\varepsilon^{\alpha_1\gamma_1} \cdots \varepsilon^{\alpha_j\gamma_j}\,
\overset{_{(m)}}{e}^{(\dot{\beta_1}...\dot{\beta_j})}_{(\gamma_1...\gamma_j)}(k)=\\[0.5cm]=
(\sigma_{r_3})_{\alpha_3 \dot{\beta}_3}\cdots (\sigma_{r_j})_{\alpha_j \dot{\beta}_j}\,
\varepsilon^{\alpha_1\gamma_1} \cdots \varepsilon^{\alpha_j\gamma_j}\varepsilon_{\alpha_1\alpha_2}\varepsilon_{\dot{\beta}_1 \dot{\beta}_2}\,
\overset{_{(m)}}{e}^{(\dot{\beta_1}...\dot{\beta_j})}_{(\gamma_1...\gamma_j)}(k)=0
\end{array}
\end{equation}
where we apply formula (\ref{stsg1}).\hfill \qed


\section{Spin projection operators
for integer and half-integer spins.\label{PrBeFr}}
\setcounter{equation}0

In this Section, we find an explicit expression for the projection operator
$\Theta^{n_1\cdots n_j}_{r_1\cdots r_j}(k)$ (see (\ref{spp1}))
for any integer spin $j > 1$,
 in terms of the operator $\Theta^{n}_{r}(k)$.
 The operator $\Theta^{n}_{r}(k)$ is the projection operator
 for spin $j=1$ and it was calculated in (\ref{theta1}) .
  For the four-dimensional $D=4$ space-time, the Behrends-Fronsdal projection operator $\Theta(k)$
was explicitly constructed in \cite{Fronsd}, \cite{BF}.
Here we find a generalization of the Behrends-Fronsdal operator to the case of an
arbitrary number of dimensions $D>2$. Also in this Section
we prove an important formula which connects the projection operators
 for half-integer spins $j$
with the projection operators for integer spins $j+1/2$.
The construction will be based on the properties of this operator,
which are listed in Proposition {\bf \ref{svop}}.

 \vspace{0.2cm}

Instead of the tensor $\Theta^{n_1 \dots n_j}_{r_1 \dots r_j}(k)$
symmetrized in the upper and lower indices,
it is convenient to consider the generating function
 \be
 \lb{genT01}
 \Theta^{(j)}(x,y) = x^{r_1} \cdots x^{r_j} \,
 \Theta^{n_1 \dots n_j}_{r_1 \dots r_j}(k) \,
 y_{n_1} \cdots y_{n_j} \; .
 \ee
For concreteness, we shall assume that the tensor with the components
 $\Theta^{n_1 \dots n_j}_{r_1 \dots r_j}(k)$ is defined in the pseudo-Euclidean $D$-dimensional
 space  $\mathbb{R}^{s,t}$
 $(s+t=D)$ with an arbitrary metric
 $\eta = ||\eta_{mn}||$,
 having the signature $(s,t)$. Indices
 $n_\ell$ and $r_\ell$ in (\ref{genT01})
 run through values $0,1,\dots,D-1$ and
 $(x_0,...,x_{D-1})$,
 $(y_0,...,y_{D-1}) \in \mathbb{R}^{s,t}$.

\begin{proposition}\label{svop1} The generating function (\ref{genT01})
of the covariant projection operator
 $\Theta^{n_1 \dots n_j}_{r_1 \dots r_j}$
 (in $D$-dimensional space-time),
satisfying properties 1)-4), listed in Proposition {\bf \ref{svop}}, has the form
 \be
 \lb{genT02}
 \Theta^{(j)}(x,y) = \sum_{A=0}^{[\frac{j}{2}]} a^{(j)}_A \;
 \bigl(\Theta^{(y)}_{(y)} \, \Theta^{(x)}_{(x)} \bigr)^A  \;
 \bigl(\Theta^{(y)}_{(x)} \bigr)^{j -2A}  \; ,
 \ee
 where $[\frac{j}{2}]$ -- integer part of $j/2$,
  \be
 \lb{genT11}
 a^{(j)}_{A} = \Bigl( - \frac{1}{2}\Bigr)^A \frac{j!}{
 (j -2A)! \, A! \, (2j +D-5)(2j +D-7)\cdots (2j +D -2A -3)}  \; ,
 \;\;\; (A \geq 1) \; ,
 \ee
 $a^{(j)}_{0}=1$, and the function
 $\Theta^{(y)}_{(x)}$ is defined as follows
 ($\eta_{r n}$ -- the metric of space $\mathbb{R}^{s,t}$):
 \be
 \lb{genT}
 \Theta^{(y)}_{(x)} \equiv  \Theta^{(1)}(x,y) =  x^r \, y_n \, \Theta^{n}_{r} \; ,
 \;\;\;\; \Theta_{r}^{n}=\eta_{r}^{n}-\frac{k_r k^n}{k^2} \; .
 \ee
 The generating function (\ref{genT02})
 satisfies the differential equation
 \be
 \lb{genT12}
 \frac{\partial}{\partial x^r} \frac{\partial}{\partial y_r}
  \Theta^{(j)}(x,y) =
 \frac{j (j+D-4)(2j+D-3)}{(2j+D-5)} \, \Theta^{(j-1)}(x,y) \; .
 \ee
\end{proposition}
{\bf Proof. }
We recall (see (\ref{thepr}), (\ref{theta1})) that the matrix
$\Theta^{n}_{r}$ which is defined in (\ref{genT})
 is a projection operator onto the subspace orthogonal to the
$D$-dimensional vector with the components $k^{r}$:
\begin{equation}\label{pr1}
k^r \, \Theta_{r}^{n}= 0 = \Theta_{r}^{n} \, k_n \; , \;\;\;\;
\Theta^r_r=\eta_{r n}\, \Theta^{rn} = D-1 \; , \;\;\;\;
 \Theta_{r}^{n} \, \Theta^{r}_{m}  = \Theta_{m}^{n} \; .
\end{equation}
Taking into account this fact, the most general covariant operator
 $\Theta^{n_1 \dots n_j}_{r_1 \dots r_j}(k)$,
 satisfying properties 2) and 3) from Proposition {\bf \ref{svop}},
is written as follows:
\begin{equation}\label{eq1b}
\Theta^{n_1 \dots n_j}_{r_1 \dots r_j}(k) =
 \frac{1}{(j!)^2}  \sum_{\sigma,\mu \in S_j}
\Theta^{n_{\sigma(1)}}_{\ell_1}
...\Theta^{n_{\sigma(j)}}_{\ell_j} \,
 B^{\ell_1 \dots \ell_j}_{m_1\dots m_j}(k) \,
\Theta^{m_1}_{r_{\mu(1)}} ...\Theta^{m_j}_{r_{\mu(j)}}\; ,
\end{equation}
where $\sigma,\mu \in S_{j}$ are permutations of the indices $\{1,2,\dots,j\}$,
and the components $B^{\ell_1 \dots \ell_j}_{m_1\dots m_j}(k)$
 are any covariant combinations of the metric $\eta_{rm}$ and coordinates $k_r$
 of the $D$-vector of momentum.
 Since the matrices $\Theta^{m}_{r}$ used in the
 right-hand side of (\ref{eq1b}) are transverse to the
 $D$-momentum $k$ (see (\ref{pr1})), the external indices of the tensor
 $B^{\ell_1 \dots \ell_j}_{m_1\dots m_j}(k)$
  can be associated only with the
  indices of the metric $\eta$ and, therefore, this tensor is represented
  in the form
 \be
 \lb{eq2b}
 \begin{array}{c}
 B^{\ell_1 \dots \ell_j}_{m_1\dots m_j}(k) =
 a^{(j)}_0(k) \; \eta^{\ell_1}_{m_1} \cdots \eta^{\ell_j}_{m_j} +
 a^{(j)}_1(k) \; \eta^{_{\ell_1\ell_2}} \eta_{_{m_1m_2}}
 \eta^{\ell_3}_{m_3}\cdots \eta^{\ell_j}_{m_j} + \\ [0.2cm]
 + a^{(j)}_2(k) \; \eta^{_{\ell_1\ell_2}} \eta_{_{m_1m_2}}
 \eta^{_{\ell_3\ell_4}} \eta_{_{m_3m_4}}
 \eta^{\ell_5}_{m_5}\cdots \eta^{\ell_j}_{m_j} + \dots \; ,
 \end{array}
 \ee
where the functions $a^{(j)}_A(k)$ depend on
  the invariants $(k)^2=k^r k_r$.
Substitution (\ref{eq2b}) into (\ref{eq1b}) gives
 \begin{equation}\label{anz1}
\Theta_{n_1 n_2...n_j}^{
r_1 r_2... r_j} =
\frac{1}{(j!)^2} \sum_{A=0}^{[\frac{j}{2}]} \;
\sum_{\sigma,\mu \in S_j}
a^{(j)}_A \Bigl( \prod_{\ell =1}^A
\Theta_{n_{\mu(2\ell-1)}n_{\mu(2\ell)}}
\Theta^{r_{\sigma(2\ell-1)}r_{\sigma(2\ell)}}
\prod_{i=2A+1}^{j}\Theta_{n_{\mu(i)}}^{r_{\sigma(i)}}
\Bigr) \; .
\end{equation}
 Finally, using expression (\ref{anz1})
 in (\ref{genT01}), we obtain formula (\ref{genT02})
for the generating function of the projection operator.
The coefficients  $a^{(j)}_A$ in (\ref{genT02}),
 as we will see below, do not depend on $(k)^2$
and their explicit form is fixed by properties 1) and 4)
 from Proposition {\bf \ref{svop}}.

 \vspace{0.2cm}

Property 4) (traceless) in Proposition {\bf \ref{svop}} for the tensor
 $\Theta$ is equivalent to the harmonic equation for the generating function
 (\ref{genT02}):
 \be
 \lb{genT04}
(\partial)^2 \, \Theta^{(j)}(x,y) = 0 \; ,
 \ee
 where $(\partial)^2= \partial_{x^r} \partial_{x_{_r}}$
 and $\partial_{x^r} = \partial/\partial x^r$. Let us substitute
  expression (\ref{genT02}) into the equation
  (\ref{genT04}) to find the conditions which fix the coefficients $a^{(j)}_A$.
It is convenient to rewrite the series (\ref{genT02}) in the form
 \be
 \lb{genT02a}
 \Theta^{(j)}(x,y) =
 \sum_{A=0}^{[\frac{j}{2}]} \tilde{a}_A \;
 (\Theta^{(x)}_{(x)})^A  (\Theta^{(y)}_{(x)})^{j -2A} \; ,
 \;\;\;\; \tilde{a}_A \equiv
  \bigl(\Theta^{(y)}_{(y)}\bigr)^A  a^{(j)}_A \; .
 \ee
As a result of the substitution (\ref{genT02}) into equation
  (\ref{genT04}), we have
 \be
 \lb{genT05}
 \begin{array}{c}
 (\partial)^2 \, \Theta^{(j)}(x,y) =
 \sum\limits_A \tilde{a}_A
(\partial)^2 \Bigl(  (\Theta^{(x)}_{(x)})^A  (\Theta^{(y)}_{(x)})^{j -2A}\Bigr) =
\\ [0.4cm]
= \sum\limits_A \tilde{a}_A
\Bigl( 2 A \bigl( 2 j+D-3-2A \bigr) \; (\Theta^{(x)}_{(x)})^{A-1}
 \cdot  (\Theta^{(y)}_{(x)})^{j -2A} \; + \\ [0.2cm]
 + \; (j -2A)(j -2A-1)
 \Theta^{(y)}_{(y)} \, (\Theta^{(x)}_{(x)})^A (\Theta^{(y)}_{(x)})^{j -2A-2} \Bigr) \; ,
\end{array}
 \ee
where in the second equality we used the relations
 \be
 \lb{genT06}
 \begin{array}{c}
 (\partial)^2 \, (\Theta^{(x)}_{(x)})^A =
 2 \, A(2 A+D-3)  \, (\Theta^{(x)}_{(x)})^{A-1} \; ,
 \end{array}
 \ee
 \be
 \lb{genT07}
 \begin{array}{c}
 2 [\partial_{x^r} \, (\Theta^{(x)}_{(x)})^A]
[\partial_{x_r} (\Theta^{(y)}_{(x)})^{j -2A}]  =
  4 A \, (j -2A) \,  (\Theta^{(x)}_{(x)})^{A-1}
 \cdot  (\Theta^{(y)}_{(x)})^{j -2A}  \; ,
 \end{array}
 \ee
 \be
 \lb{genT08}
 \begin{array}{c}
 (\partial)^2 \,(\Theta^{(y)}_{(x)})^{j -2A} = (j -2A)(j -2A-1)
 \Theta^{(y)}_{(y)} \,(\Theta^{(y)}_{(x)})^{j -2A-2} \; .
  \end{array}
 \ee

 Thus, to fulfill the identity (\ref{genT04}),
according to (\ref{genT05}),  it is necessary to require the recurrence
relation for the coefficients $\tilde{a}_{A}$:
 \be
 \lb{genT09}
 \tilde{a}_{A +1} = - \frac{1}{2} \frac{(j -2A)(j -2A-1)}{
 (A+1)(2j -2A+D-5)} \; \Theta^{(y)}_{(y)} \; \tilde{a}_{A} \; ,
 \ee
 which in turn gives the relations for the coefficients $a^{(j)}_A$:
 \be
 \lb{genT10}
 a^{(j)}_{A} = - \frac{1}{2} \; \frac{(j -2A+2)(j -2A+1)}{
 A \; (2j -2A+D-3)} \; a^{(j)}_{A-1} \; .
 \ee
The solution of equation (\ref{genT10}) has the form
 \be
 \lb{genT11b}
 a^{(j)}_{A} = \Bigl( - \frac{1}{2}\Bigr)^A \frac{j! \; a^{(j)}_{0}}{
 (j -2A)! \, A! \, [(2j +D-5)(2j +D-7)\cdots (2j +D-3-2A)]}  \; ,
 \ee
i.e., the condition (\ref{genT04}) determines the coefficients
 $a^{(j)}_{A}$ up to a single arbitrary factor $a^{(j)}_{0}$.
 Note that, firstly, the product of factors in square brackets in the denominator
  (\ref{genT11b}) must be considered equal to unity for $A=0$ and,
secondly, when substituting the coefficients (\ref{genT11b})
 into the sum (\ref{genT02}),
it is clear that this sum is automatically terminated for $A > j/2$
since in this case an infinite factor  $(j -2A)!=\infty$ appears in the denominator.

\vspace{0.2cm}

 Let us now verify condition 1) from Proposition
 {\bf \ref{svop}}. First of all, the property  $\Theta(x,y)=\Theta(y,x)$
 for the function (\ref{genT02}) and reality condition $\Theta^*=\Theta$
 are equivalent to $\Theta^\dagger=\Theta$
 for the matrices (\ref{anz1}).
The projector condition $\Theta^2 = \Theta$ (see property 1
in Proposition {\bf \ref{svop}})
for the matrix (\ref{anz1}) can be checked directly:
 $$
 \begin{array}{c}
 (\Theta^2)_{n_1 n_2...n_j}^{
r_1 r_2... r_j} = \Theta_{m_1 m_2...m_j}^{
r_1 r_2... r_j}
\frac{1}{(j!)^2} \sum\limits_{A=0}^{[\frac{j}{2}]} \;
\sum\limits_{\sigma,\mu \in S_j}
a^{(j)}_A \Bigl( \prod\limits_{\ell =1}^A
\Theta^{m_{\sigma(2\ell-1)}m_{\sigma(2\ell)}}
\Theta_{n_{\mu(2\ell-1)}n_{\mu(2\ell)}}
\prod\limits_{i=2A+1}^{j}\Theta_{n_{\mu(i)}}^{m_{\sigma(i)}}
\Bigr) =  \\ [0.3cm]
= \Theta_{m_1 m_2...m_j}^{
r_1 r_2... r_j}
\frac{1}{j!}  \,
\sum\limits_{\mu \in S_j}
a^{(j)}_0 \Bigl( \Theta_{n_{\mu(1)}}^{m_1} \cdots
\Theta_{n_{\mu(j)}}^{m_j} \Bigr) =
a^{(j)}_0 \, \Theta_{n_1 n_2...n_j}^{r_1 r_2... r_j} \; ,
\end{array}
 $$
where in the second and third equalities we used the identities
 $$\Theta_{m_1 m_2...m_j}^{
r_1 r_2... r_j} \Theta^{m_i m_\ell}=0 \; , \;\;\;
\Theta_{m_1... m'_i...m_j}^{
r_1 r_2... r_j} \Theta^{m'_i}_{m_i}=
\Theta_{m_1... m'_i...m_j}^{
r_1 r_2... r_j} \eta^{m'_i}_{m_i} =
\Theta_{m_1 m_2... m_j}^{r_1 r_2... r_j} \; ,
$$
that follow from conditions 3) and 4) of Proposition {\bf \ref{svop}}.
Thus, to fulfill the projector condition,
 we must fix the initial coefficient in the expansion
(\ref{genT02}) as $a^{(j)}_{0}=1$. With this value
formula (\ref{genT11b}) turns into formula (\ref{genT11}).

\vspace{0.2cm}

Finally, we prove the identity (\ref{genT12}).
 For this we calculate
 \be
 \lb{genT16}
 \begin{array}{c}
 \partial_{x^r} \partial_{y_r} \, \Theta^{(j)}(x,y) =
 \sum\limits_{A=0}^{[\frac{j}{2}]} a^{(j)}_A \,
 \partial_{x^r} \partial_{y_r}
  \Bigl( (\Theta^{(x)}_{(x)} \Theta^{(y)}_{(y)})^A
  (\Theta^{(y)}_{(x)})^{j -2A} \Bigr)=
 \\ [0.2cm]
= \sum\limits_{A=0}^{[\frac{j}{2}]}
\Bigl(a^{(j)}_{A+1}  4 (A+1)^2  +
 a^{(j)}_{A} (j -2A)(D-2 + j+ 2A) \Bigr)
 (\Theta^{(x)}_{(x)} \Theta^{(y)}_{(y)})^{A}
 (\Theta^{(y)}_{(x)})^{j -2A-1} \; ,
\end{array}
 \ee
 where we used the equalities
 $$
 \begin{array}{c}
 [\partial_{x^r} \partial_{y_r} (\Theta^{(x)}_{(x)} \Theta^{(y)}_{(y)})^A]
  = 4 \, A^2 \, \Theta^{(y)}_{(x)} \,
  (\Theta^{(x)}_{(x)} \Theta^{(y)}_{(y)})^{A-1} \; ,
  \\ [0.2cm]
 [\partial_{x^r} \, (\Theta^{(x)}_{(x)}\Theta^{(y)}_{(y)})^A]
[\partial_{y_r} (\Theta^{(y)}_{(x)})^{j -2A}] =
2 (j-2A) A \,  (\Theta^{(x)}_{(x)}\Theta^{(y)}_{(y)})^A
\, (\Theta^{(y)}_{(x)})^{j -2A-1} = \\ [0.2cm]
= [\partial_{y_r} \, (\Theta^{(x)}_{(x)}\Theta^{(y)}_{(y)})^A]
[\partial_{x^r} (\Theta^{(y)}_{(x)})^{j -2A}] \; ,  \\ [0.2cm]
[\partial_{x^r} \partial_{y_r} \,(\Theta^{(y)}_{(x)})^{j -2A}] =
(j -2A)(D +j-2A-2) \,(\Theta^{(y)}_{(x)})^{j -2A-1} \; .
\end{array}
 $$
Taking into account the explicit formula for the coefficients
 (\ref{genT11}), we obtain the relation
  $$
  a_{A+1}^{(j)}4(A+1)^2 +  a_A^{(j)} (j -2A)(D +j+2A-2) =
  \frac{j (j+D-4)(2j+D-3)}{2j+D-5} \;
  a_A^{(j-1)} \; ,
 $$
 substituting this into (\ref{genT16}),
  we immediately derive the identity (\ref{genT12}). \hfill \qed

  \vspace{0.2cm}

 \noindent
 {\bf Remark 1.} Identity (\ref{genT12})
 for the generating functions (\ref{genT01})
gives the equality that connects the projection operators (\ref{anz1})
for the spins $j$ and $(j-1)$:
 \be
 \lb{genT15}
 \eta^{r_1}_{n_1} \; (\Theta^{(j)})^{n_1 n_2 \dots n_j}_{r_1 r_2 \dots r_j} =
  (\Theta^{(j)})^{r_1 n_2 \dots n_j}_{r_1 r_2 \dots r_j} =
 \frac{(j+D-4)(2j+D-3)}{j(2j+D-5)}
  (\Theta^{(j-1)})^{n_2 \dots n_j}_{r_2 \dots r_j}\; .
 \ee
In other words the trace of the matrix $\Theta^{(j)}$
 over the pair of indices is proportional to the matrix $\Theta^{(j-1)}$.
Using formula (\ref{genT15}), we can calculate the complete trace
of the Behrends-Fronsdal projector $\Theta^{(j)}$ in the case of
  $D$-dimensional space-time ($D\geq 3$):
 \be
 \lb{genT17}
  (\Theta^{(j)})^{r_1 r_2 \dots r_j}_{r_1 r_2 \dots r_j} =
 \frac{(D-4+j)!}{j!\, (D-3)!}\; (2j+D-3) \; .
 \ee
 This trace is equal to the dimension of the subspace,
which is cut out
from the space of vector-tensor wave functions
  $f_{n_1 ... n_j}(k)$ by the projector $\Theta^{(j)}$.
  In other words, the trace (\ref{genT17})
is equal to the number $N$ of independent components
of symmetric vector-tensor wave functions
  $f_{(n_1 ... n_j)}(k)$ that satisfy the conditions
   $$
   k^{n_1} f_{(n_1 ... n_j)}(k) = 0 \; , \;\;\;
   \eta^{n_1 n_2}f_{(n_1 n_2 ... n_j)}(k) = 0 \; .
   $$
   On the space of these functions
   an irreducible massive representation of the
 $D$-dimensional rotation group is realized.
For example, for $D=3$ we have $N=2$ ($\forall j$), and for $D=4$
 formula (\ref{genT17}) gives the well-known result
 $N=(2j+1)$, which coincides with the number of polarizations of the massive particles with spin $j$
 (it coincides with the dimension of the irreducible representation
  with spin  $j$ of a small subgroup
 $SU(2) \subset SL(2,\mathbb{C})$).

 \noindent
 {\bf Remark 2.} From relation (\ref{genT02}) the useful identity \cite{BF}
 immediately follows:
 \be
 \lb{genT18}
  \Theta^{(j)}(x,x) =
 \bigl( \Theta^{(1)}(x,x) \bigr)^j \;
 \sum_{A=0}^{[\frac{j}{2}]} a^{(j)}_A = (k)^{-2j}
 \Bigl( (k)^2 (x)^2  -  (k^n x_n)^2 \Bigr)^j \;
 \sum_{A=0}^{[\frac{j}{2}]} a^{(j)}_A \; ,
 \ee
 where the coefficients $a^{(j)}_A$ are defined in (\ref{genT12}).

The sum of the coefficients $a^{(j)}_A$ in the right-hand side
 of (\ref{genT18}) can be calculated explicitly
 by using relations (\ref{genT12}) and (\ref{genT18}).
Indeed, we put $x_r=y_r$  in (\ref{genT12}),
take into account (\ref{genT04})
and apply the operator
$\partial_{x^r}\partial_{x_r}$ to both sides of equality (\ref{genT18}).
After that, comparing the results
obtained in both sides of (\ref{genT18}),
we deduce the recurrence equation:
$$
S^{(j)}_D=\frac{(j+D-4)}{(2j+D-5)}S^{(j-1)}_D \; ,
$$
where $S^{(j)}_D= \sum_{A=0}^{[\frac{j}{2}]} a^{(j)}_A$.
Solving this equation with the initial condition $S^{(1)}_D=a^{(j)}_0=1$, we find:
\begin{equation}
\lb{suma}
S^{(j)}_D=\frac{(j+D-4)!}{(D-3)!\;(2j+D-5)(2j+D-7) \cdots (D-1)} \; , \;\;\;\;
j >1 \; .
\end{equation}
For $D=4$ the expression for the sum (\ref{suma}) is simplified and we have
$S^{(j)}_4= \frac{j!}{(2j-1)!!}$, where $j>0$.

 \vspace{0.4cm}

 Now we will construct the spin projection operator
 $\Theta^{(j)}(k)$
 in the case of half-integer spins $j$. Recall that
 for spin $j=1/2$ the operator $\Theta^{(1/2)}(k)$ was
explicitly calculated in (\ref{prh1}).
For $j=3/2$ we found formula (\ref{ers3p}) for
 the operator $\Theta^{(3/2)}(k)$ in terms of bispinors
 (\ref{epem}) and polarization vectors (\ref{fpsl1}).
 To obtain the general formula for any half-integer spin $j$
 (in the case of 4-dimensional
 space-time), one can use the definition of the
 spin projection operator
$\bigr((\Theta^{(j)})^{n_1 \dots n_{j-1/2}}_{r_1 \dots r_{j-1/2}}\bigl)^{\;\; B}_{A}$
 as the sum over polarizations:
\begin{equation}\label{genT20}
\bigr((\Theta^{(j)})^{n_1 \dots n_{j-1/2}}_{r_1 \dots r_{j-1/2}}\bigl)^{\;\; B}_{A}=
 \frac{(-1)^{_{j-1/2}}}{2}
 \sum_{m=-j}^j (e^{_{(m)}}_{r_1\cdots r_{j-1/2}})_A
 (\overline{e}^{_{(m)}n_1\cdots n_{j-1/2}})^B \; .
\end{equation}
Here $r_i$ and $n_i$ are the vector indices while
$A$ and $B$ are the indices of the Dirac spinors.
The polarization spin-tensors
$(e^{_{(m)}}_{r_1\cdots r_{j-1/2}})_A$ of arbitrary
half-integer spin $j$
are expressed in terms of polarizations of the integer spin
and bispinors (\ref{epem}) as follows:
\begin{equation}\label{ovpc}
\begin{array}{c}
(e^{(m)}_{r_1\cdots r_{j-1/2}})_A =
\sqrt{\frac{(j+m)}{2j}}{\sf e}^{_{(m-1/2)}}_{r_1\cdots r_{j-1/2}}
e^{(+)}_A +
\sqrt{\frac{(j-m)}{2j}}{\sf e}^{_{(m+1/2)}}_{r_1 \cdots r_{j-1/2}}
e^{(-)}_A \;, \\[0.5cm]
m=-j, \dots, j \; .
\end{array}
\end{equation}
Formula (\ref{ovpc}) is a generalization
of (\ref{vp3P}) and is obtained from equations
(\ref{npkp}) and (\ref{vtdsw}). After substitution
of (\ref{ovpc}) into (\ref{genT20}) one can
calculate the sum over polarizations $m$ in (\ref{genT20}) and deduce explicit formula for the operator $\Theta^{(j)}$.
For a special case $j=1/2$ it was done in (\ref{prh1}).
However, it is rather a long way to obtain an explicit expression
 for the operator $\Theta^{(j)}$. Here we will use another
 method which is based on the ideas of the paper \cite{BF}.
 Moreover, this method gives us a possibility to
 find the spin projection operator $\Theta^{(j)}$
 (for half-integer spins $j$) for the general case of
 arbitrary space-time dimension $D$.

 \begin{proposition}\label{svop5}  For
 arbitrary space-time
 dimension $D>2$ and any half-integer spin $j$ the projection
operator $\Theta^{(j)}$ satisfying
conditions 1)---4) of  Proposition {\bf \ref{svop}}
and additional spinorial conditions
\be
  \lb{genT21}
 (\Theta^{(j)})^{n_1 \dots n_{j-1/2}}_{r_1 \dots r_{j-1/2}}
 \cdot \gamma_{n_1} = 0  =  \gamma^{r_1} \cdot
 (\Theta^{(j)})^{n_1 \dots n_{j-1/2}}_{r_1 \dots r_{j-1/2}}
  \; ,
  \ee
has the form
 \be
 \lb{genT19}
((\Theta^{(j)})^{n_1 \dots n_{j-1/2}}_{r_1 \dots r_{j-1/2}}) ^{\;\; B}_{ A}=
 c^{(j)} \; (\Theta^{(1/2)})^{\;\; G}_{A}
\, (\gamma^{r})^{\;\; C}_{G}(\gamma_{n})^{\;\; B}_{C}
(\Theta^{(j+\frac{1}{2})})^{n\,n_1 \dots n_{j-1/2}}_{r\,r_1 \dots r_{j-1/2}} \; ,
 \ee
where $(\Theta^{(1/2)})^{\;\; G}_{A} =
\frac{1}{2m}(k^n \, \gamma_n + m\, I_4)^{\;\; G}_{A}$ ---
spin projection operator (\ref{prh1}) for spin $j=1/2$,
$\Theta^{(j+\frac{1}{2})}$ -- projection operator (\ref{anz1})
for integer spin $(j+\frac{1}{2})$ and
the factor $c^{(j)}$ is defined as
\begin{equation}
\label{coeffj}
c^{(j)}=\frac{(j+1/2)}{(2j+D-2)}   \; .
\end{equation}
\end{proposition}
{\bf Proof.}
  First of all we note that conditions 2)---4)
  of Proposition {\bf \ref{svop}} are trivially
  fulfilled for the operator (\ref{genT19}) since these conditions
  are valid for the integer spin projection operator
  $\Theta^{(j+\frac{1}{2})}$ by construction.
  Conditions (\ref{genT21}) are valid for the
  operator (\ref{genT19}) in view of the
  properties 2),4)
  of Proposition {\bf \ref{svop}}
  and the identity $\gamma_n\gamma_k +\gamma_k\gamma_n
  = 2\eta_{nk}$ for gamma-matrices, so that we have
  the first equality in (\ref{genT21})
  $$
  \gamma_n \; (\Theta^{(j+\frac{1}{2})})^{n\,n_1 \dots n_{j-1/2}}_{r\,r_1 \dots r_{j-1/2}} \;  \gamma_{n_1} =
  (\Theta^{(j+\frac{1}{2})})^{n\,n_1 \dots n_{j-1/2}}_{r\,r_1 \dots r_{j-1/2}} \; \eta_{n n_1} = 0 \; .
  $$
  The second equality in (\ref{genT21}) can be proved
  analogously. So the rest of what we have to prove is
  the projection property
\begin{equation}\label{vdkn}
((\Theta^{(j)})^{n_1 \dots n_{j-1/2}}_{r_1 \dots r_{j-1/2}}) ^{\;\; B}_{ A}
((\Theta^{(j)})^{p_1 \dots p_{j-1/2}}_{n_1 \dots n_{j-1/2}}) ^{\;\; C}_{ B}=((\Theta^{(j)})^{p_1 \dots p_{j-1/2}}_{r_1 \dots r_{j-1/2}}) ^{\;\; C}_{ A}
\end{equation}
As we will see below this property fixes the constant
$c^{(j)}$ explicitly. First we substitute into the left hand
side of (\ref{vdkn}) the representation (\ref{genT19})
and use the explicit form (\ref{anz1}) of the operator
 $\Theta^{(j+\frac{1}{2})}$. As a result,
 we obtain
\begin{equation}\label{vdkn1}
\begin{array}{c}
c^{(j)}(\Theta^{(1/2)}) \, \gamma^{r} \gamma_{n}
(\Theta^{(j+\frac{1}{2})})^{n\,n_1 \dots n_{j-1/2}}_{r\,r_1 \dots r_{j-1/2}}
\cdot (\Theta^{(j)})^{p_1 \dots p_{j-1/2}}_{
n_1 \dots n_{j-1/2}}=
c^{(j)}(\Theta^{(1/2)}) \, \gamma^{r} \gamma_{n} \; \cdot
\\[0.5cm]
 \cdot \; \frac{1}{((j+\frac{1}{2})!)^2}\!\!
{\displaystyle \sum_{_{P(n),P(r)}}}\!\! \bigl(\Theta^n_r \Theta^{n_1}_{r_1}\cdots\Theta^{n_{j-1/2}}_{n_{j-1/2}}
+a^{(j)}_1\Theta_{r r_1}\Theta^{n n_1}
{\displaystyle \prod_{i=2}^{j-1/2} \Theta^{n_i}_{r_i}}+... \bigr) \cdot
 (\Theta^{(j)})^{p_1 \dots p_{j-1/2}}_{n_1 \dots n_{j-1/2}}=
 \\[0.5cm]
 = c^{(j)}(\Theta^{(1/2)}) \, \gamma^{r} \gamma_{n} \frac{1}{((j+\frac{1}{2})!)^2} \!\!
{\displaystyle \sum_{_{P(n),P(r)}}}\!\!
 \bigl(\Theta^n_r \Theta^{n_1}_{r_1}\cdots\Theta^{n_{j-1/2}}_{r_{j-1/2}})
 \cdot (\Theta^{(j)})^{p_1 \dots p_{j-1/2}}_{
 n_1 \dots n_{j-1/2}} \; ,
 \end{array}
 \end{equation}
 where the sum is taken over all permutations
 $P(n)$ and $P(r)$ of indices $(n,n_1,...,n_{j-1/2})$
  and $(r,r_1,...,r_{j-1/2})$, and in the
  last equality we have used the conditions
  $$
  \gamma_n \; \Theta^{n n_1} \cdot
 (\Theta^{(j)})^{p_1 \dots p_{j-1/2}}_{n_1 \dots n_{j-1/2}}
  =0 \; , \;\;\;
  \Theta^{n_k n_\ell} \cdot
 (\Theta^{(j)})^{p_1 \dots p_{j-1/2}}_{n_1 \dots n_{j-1/2}}
  =0 \; .
  $$
  Finally the right hand side of (\ref{vdkn1})
  can be transformed as follows:
 \begin{equation}\label{vdkn2}
\begin{array}{c}
c^{(j)} \; (\Theta^{(1/2)}) \, \gamma^{r} \gamma_{n} \frac{1}{((j+\frac{1}{2})!)^2}
{\displaystyle \Bigl( \!\!\sum_{_{P(n),P(r)}}}
\bigl( \eta^n_r
 -\frac{k^n k_r}{\sf m^2})
 \eta^{n_1}_{r_1}\dots\eta^{n_{j-1/2}}_{r_{j-1/2}} \Bigr) \cdot
(\Theta^{(j)})^{p_1 \dots p_{j-1/2}}_{n_1 \dots n_{j-1/2}}=\\[0.5cm]=c^{(j)}\; (\Theta^{(1/2)}) \cdot
\Bigl(\frac{D}{(j+1/2)}+2(1-\frac{1}{(j+1/2)})-
 \frac{\hat{k}^2}{\sf m^2}\, \frac{1}{(j+1/2)}\Bigr) \cdot
(\Theta^{(j)})^{p_1 \dots p_{j-1/2}}_{r_1 \dots r_{j-1/2}}=\\[0.5cm]=
 {\displaystyle
 c^{(j)} \; \frac{(2j+D-2)}{(j+1/2)} \;
 (\Theta^{(j)})^{p_1 \dots p_{j-1/2}}_{r_1 \dots r_{j-1/2}}
 }  \; ,
\end{array}
\end{equation}
where we have used expression
(\ref{theta1}) for $\Theta^{n}_{r}$ and the identity
 $(\Theta^{(\frac{1}{2})}) \cdot (\Theta^{(j+\frac{1}{2})})
= (\Theta^{(j+\frac{1}{2})})$.
We have also taken into account that in view
of the sums over all permutations of indices
$(n,n_1,...,n_{j-1/2})$ and $(r,r_1,...,r_{j-1/2})$
we have
\begin{equation}\label{vdkn3}
\begin{array}{c}
\gamma^{r} \gamma_{n}\!\!  \sum\limits_{_{P(n),P(r)}} \!\!
\bigl(\eta^n_r \eta^{n_1}_{r_1}\dots
\eta^{_{n_{j-1/2}}}_{_{r_{j-1/2}}}\bigr)=
(j+1/2) \, D\!\!  \sum\limits_{_{P(n),P(r)}} \!\! \bigl(\eta^{n_1}_{r_1}
\dots\eta^{_{n_{j-1/2}}}_{_{r_{j-1/2}}} \bigr)+\\[0.5cm]+
(j+1/2)(j-1/2)\!\!  \sum\limits_{_{P(n),P(r)}} \!\! \bigl(\gamma^{n_1} \gamma_{r_1} \dots \eta^{_{n_{j-1/2}}}_{_{r_{j-1/2}}} \bigr)=
(j+1/2) \, D\!\!  \sum\limits_{_{P(n),P(r)}} \!\!
 \bigl(\eta^{n_1}_{r_1}\dots
 \eta^{_{n_{j-1/2}}}_{_{r_{j-1/2}}}\bigr)
 +\\[0.5cm]+
(j+1/2)(j-1/2)\!\!  \sum\limits_{_{P(n),P(r)}} \!\!
\bigl(2\eta^{n_1}_{r_1} \dots
\eta^{_{n_{j-1/2}}}_{_{r_{j-1/2}}}+
\gamma_{r_1} \gamma^{n_1}\dots
 \eta^{_{n_{j-1/2}}}_{_{r_{j-1/2}}}\bigr) \; ,
\end{array}
\end{equation}
where in the second equality we changed the order
of gamma-matrices. Finally, from the right-hand
side of (\ref{vdkn2}), we see that the
operator $(\Theta^{(j+\frac{1}{2})})$ is
 a projector only if the constant $c^{(j)}$
 is given by formula (\ref{coeffj}).
\hfill \qed

\vspace{0.3cm}

\noindent
{\bf Remark 3.} For the special case $D=4$
 formula (\ref{genT19}) for the projection spin operator
was obtained in \cite{BF} (see also e.g. \cite{ShizH}).
One can check that for
spin $j=1/2$ relation (\ref{genT19}) is
a trivial identity, while
 spin $j=3/2$ it
 takes the form
\begin{equation}
 \label{app32}
 \Theta^{(\frac{3}{2})}_{n_2 r_2} = \frac{2}{(D+1)} \;
\Theta^{(1/2)} \cdot \gamma^{r_1}\, \gamma^{n_1}
\, \Theta^{(2)}_{n_1 n_2 r_1 r_2} \; .
\end{equation}
For the case $D=4$
this expression can be verified by direct calculation
 of the sum over polarizations.
 It can be done by
using formulas (\ref{prh1}), (\ref{sp01}),
(\ref{ers3p}) and the representation (\ref{rios05})
for the density matrix of massive particles with spin $j=2$
which also follows from general formulas (\ref{rcsd}) and (\ref{spp1}).
For $j = 2$ formula (\ref{anz1}) gives:
\begin{equation}\label{grfr2}
\Theta^{(2)}_{n_1 n_2 r_1 r_2}=\frac{1}{2}(\Theta^{(1)}_{n_1 r_1 }\Theta^{(1)}_{n_2 r_2 }+\Theta^{(1)}_{n_1 r_2 }\Theta^{(1)}_{ r_1 n_2})-\frac{1}{3}\Theta^{(1)}_{n_1 n_2 }
\Theta^{(1)}_{ r_1 r_2} \; .
\end{equation}
By using this formula and the explicit
form (\ref{theta1}) of the
operator $\Theta^{(1)}_{n r }$, as well
as the projection property
$(\Theta^{(1/2)}) \cdot (\gamma^{r}\, k_r) =
{\sf m}\,(\Theta^{(1/2)})$,  we
obtain for (\ref{app32}):
\begin{equation}\label{rsHf}
 \Theta^{(\frac{3}{2})}_{n_2 r_2}=\frac{1}{3}\Theta^{(1/2)}\bigr(\eta_{n_2 r_2}+\gamma_{n_2} \gamma_{r_2}+\frac{1}{\sf m}(\gamma_{n_2}\,k_{r_2}
-\gamma_{r_2}\,k_{n_2})-
 2 \frac{k_{n_2} k_{r_2}}{{\sf m}^2}\bigl) \; .
\end{equation}

\section{Conclusion}
\setcounter{equation}0

In this paper, on the basis of unitary representations of the covering group $ISL(2,\mathbb{C})$ of the Poincar\'{e} group,
we have constructed explicit solutions of the wave equations for free massive particles of arbitrary spin $j$ (the Dirac-Pauli-Fierz equations). Then we
proposed the method for decomposing of these solutions into
a sum over independent components corresponding to different polarizations.
The sum over the polarizations for the density matrix of
particles with arbitrary integer spin is
calculated explicitly.
 This density matrix (spin projection operator)
 coincides with the Behrends-Fronsdal projection
 operator for space-time dimension $D = 4$. The generalization of the
Behrends-Fronsdal projection operator for any number of
space-time dimensions $D>2$ was found. We also
found the generalization of the explicit formula for the density matrix
(spin projection operator) of particles with half-integer spins.
The most interesting examples corresponding to spins $j=1/2,1,3/2$ and $j=2$ were discussed in detail.

We have to stress that the massless case can also be
considered in a similar manner.
Certain steps in this direction were made in \cite{IR}.
Just as in the massive case, the spin-tensor wave functions of free massless particles with arbitrary
helicity are constructed from the vectors of spaces of the unitary massless Wigner representations
for the covering group $ISL (2, \mathbb {C})$ of the Poincare group.

Moreover, formula (\ref{tp}) is carried over to the massless case practically unchanged
 (we need to remove the normalizing factor ${\sf m}^{-r}$
 and choose the test momentum as $q=(E,0,0,E)$). The
 corresponding spin-tensor wave functions satisfy the
 Penrose equations
(these equations for fields of massless
particles were formulated by Penrose in the coordinate representation; see \cite{PenR} and \cite{PenR3})
instead of the Dirac-Pauli-Fierz equations.
It is remarkable that
 instead of the two-spinor approach, which
  is suitable for the massive case and used in this paper,
we arrive in the massless case at the twistor formalism \cite{PenR}.

 We hope that the two-spinor formalism considered in this
 paper for describing massive particles
of arbitrary spin will be useful in the construction
 of scattering amplitudes of massive particles in a similar way to
   the construction of
 spinor-helicity scattering amplitudes
 for massless particles \cite{PaTa}, \cite{BeGi}, \cite{Witt} (see
 also \cite{ElHua} and references therein). Some steps in this direction
 have already been done in papers \cite{CoMa},\cite{CJM},\cite{Ma} where
 the analogous two-spinor formalism and its special generalization were used.

 \vspace{0.2cm}

 The authors are grateful to V.A.~Rubakov and S.A.~Fedoruk for numerous useful
  discussions. The work of API was supported by the Russian Science
  Foundation, grant 14-11-00598. The work of
  MAP was supported by the RFBR, grant 16-01-00562.

 \vspace{1cm}
\begin{otherlanguage}{english}

\end{otherlanguage}

\end{document}